%% file: main_arxiv.tex
\AtBeginDocument{%
	\providecommand\BibTeX{{%
			\normalfont B\kern-0.5em{\scshape i\kern-0.25em b}\kern-0.8em\TeX}}}

\documentclass[10pt,letterpaper]{article}
\usepackage[margin=1in]{geometry}
\usepackage{times}
\usepackage[round]{natbib}
\usepackage{xcolor}
\usepackage[compact]{titlesec}
\usepackage[T1]{fontenc}
\usepackage[utf8]{inputenc}
\usepackage{url}
\usepackage{amsmath,amsfonts}
\usepackage{algorithmic}
\usepackage{graphicx}
\usepackage{textcomp}
\usepackage{float}
\usepackage{booktabs}
\usepackage[compact]{titlesec}
\usepackage{enumitem}
\usepackage{dsfont}
\usepackage[bottom]{footmisc}
\usepackage[labelformat=simple]{subcaption}

\usepackage{xspace}
\usepackage{amsthm}
\usepackage{multirow}
\theoremstyle{definition}
\newtheorem{definition}{Definition}[]
\usepackage[ruled, vlined, linesnumbered]{algorithm2e} 
\usepackage{mathtools}
\usepackage{array}
\usepackage{lipsum}
\usepackage{tablefootnote}
\usepackage[export]{adjustbox}
\usepackage{footmisc}
\usepackage{cleveref}
\crefformat{footnote}{#2\footnotemark[#1]#3}
\usepackage{balance}
\usepackage{amsmath,amssymb,amsfonts}%
\usepackage{bm}
\newcommand\bolden[1]{{\boldmath\bfseries#1}}
\usepackage{contour}

\makeatletter
\newcommand{\AlignFootnote}[1]{%
	\ifmeasuring@
	\else
	\iffirstchoice@
	\footnote{#1}%
	\fi
	\fi}
\makeatother

\date{}

\begin{document}

\title{On Improving the Cohesiveness of Graphs by Merging Nodes: Formulation, Analysis, and Algorithms}

\author{Fanchen Bu\thanks{School of Electrical Engineering, KAIST, Daejeon, South Korea, boqvezen97@kaist.ac.kr} \	
	and Kijung Shin\thanks{Kim Jaechul Graduate School of AI and School of Electrical Engineering, KAIST, Seoul, South Korea, kijungs@kaist.ac.kr}}

\input{dfn.tex}

\maketitle

\input{000abs.tex}

\input{010intro.tex}

\input{0A0relwk.tex}
\input{020prelim.tex}
\input{030anlys.tex}

\input{040mthd.tex}
\input{050exper.tex}

\input{0B0concl.tex}

\bibliographystyle{ACM-Reference-Format}
\bibliography{BIB/ref}

\end{document}

%% file: dfn.tex
\definecolor{myred}{RGB}{195, 79, 82}
\definecolor{mygreen}{RGB}{86, 167 104}
\definecolor{myblue}{RGB}{74, 113 175}
\definecolor{pltblue}{RGB}{31, 119, 180}
\definecolor{pltorange}{RGB}{255, 127, 14}
\definecolor{pltgreen}{RGB}{44, 160, 44}
\newcolumntype{d}[1]{D..{#1}}
\newcommand\mc[1]{\multicolumn{1}{c}{#1}} 

\newcommand\crr[1]{{#1}}
\newcommand\red[1]{{#1}}
\newcommand\blue[1]{\textcolor{blue}{#1}}
\newcommand\orange[1]{\textcolor{orange}{#1}}
\newcommand\green[1]{\textcolor{green}{#1}}
\newcommand\myred[1]{\textcolor{myred}{#1}}
\newcommand\myblue[1]{\textcolor{NavyBlue}{#1}}
\newcommand\mygreen[1]{\textcolor{mygreen}{#1}}

\newcommand\pltblue[1]{\textcolor{pltblue}{#1}}
\newcommand\pltorange[1]{\textcolor{pltorange}{#1}}
\newcommand\pltgreen[1]{\textcolor{pltgreen}{#1}}

\newcommand{\smallsection}[1]{{\noindent {\bf{\underline{\smash{#1}}}}}}

\definecolor{peace}{RGB}{228, 26, 28}
\definecolor{love}{RGB}{55, 126, 184}
\definecolor{joy}{RGB}{77, 175, 74}
\definecolor{kindness}{RGB}{152, 78, 163}
\newcommand\love[1]{\textcolor{love}{#1}}
\newcommand\kijung[1]{\textcolor{peace}{[Kijung: #1]}}
\newcommand\fanchen[1]{\textcolor{love}{[Fanchen: #1]}}

\newcommand{\cmark}{\ding{51}}%
\newcommand{\xmark}{\ding{55}}%

\newcommand{\mybox}{%
    \collectbox{%
        \setlength{\fboxsep}{1pt}%
        \fbox{\BOXCONTENT}%
    }%
}


\newcommand{\Q}{\mathbb{Q}}
\newcommand{\R}{\mathbb{R}}
\newcommand{\N}{\mathbb{N}}
\newcommand{\Z}{\mathbb{Z}}
\newcommand{\E}{\mathbb{E}}
\newcommand{\F}{\mathcal{F}}
\newcommand{\ceil}[1]{\lceil #1 \rceil}
\newcommand{\set}[1]{\{#1\}}
\newcommand{\bred}[1]{{\color{red}\textbf{#1}}}
\newcommand{\textub}[1]{\underline{\textbf{#1}}}


\newtheorem{lem}{Lemma}
\renewcommand{\thelem}{\arabic{lem}}

\newtheorem{thm}{Theorem}
\renewcommand{\thethm}{\arabic{thm}}

\newtheorem{expl}{Example}
\renewcommand{\theexpl}{\arabic{expl}}

\newtheorem{rem}{Remark}
\renewcommand{\therem}{\arabic{rem}}

\newtheorem{pro}{Problem}
\renewcommand{\thepro}{\arabic{pro}}

\newtheorem{obs}{Observation}
\renewcommand{\theobs}{\arabic{obs}}

\newtheorem{app}{Application}
\renewcommand{\theapp}{\arabic{app}}

\newtheorem{prop}{Proposition}
\renewcommand{\theprop}{\arabic{prop}}

\newtheorem{cor}{Corollary}
\renewcommand{\thecor}{\arabic{cor}}

\newtheorem{claim}{Claim}
\renewcommand{\theclaim}{\arabic{claim}}










\makeatletter
\newenvironment{pf}[1][\proofname]{\par
  \pushQED{\qed}%
  \normalfont \topsep0\p@\relax
  \trivlist
  \item[\hskip\labelsep\itshape
  #1\@addpunct{.}]\ignorespaces
}{%
  \popQED\endtrivlist\@endpefalse
}
\makeatother

\newcommand{\argmax}{\mathop{\mathrm{arg\,max}}}
\newcommand{\argmin}{\mathop{\mathrm{arg\,min}}}
\newcommand{\setadd}[2]{${#1} \leftarrow {#1} \cup \set{#2}$}
\newcommand{\bus}[1]{\textbf{\underline{\smash{#1}}}}

\newcommand{\bmshort}{BATMAN\xspace}
\newcommand{\bmlong}{\bus{B}est-merger se\bus{A}rcher for \bus{T}russ \bus{MA}ximizatio\bus{N}\xspace}
\newcommand{\bmfull}{\textbf{BATMAN} (\bus{B}est-merger se\bus{A}rcher for \bus{T}russ \bus{MA}ximizatio\bus{N})\xspace}

\newcommand{\bolita}[1]{\bolden{\textit{#1}}}

%% file: 000abs.tex
\begin{abstract}
Graphs are a powerful mathematical model, and they are used to represent real-world structures in various fields. In many applications, real-world structures with high connectivity and robustness are preferable. For enhancing the connectivity and robustness of graphs, two operations, adding edges and anchoring nodes, have been extensively studied. However, merging nodes, which is a realistic operation in many scenarios (e.g., bus station reorganization, multiple team formation), has been overlooked. In this work, we study the problem of improving graph cohesiveness by merging nodes. First, we formulate the problem mathematically using the size of the $k$-truss, for a given $k$, as the objective. Then, we prove the NP-hardness and non-modularity of the problem. After that, we develop \bmshort, a fast and effective algorithm for choosing sets of nodes to be merged, based on our theoretical findings and empirical observations. Lastly, we demonstrate the superiority of \bmshort over several baselines, in terms of speed and effectiveness, through extensive experiments on fourteen real-world graphs.
\end{abstract}

%% file: 010intro.tex
\section{Introduction}\label{sec:intro}
As a powerful mathematical model, graphs have been widely used in various fields to represent real-world structures.
Some typical applications of graphs are recommendation systems \citep{silva2010graph},
social network analysis \citep{scott1988social},
and biological system analysis on molecular graphs \citep{manolopoulos1992molecular} and protein-protein interactions \citep{brohee2006evaluation}.
Moreover, many optimization problems on real-world structures have been formulated as ones on the abstracted graphs.

In many real-world applications, it is desirable to have a well-connected and robust structure.
For example, in transportation systems, it is preferable that stations are connected tightly with each other so that traffic routes are resilient even if some accidents happen~\citep{jin2014enhancing, zhou2019resilience};
in organizations like companies, often several interconnected projects or tasks are carried out at the same time, and thus several teams are supposed to form dense and highly-connected communities in an underlying graph so that the teams can closely collaborate with each other~\citep{gutierrez2016multiple,baghel2018multiple,addanki2020multi}.

A straightforward operation to enhance the connectivity and robustness of graph structures is adding edges~\citep{beygelzimer2005improving, sun2021budget}.
Besides, anchoring nodes (i.e., forcefully including some nodes in a cohesive subgraph)~\citep{bhawalkar2015preventing, zhang2017finding, zhang2018finding, zhang2018efficiently, laishram2020residual, linghu2020global} has also been widely studied.

However, merging nodes,  which is another realistic operation in many applications, has been overlooked.
Merging nodes, or formally \textit{vertex identification} \citep{oxley2006matroid}, is the operation where we merge two nodes into one, and any other node adjacent to either of the two nodes will be adjacent to the ``new'' node.
Merging nodes may strike you as too radical at first sight, but it is indeed a very realistic and helpful operation in several real-world examples such as:
\begin{enumerate}[leftmargin=*]
    \item \textbf{Bus station reorganization.} Merging some nearby stations not only makes traffic networks more compact and systematic but also reduces maintenance expenses since the total number of stations is reduced~\citep{wei2020using}.
    \red{For example, CTtransit, a bus-system company in the united states, proposed to merge multiple bus stations in New Haven and discussed the benefits~\citep{cttransit_2010}.}
    \item \textbf{Multiple Team formation.} Forming teams (i.e., ``merging'' individuals) within an organization can increase individual performance and cultivate a collaborative environment \citep{chhabra2013team}.
    How to form well-performing and synergic teams is an important research topic~\citep{gutierrez2016multiple,baghel2018multiple,addanki2020multi} 
    in business and management~\citep{moreland2002training,kozlowski2013work}.
\end{enumerate}

In this paper, we study the problem of improving the connectivity and robustness of graphs by merging nodes.
To the best of our knowledge, we are the first who study this problem.
We propose to use the size (spec., the number of edges) of a $k$-truss \citep{cohen2008trusses} as the objective quantifying the connectivity and robustness.
Given a graph $G$ and an integer $k$, the $k$-truss of $G$ is the maximal subgraph of $G$ where each edge is in at least $k-2$ triangles;
and we say that an edge has trussness $k$ if the edge is in the $k$-truss but not the $(k+1)$-truss.
Specifically, $k$-trusses have the following merits:
\begin{enumerate}[leftmargin=*]
    \item \textbf{Cohesiveness.} $k$-Trusses require both engagements of the nodes and interrelatedness of the edges compared to some other cohesive subgraph models. Specifically, given any graph, a $k$-truss is always a subgraph of the $(k - 1)$-core~\citep{seidman1983network} but not vice versa, and each connected component of a $k$-truss is $(k-1)$-edge-connected~\citep{jordan1869assemblages, cai1989minimum} with bounded diameter \citep{huang2014querying}.
    \item \textbf{Computational efficiency.} $k$-Trusses can be computed efficiently with time complexity $O(m^{1.5})$~\citep{wang2012truss}, where $m$ is the number of edges; in contrary, given a graph, enumerating all the cliques or many variants ($n$-cliques~\citep{luce1950connectivity}, $k$-plexes~\citep{seidman1978graph}, $n$-clans and $n$-clubs~\citep{mokken1979cliques}) is NP-hard.
    \item \textbf{Applicability.} $k$-Trusses, especially their sizes, ably capture connectivity and robustness of 
    transportation \citep{diop2020local},
    social networks \citep{zhu2019pivotal, zhao2020community}, communication \citep{ghalmane2018k}, 
    and recommendation \citep{yang2022k}.
    Specifically, $k$-trusses also have realistic meanings in the two aforementioned real-world examples (bus station~\citep{zhu2022higher,derrible2010complexity} reorganization and multiple team formation~\citep{brewer2016better,durak2012degree}).
\end{enumerate}
Due to the desirable theoretical properties and practical meaningfulness of $k$-trusses, several existing works \citep{zhang2020finding, chen2021breaking, chen2022locating, zhu2019pivotal, sun2021budget} used the size of a $k$-truss as the objective.

Therefore, we consider the problem of maximizing the size of a $k$-truss in a given graph by merging nodes.
In Figures~\ref{fig:effective_merge} and \ref{fig:effective_truss}, we show the effectiveness of merging nodes (spec., its superiority over adding edges) and maximizing the size of a $k$-truss (spec., the correlations between the truss size and various robustness measures), respectively
(see Section~\ref{subsec:effective_study} for more details).
We mathematically formulate the problem as an optimization problem on graphs named \textbf{TIMBER} (\bus{T}russ-s\bus{I}ze \bus{M}aximization \bus{B}y m\bus{ER}gers), and prove the NP-hardness and non-modularity of the problem.

\input{FIG/effective_truss_1.tex}

\input{FIG/effective_truss_2.tex}

For the TIMBER problem, we develop \bmshort (\bmlong), a fast and effective algorithm equipped with
\textbf{(1)} search-space pruning based on our theoretical analysis, and 
\textbf{(2)} simple yet powerful heuristics for choosing promising mergers.
Starting from a computationally prohibitive naive greedy algorithm, we theoretically analyze the changes on a graph after mergers and use the findings to design speed-improving heuristics.
For example, we prove that after merging two nodes, the trussness of an edge that is not incident to either of the merged nodes changes by at most one. Hence, we only need to consider the edges with original trussness at least $k - 1$ for an input $k$.
We first reduce the search space by
\textbf{(1)} losslessly pruning the space of \textit{outside nodes} (nodes that are not in the $(k-1)$-truss) using a maximal-set-based algorithm,
\textbf{(2)} proposing and using a new heuristic to efficiently find promising \textit{inside nodes} (nodes that are in the $(k-1)$-truss), and
\textbf{(3)} excluding the mergers of two outside nodes with the rationality of doing so.
Our fast and effective heuristics for finding promising pairs among the selected nodes are based on the number of edges with trussness $k - 1$ gaining (and losing) support.

Through extensive experiments on 14 real-world graphs, we compare our proposed algorithm, \bmshort, to several baseline methods and show that \bmshort consistently performs best w.r.t the final increase in the size of $k$-trusses, 
achieving $1.38 \times$ to $10.08 \times$ performance superiority over the baseline methods on all the datasets.

In short, our contributions are four-fold:
\begin{enumerate}[leftmargin=*]
    \item \textbf{A novel Problem:} We introduce and formulate TIMBER
     (Problem~\ref{prob:best_mergers}), a novel optimization problem on graphs with several potential real-world applications, as listed above.    
    \item \textbf{Theoretical Analysis:} We prove the NP-hardness (Theorem~\ref{thm:NP_hard}) and non-submodularity (Theorem~\ref{thm:non_mod}) of TIMBER.
    \item \textbf{A fast Algorithm:} We design \bmshort (Algorithm~\ref{alg:overall}), a fast and effective algorithm for TIMBER, based on our theoretical (Lemmas~\ref{lem:limited_truss_change}-\ref{prop:no_inside_no_help}) and empirical findings (Section~\ref{subsec:exp_empirical_support}). We also theoretically analyze the time complexity of \bmshort (Theorem~\ref{thm:overall_time}).
    \item \textbf{Extensive Experiments:} We compare \bmshort with several baseline methods and demonstrate the advantages of \bmshort and its components using 14 real-world graphs (Section~\ref{sec:exp}).
\end{enumerate}
For \textbf{reproducibility}, the code and datasets are available online \citep{onlineSuppl}.\footnote{\url{https://github.com/bokveizen/cohesive-truss-merge}}

\section{On real-world examples}\label{sec:on_rw_examples}
In this section, we provide more discussions on the real-world examples (bus station reorganization and multiple team formation) used in this work.
Specifically, we provide more details on how both merging nodes and $k$-trusses have realistic meanings.

\subsection{Bus station reorganization}
Regarding bus station reorganization (or transportation systems in general), we can consider the following specific real-world scenario: we are managing the bus transportation system of a city, and we want to reduce the total number of bus stations to decrease the expenses of maintaining the bus stations due to some financial reasons. 
We model the transportation system as a graph with bus stations as nodes and routes as edges, and we aim to do so by merging bus stations while maximizing connectivity among the stations.
As a measure of (higher-order) connectivity~\citep{huang2014querying,yin2017local,chang2019cohesive}, the size of a $k$-truss is a reasonable choice as a numerical metric for this purpose. Moreover, k-trusses are highly related to triangles, while triangles are important motifs for indicating higher-order connectivity~\citep{zhu2022higher,zhang2019wavelet} and robustness~\citep{derrible2010complexity,eraso2021evolution,ping2006topological} in transportation/traffic systems. 
In such a scenario, our proposed algorithm, which aims to maximize the k-truss size by merging nodes, can be used as a solution for finding stations to merge.

\subsection{Multiple team formation}
Regarding multiple team formation, we can consider the following specific real-world example: as the CEO of a company, we have employees (nodes) and social relations (edges) between them (which corresponds to the input graph), and we want to form multiple small-scale teams (merge nodes) among the employees. 
We aim to maximize the communication between teams (which is known to be beneficial to teams’ performance~\citep{brewer2016better,hillier2013groups,macht2014structural}, where we suppose that two teams A and B can communicate well with each other if at least one member in A and one member in B have social connections (i.e., communication is correlated with connectivity). 
In general, when we merge nodes into teams, the formed teams constitute a social network, where cohesive subgraphs such as $k$-trusses are indicative of high connectivity and robustness~\citep{wang2012truss}. Moreover, the abundance of triangles and the bounded diameter (i.e., teams can reach each other within a bounded number of hops) indicated by $k$-trusses are both helpful for better communication between nodes (teams)~\citep{durak2012degree}. 
In such a scenario, our proposed algorithm, which aims to maximize the $k$-truss size by merging nodes, can be used as a solution for forming teams.

\subsection{Limitations and more discussions}
Definitely, in real-world scenarios, more conditions and factors might be considered, and we would like to emphasize that we are considering a more general problem, while additional real-world constraints can be considered ad hoc in practical usage. 
For example, for the bus station reorganization application, where we consider the constraints that only bus stations within a distance threshold can be merged (and indeed we have such information), when we choose candidate pairs (in Algorithms~\ref{alg:iom_heu} and \ref{alg:iim_heu}), for each candidate merger, we can simply check the distance between the two nodes (stations), and include the merger in the final returned set of candidates only if the distance is within the distance threshold.\footnote{See Section~\ref{app:add_exp_bus} for related experiments on real-world bus station datasets, where we take distance constraints into consideration.}

%% file: FIG/effective_truss_1.tex
\begin{figure}[t!]
    \centering
    \includegraphics[scale=0.24]{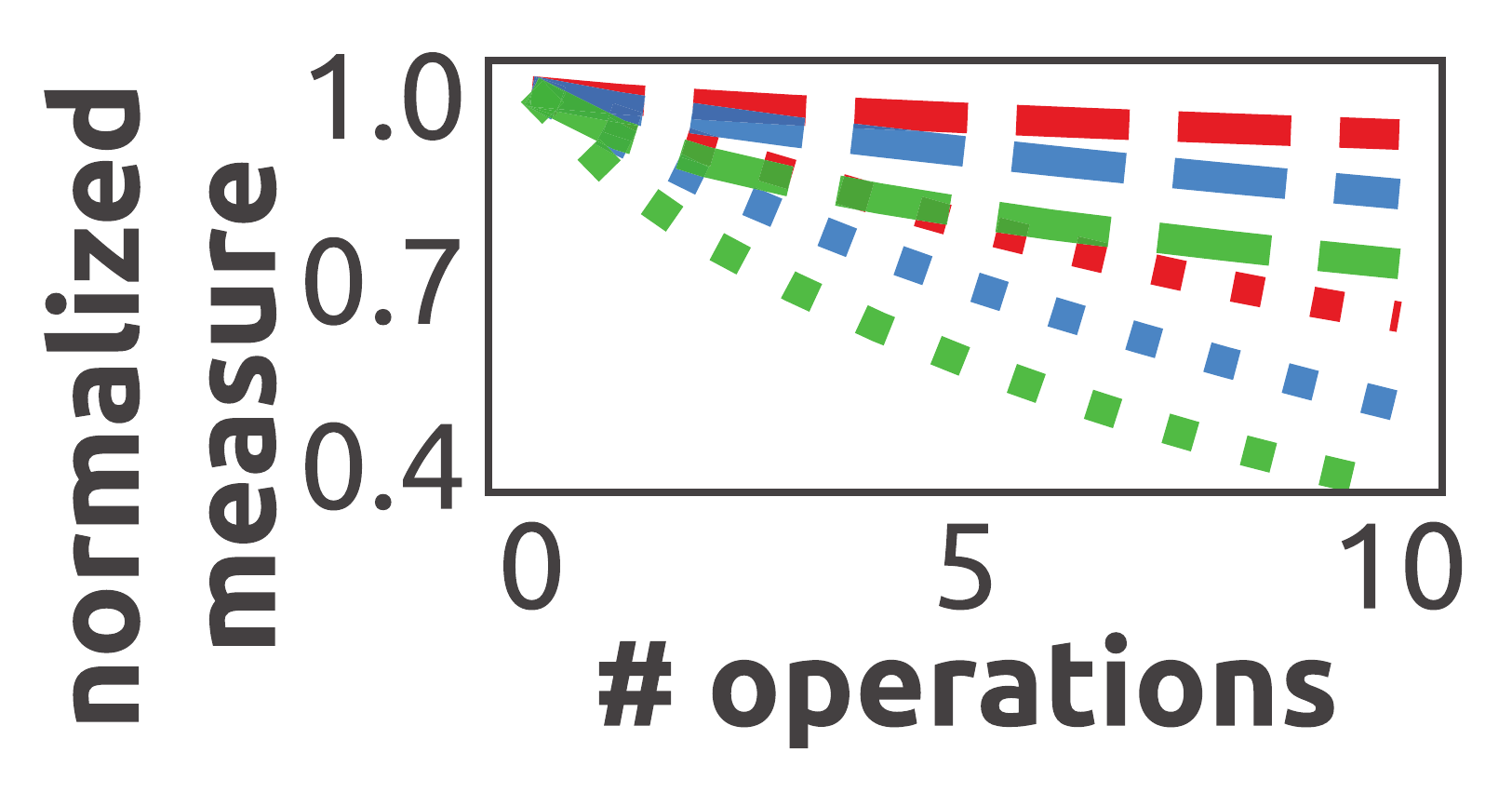}
    \includegraphics[scale=0.24]{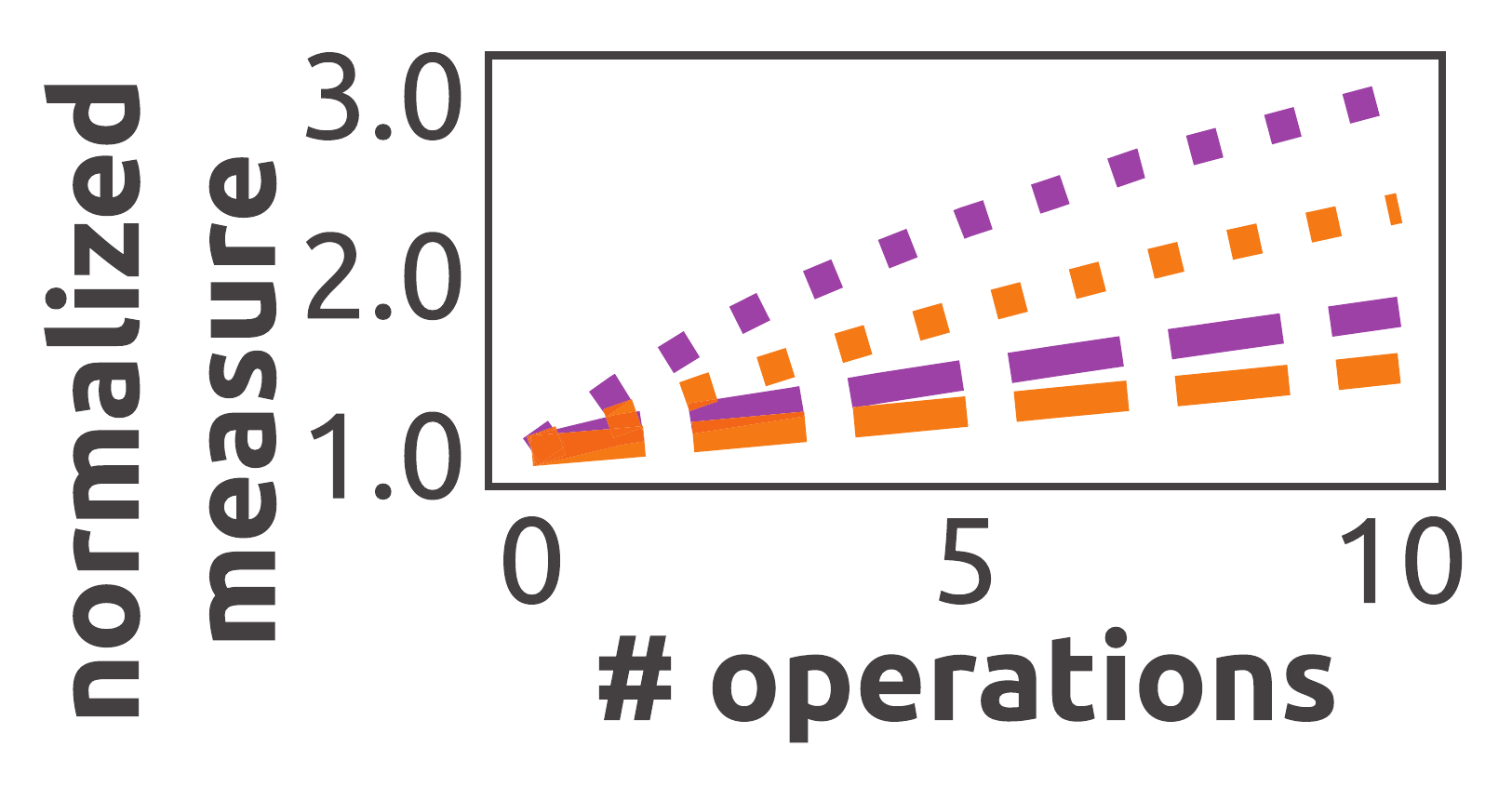}    
    \includegraphics[scale=0.45]{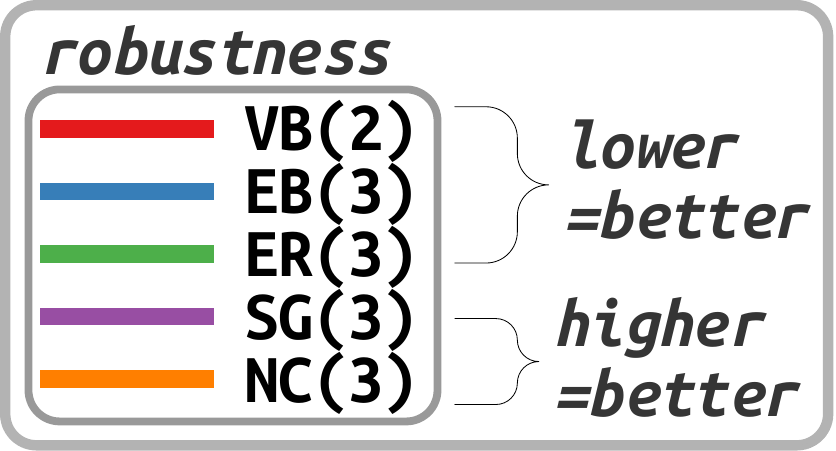}
    \caption{{Merging nodes is much more effective than adding edges in enhancing graph robustness.}
    \red{For each robustness measure, we do 10 rounds of merging nodes (dotted) or adding edges (dashed).
    In each round, we greedily choose the node pair or edge that improves the measure most.
    In the legend, we include the minimum times (\red{$\leq$ 3} for all measures) of merging nodes that are needed to achieve a better improvement achieved by adding 10 edges.}
    \red{See Section~\ref{subsec:effective_study} for the details.}}
    \vspace{-1mm}
    \label{fig:effective_merge}
\end{figure}

%% file: FIG/effective_truss_2.tex
\begin{figure}[t!]
    \centering
    \includegraphics[scale=0.24]{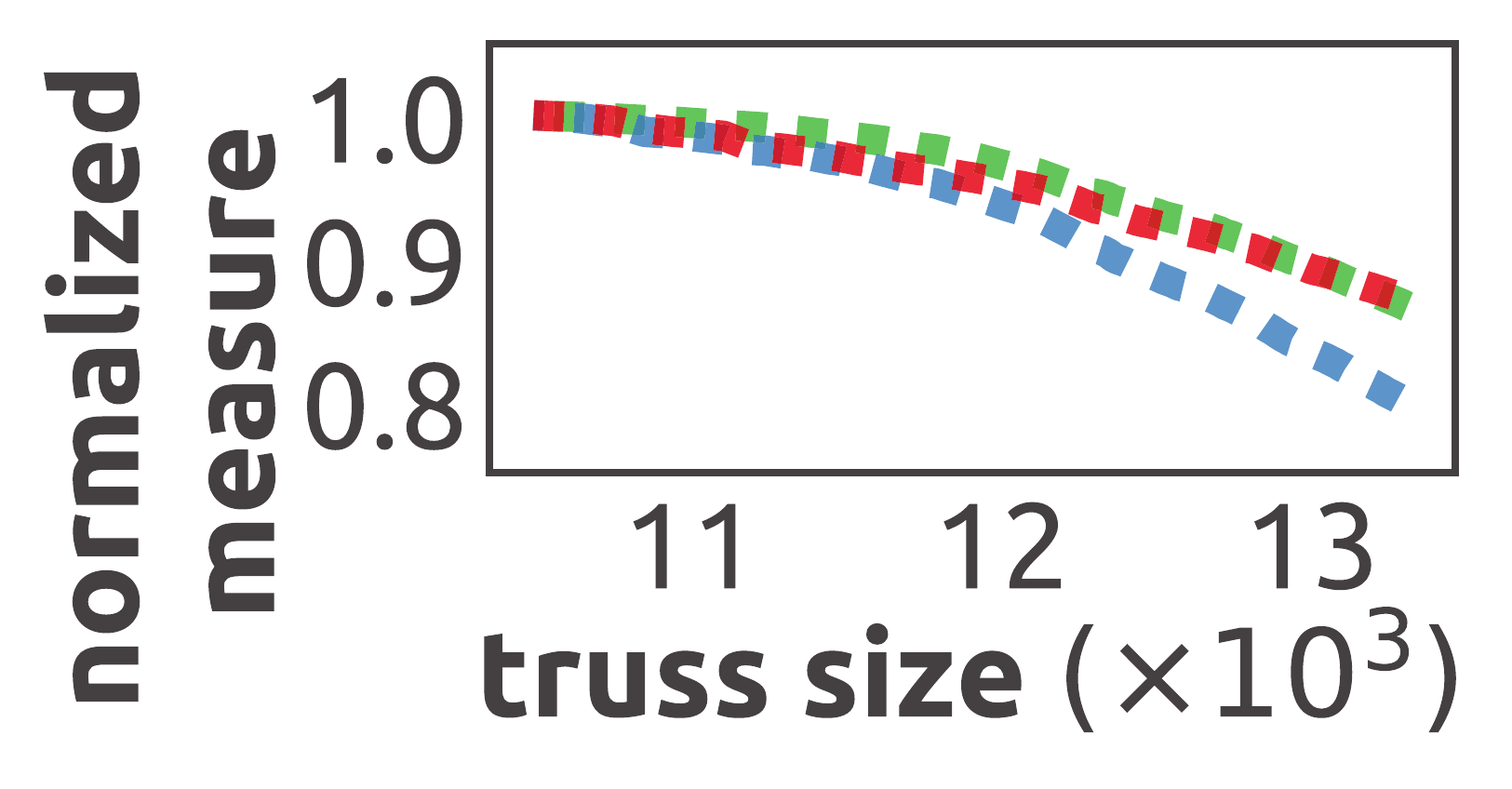}
    \includegraphics[scale=0.24]{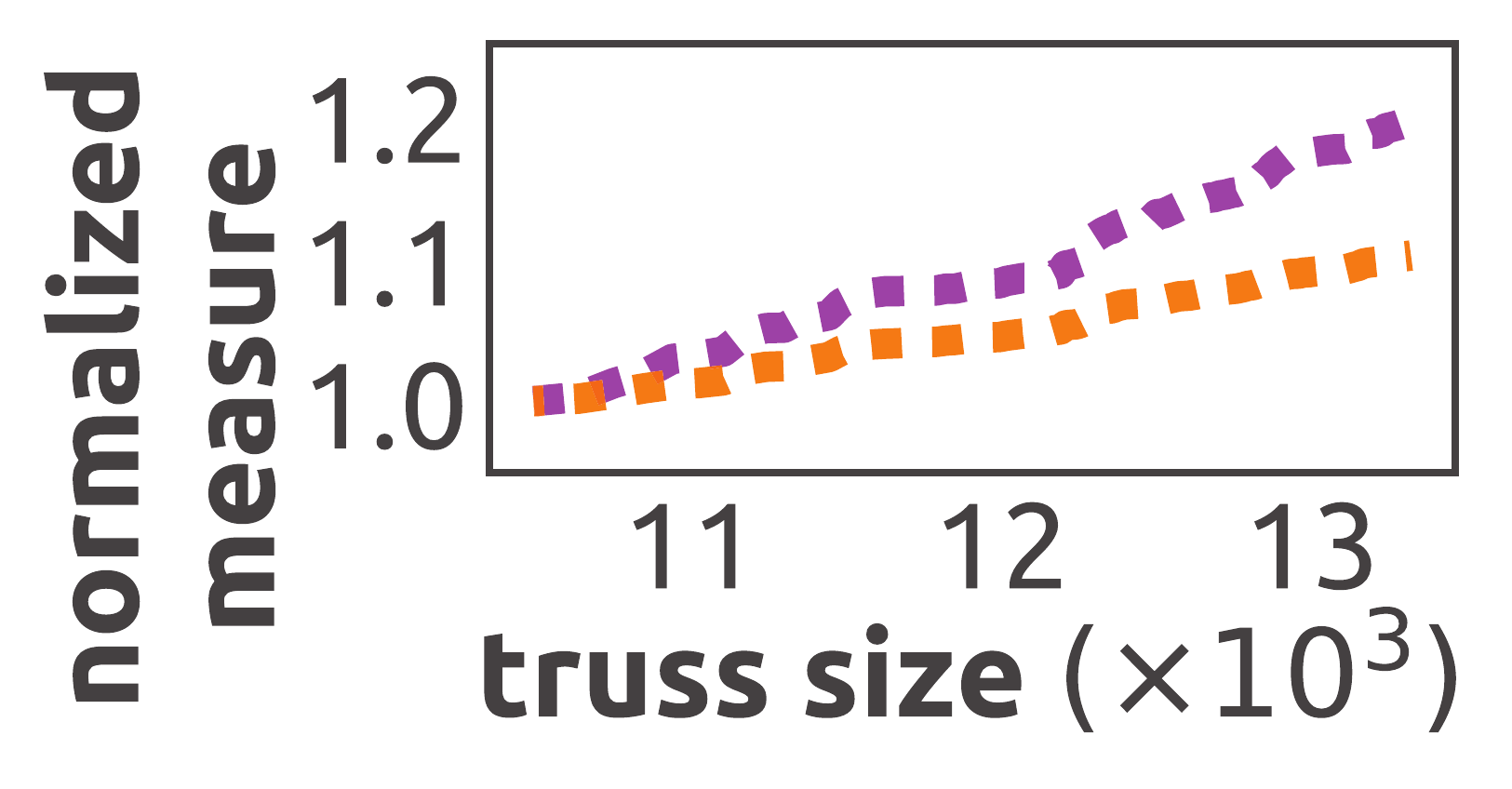}    
    \includegraphics[scale=0.45]{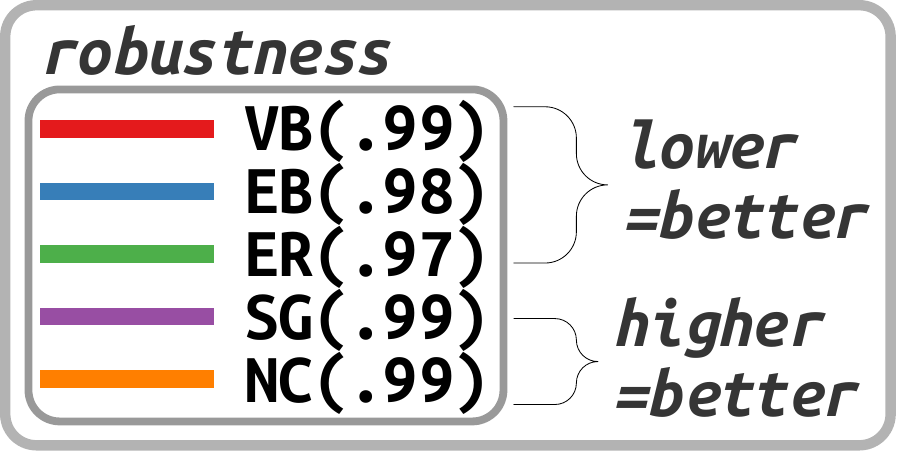}
    \caption{{Maximizing the size of a $k$-truss is effective: graph robustness improves when we enlarge a $k$-truss.}
    For each robustness measure, we report the relation between it and the truss size along the process of enlarging a $k$-truss by merging nodes using our proposed method \bmshort.
    We include the absolute value (0.97-0.99 for all measures) of Pearson's $r$ in the legend.    
    \red{See Section~\ref{subsec:effective_study} for the details.}}
    \vspace{-1mm}
    \label{fig:effective_truss}
\end{figure}

%% file: 0A0relwk.tex
\input{TAB/notation.tex}
\section{Related Work}\label{sec:relwk}

\smallsection{$k$-Trusses.}
Based on the concept of $k$-cores \cite{seidman1983network}, the concept of $k$-trusses was introduced by \cite{cohen2008trusses}.
\cite{wang2012truss} proposed an efficient truss decomposition algorithm with time complexity $O(m^{1.5})$, where $m$ is the number of edges in the input graph.
\cite{huang2014querying} used $k$-trusses to model the communities in graphs (see also~\citep{akbas2017truss}) and studied the update of $k$-trusses in dynamical graphs (see also~\citep{zhang2019unboundedness, luo2020batch}).
Related problems are also studied for weighted graphs~\citep{zheng2017finding},
signed graphs~\citep{zhao2020community}, directed graphs~\citep{liu2020truss},
uncertain graphs~\citep{huang2016truss, sun2021efficient},
and simplicial complexes~\citep{preti2021strud}.
In \citep{chen2021higher}, higher-order neighbors are considered to generalize the concept of $k$-trusses.

\smallsection{Graph structure enhancement and attacks.}
Several studies of graph structure enhancement or attacks are conducted based on cohesive subgraph models.
Specifically, the problems of maximizing the size of a $k$-truss by anchoring nodes~\citep{zhang2018efficiently, zhang2018finding} and by adding edges \citep{sun2021budget, chen2022locating} have been studied;
and the opposite direction, i.e., minimizing the size of a $k$-truss, has also been considered~\citep{chen2021breaking, chen2022locating}.
There are also a series of counterparts considering the model of $k$-cores~\citep{bhawalkar2015preventing, zhang2017finding, zhang2017olak, zhu2018k, zhang2018efficiently, liu2021efficient, zhou2021vek, linghu2020global, linghu2022anchored, laishram2020residual, medya2020game, zhang2020finding, zhao2021finding, sun2022fast} using the operations of adding (or deleting) edges (or nodes) and anchoring nodes.
However, no existing work studies graph structure enhancement or attacks by merging nodes, while merging nodes is indeed a basic operation on graphs~\citep{oxley2006matroid} and practically meaningful.

\smallsection{Related work on real-world examples.}
For designing and optimizing traffic networks, many tools and methods have been used, e.g., mixed-integer programming~\citep{jin2014enhancing}, linear programming~\citep{liang2019bus}, and genetic algorithm~\citep{cao2022optimization,li2019resilience}.
The multiple team formation problems have been widely studied in the field of operations research, where many methods, such as variable neighborhood local search metaheuristics~\citep{gutierrez2016multiple} integer programming~\citep{campelo2021integer}, and evolutionary algorithm~\citep{baghel2018multiple}, have been used. In this work, we study the problems from the perspective of social networks.

%% file: TAB/notation.tex
\begin{table}[t!]
	\begin{center}
		\caption{Notations.}\label{tab:notations}
            {
            \begin{adjustbox}{max width=\linewidth}
			\begin{tabular}{p{0.3\linewidth}  p{0.8\linewidth}}
				\toprule
				\textbf{Notation}       & \textbf{Definition}\\
				\midrule
				$G = (V, E)$            & a graph with node set $V$ and edge set $E$ \\
				$N(v; G)$               & the set of neighbors of $v\in V$ \\
				$d(v; G)$               & the degree of $v\in V$ \\
				$G[V']$                 & the induced subgraph of $G$ on $V'\subseteq V$ \\
				$s(e; G)$               & the support of $e\in E$ \\
                $T_k(G)$ & the $k$-truss of $G$ \\
				$t(e; G)$, $t(v; G)$             & the trussness of $e\in E$ and $v\in V$\\
                $\hat{E}_k(G)$ & the shell edges {with trussness $k$}, i.e., $E(T_{k-1}) \setminus E(T_k)$ \\
				$PM(v_1, v_2; G)$       & the graph after merging $v_1$ and $v_2\in V$ into $v_1$ in $G$ \\
                $\Tilde{N}_k(v;G)$ & the inside neighbors {of $v\in V$}, i.e., $N(v) \cap V(T_{k-1})$ \\ 
				\bottomrule
			\end{tabular}
            \end{adjustbox}
			}
	\end{center}
\end{table}

%% file: 020prelim.tex
\section{Preliminaries}\label{sec:prelim}
Let $G = (V, E)$ be an unweighted, undirected graph without self-loops, multiple edges, or isolated nodes.
Let $\N$ denote the set of positive integers and let $\setminus$ denote the set subtraction operation.
We call $V = V(G)$ the \textit{node set} of $G$ and $E = E(G) \subseteq \binom{V}{2}$ the \textit{edge set} of $G$.
Each edge $e = (v_1, v_2) = (v_2, v_1) \in E$ {joins} two nodes $v_1$ and $v_2$ and is treated as a 2-set without order.
The set $N(v; G)$ of \textit{neighbors} of a node $v$ consists of the nodes {adjacent to} $v$, i.e., $N(v; G) = \set{v' \in V: (v, v') \in E}$;
and the \textit{degree} $d(v; G)$ of $v$ in $G$ is the number of neighbors of $v$, i.e., $d(v; G) = |N(v; G)|$.
Given a subset $V' \subseteq V$ of nodes, the \textit{induced subgraph} $G[V'] = (V', E')$ of $G$ induced on $V'$ is defined by $E' = \set{e \in E: e \subseteq V'}$, and we use $G \setminus V'$ to denote the graph $G[V \setminus V']$\footnote{We use $\setminus$ to denote the set subtraction operation.} obtained by removing the nodes in $V'$ from $G$.
The \textit{support} $s(e; G)$ of $e = (v_1, v_2)$ is the number of triangles that contain both $v_1$ and $v_2$.

\begin{definition}[$k$-truss and trussness]
    Given a graph $G = (V, E)$ and $k \in \N$,\footnote{We use $\N$ to denote the set $\set{1, 2, 3, \ldots}$ of positive integers.} the \bolita{$k$-truss} of $G$, denoted by $T_k = T_k(G)$, is the maximal subgraph of $G$ where each edge in $T_k$ has support at least $k - 2$ {within $T_k$}, i.e., $s(e; T_k) \geq k - 2, \forall e \in E(T_k)$.\footnote{$k$-Trusses are meaningful only when $k \geq 3$ since otherwise the $k$-truss is just the whole graph.    
    In this paper, we assume that $k \geq 3$ without further clarification.}
    We call the number $|E(T_k)|$ of edges in $T_k$ its \bolita{size}.
    The \bolita{trussness} $t(e; G)$ of an edge $e$ (w.r.t $G$) is the largest $k$ such that $e$ is in $T_k(G)$, i.e., $t(e; G) = \max \set{k \in \N: e \in E(T_k(G))}$.
    The trussness $t(v; G)$ of a node $v$ is the largest trussness among the trussness of all the edges containing {(i.e. incident to)} $v$, i.e., $t(v; G) = \max \set{t(e; G): v \in e}$.
\end{definition}
 For example, in Figure~\ref{fig:example_1}, in the original graph in the middle, the degree of the node $vb$ is $6$, the support of the edge $(va, vb)$ is $3$, the $5$-truss is the subgraph {formed by} the five nodes {($va$, $vb$, $vc$, $vd$, and $ve$)} and the ten edges {between} them (the size is $10$), and the trussness of each edge is explicitly demonstrated.

\input{FIG/example_1.tex}

In this paper, merging two nodes in a graph means identifying the two nodes \citep{oxley2006matroid} into one node, as described in Definition~\ref{defn:mergers}.
Any other node adjacent to either of the two nodes will be connected to the merged node, without adding any self-loop or multi-edge.
Since we focus on simple graphs, we neither add a multi-edge even if some node is adjacent to both pre-merger nodes, nor add a self-loop even if the two pre-merger nodes are adjacent to each other.
\begin{definition}[mergers] \label{defn:mergers}
    Given a graph $G = (V, E)$ and two nodes $v_1, v_2 \in V$.
    If we merge $v_1$ and $v_2$ into $v_1$ in $G$, then the \bolita{post-merger graph} $PM(v_1, v_2; G) = (V', E')$ after the \bolita{merger} between $v_1$ and $v_2$ is defined by $V' = V \setminus \set{v_2}$ and 
    $E'$ derives from $E$ by ``shifting'' the edges incident to $v_2$ to $v_1$ \textit{without adding multiple edges or self-loops},
    i.e., $E' = E \cup \set{(v_1, u): u \in N(v_2), u \neq v_1} \setminus \set{(v_2, u): u \in N(v_2)}$. 
    We use $PM(P; G)$ to denote the post-merger graph when we merge multiple pairs in $P$ in $G$ (note that the order does not matter).    
\end{definition}
Recall the example in Figure~\ref{fig:example_1}. Let $G_o$ denote the original graph in the middle, then the two post-merger graphs on the left and right are $PM(vd, vb; G)$ and $PM(vf, vz; G)$, respectively.

We summarize the notations in Table~\ref{tab:notations}.
In the notations, the input graph $G$ can be omitted when the context is clear.

%% file: FIG/example_1.tex
\begin{figure}[t!]
    \centering
    \includegraphics[scale=0.34]{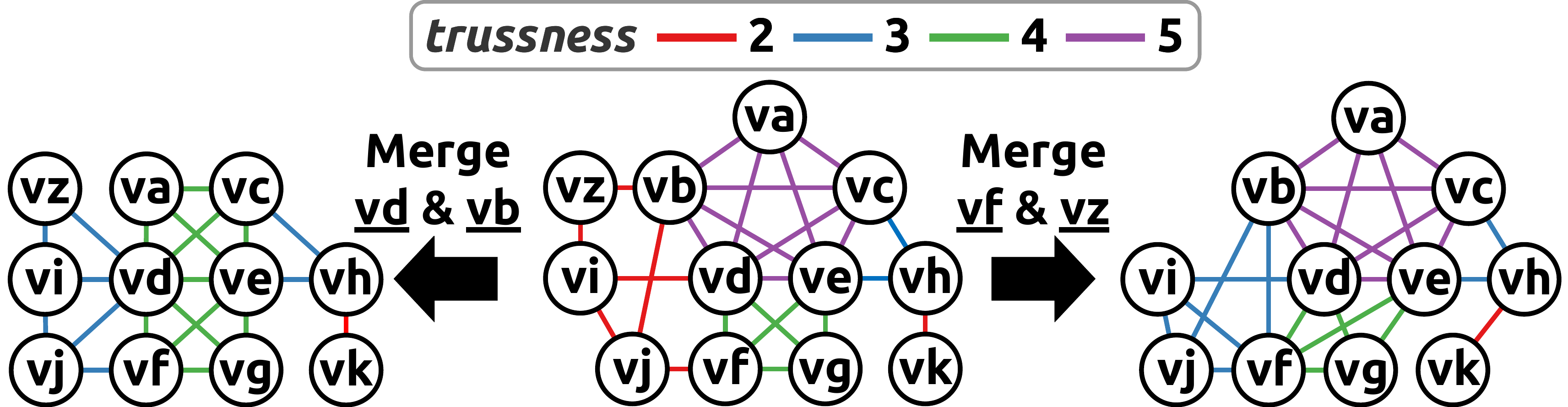}
    \caption{{Two} different ways of merging a pair of nodes in a graph with {different} consequences.}
    \label{fig:example_1}
\end{figure}

%% file: 030anlys.tex
\section{Problem Statement and Hardness}\label{sec:anlys}
\label{subsec:problem_statement}
In this section, we give the formal problem statement and analyze the theoretical hardness of our problem.

\subsection{Problem statement}
\begin{pro} \label{prob:best_mergers} \textsc{\textnormal{(TIMBER: \bus{T}russ-s\bus{I}ze \bus{M}aximization \bus{B}y m\bus{ER}gers)}} 
\begin{itemize}[leftmargin=*]
        \item \textbf{Given:} a graph $G = (V, E)$, $k\in \N$, and $b \in \N$,
        \item \textbf{Find:} a set $P$ of up to $b$ node mergers   in $G$, i.e., $P \subseteq \binom{V}{2}$ and $|P| \leq b$,
        \item \textbf{to Maximize:} the size of the $k$-truss after the mergers, i.e., $$f(P) = f(P; G, k)=|E(T_k(PM(P;G)))|.$$
\end{itemize}
\end{pro}
As mentioned before, for the example in Figure~\ref{fig:example_1}, merging $vf$ and $vz$ maximizes the size of the $3$-truss, i.e., with the original graph in Figure~\ref{fig:example_1}, $k = 3$, and $b = 1$ as the inputs, {$P=\set{(vf, vz)}$ is the solution that maximizes our objective function $f(P)=f(P;G,k)$.}

\subsection{On the counterpart problem using $k$-cores}\label{subsec:on_counterpart_k_core}
\begin{figure}[t!]
	\centering
	\includegraphics[scale=0.4]{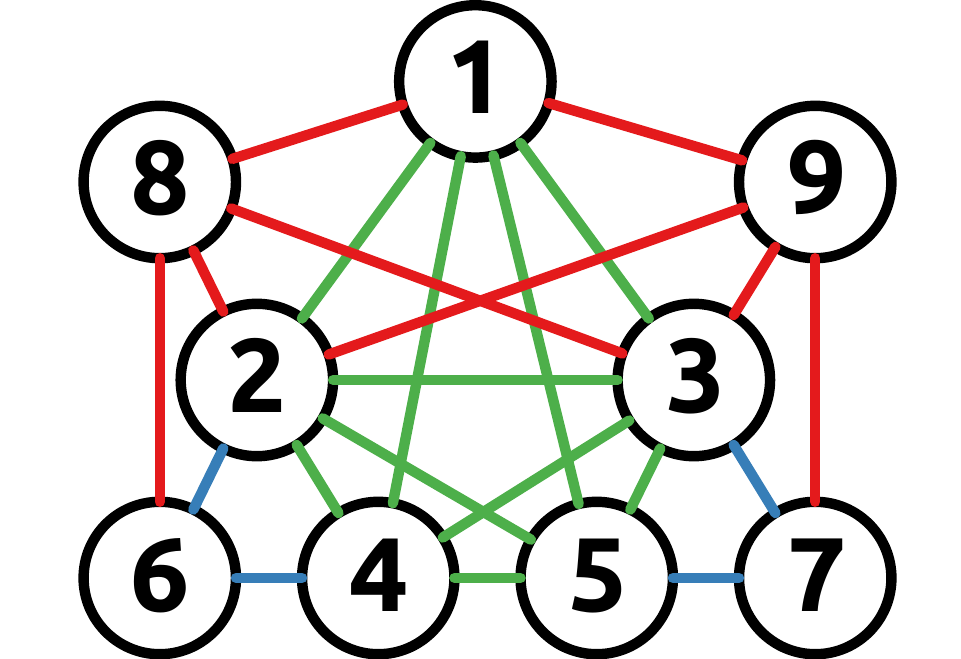}
	\caption{An example for the counterpart problem.}
	\label{fig:example_app_kcore}
\end{figure}

We would like to discuss the counterpart problem using $k$-cores and analyze the technical similarity between this problem and the anchored $k$-core problem~\citep{bhawalkar2015preventing}.
The counterpart problem using $k$-cores is defined below.
Note that the size of a $k$-core is usually defined as the number of nodes in the $k$-core.
\begin{pro} \label{pro:best_merger_core} \textsc{\textnormal{(The counterpart problem of TIMBER using $k$-cores)}}
\begin{itemize}[leftmargin=*]
	\item \textbf{Given:} a graph $G = (V, E)$, $k\in \N$, and $b \in \N$,
	\item \textbf{Find:} a set $P$ of up to $b$ node mergers in $G$, i.e., $P \subseteq \binom{V}{2}$ and $|P| \leq b$,
	\item \textbf{to Maximize:} the size of the $k$-core after the mergers, i.e., $$f(P) = f(P; G, k)=|V(C_k(PM(P;G)))|,$$
	where $C_k(G_0)$ is the $k$-core of a graph $G_0$.
\end{itemize}
\end{pro}

We also provide the problem statement of the anchored $k$-core problem here for the sake of completeness.
We first define the anchored $k$-core.
\begin{definition}[anchored $k$-cores]
Given $G = (V, E)$, $k \in \N$ and a set $A \subseteq V$ of anchors,
the anchored $k$-core of $G$ w.r.t the anchor set $A$ is the maximum subgraph
$\tilde{C}_k(G; A) = (V', E')$ of $G$ where $A \subseteq V'$ and $d(v'; \tilde{C}_k(G; A)) \geq k, \forall v' \in V' \setminus A$.
\end{definition}

\begin{pro} \label{pro:anchord_core} \textsc{\textnormal{(The anchored $k$-core problem)}}
\begin{itemize}[leftmargin=*]
	\item \textbf{Given:} a graph $G = (V, E)$, $k\in \N$, and $b \in \N$,
	\item \textbf{Find:} a set $A$ of up to $b$ nodes in $G$, i.e., $A \subseteq V$ and $|A| \leq b$,
	\item \textbf{to Maximize:} the size of the $k$-core after anchoring the chosen nodes in $A$, i.e., $$f(A) = f(A; G, k)=|V(\tilde{C}_k(G;A))|.$$
\end{itemize}
\end{pro}

We claim the technical similarity between the counterpart problem of TIMBER using $k$-cores and the anchored $k$-core problem, stated as follows.
\begin{claim}
Problem~\ref{pro:best_merger_core} and Problem~\ref{pro:anchord_core} are technically similar.
Specifically, merging two nodes in Problem~\ref{pro:best_merger_core} is similar to anchoring both of the nodes in Problem~\ref{pro:anchord_core}.
\end{claim}
Formally, given $G = (V, E)$ and $k \in \N$, if two nodes $v_1$ and $v_2$ are not in the current $k$-core and have no common neighbor in the current $(k-1)$- and $(k-2)$-shell,\footnote{A weaker but still sufficient condition is that $|V(\tilde{C}_k(PM(\set{(v_1, v_2)};G); \set{v_1}))| = |V(\tilde{C}_k(PM(\set{(v_1, v_2)};G); \set{v_1}))| - 1$, i.e., no node other than $v_1$ and $v_2$ in the anchored $k$-core after anchoring $v_1$ and $v_2$ has exactly degree $k$ and are adjacent to both $v_1$ and $v_2$.}
and after the merger between them, $v_1$ is the new $k$-core, then $|V(\tilde{C}_k(G;\set{v_1, v_2}))| = |V(C_k(PM(\set{(v_1, v_2)};G)))| + 1$, where the difference of one node comes from the merger itself which reduces the number of nodes by one.
See the example in Figure~\ref{fig:example_app_kcore}. Let $k = 4$, the current $k$-core contains the five nodes $1, 2, 3, 4, 5$. 
Both merging $6$ and $7$ and anchoring $6$ and $7$ brings $8$ and $9$ into the $k$-core.
See also Section~\ref{subsec:effective_study} for empirical comparison between $k$-trusses and $k$-cores as measures of graph cohesiveness and robustness.

\input{FIG/NP.tex}

\subsection{Hardness Analysis}
\begin{thm}\label{thm:NP_hard}
    The TIMBER problem is NP-hard for all $k \geq 3$.\footnote{That is, for all meaningful $k$ values.}
\end{thm}
\begin{pf}
    \begin{pf}
    	We show the NP-hardness by reducing the NP-hard maximum coverage (MC) problem to the TIMBER problem.
    	Consider the MC problem with the collection of $n$ sets $\mathcal{S} = \set{S_1, S_2, \ldots, S_n}$ and budget $b$.
    	Let $T = \set{t_1, t_2, \ldots, t_m} = \bigcup_{i = 1}^{n} S_i$.
    	Consider the decision version where we shall answer whether there is a subset $\mathcal{S}' \subseteq \mathcal{S}$ with $|\mathcal{S}'| \leq b$ such that at least $X$ elements in $T$ are covered by $\mathcal{S}'$.    	
    	We shall construct a corresponding instance of the TIMBER problem.    	
    	We construct the graph $G$ as follows.
    	For each $t_j \in T$, we create $2d$ nodes $t_{jp1}$ and $t_{jp2}, \forall 1 \leq p \leq d$, where $d$ is sufficiently large ($d > 10kmn$), and add edges $(t_{jp1}, t_{jp'2})$ for all $p \neq p'$.    	
    	For each $S_i \in \mathcal{S}$, we create two nodes $s_{i1}$ and $s_{i2}$, and for each $t_j \in S_i$, we add edges $(s_{i1}, t_{jp1})$ and $(s_{i2}, t_{jp2}), \forall 1 \leq p \leq d$.    	
    	Fix any $k \geq 3$, we create $k - 3$ nodes $r_1, r_2, \ldots, r_{k-3}$, each of which is connected with all $t$-nodes (i.e., $t_{jp1}$ and $t_{jp2}, \forall j, p$).    	
    	See Figure~\ref{fig:NP_example} for an example of the construction.
    	We also consider the decision version of the TIMBER problem where we shall answer whether there is a set $P'$ of pairs of nodes with $|P'| \leq b$ such that $f(P') \geq Xd^2$.
    	    	
    	\noindent $\Rightarrow)$ Given a YES-instance $\mathcal{S}' = \set{S_{i_1}, S_{i_2}, \ldots, S_{i_{b'}}}$ with $|\mathcal{S}'| = b' \leq b$ for the MC problem,     	
    	we claim that $P' = \set{(s_{i_1 1}, s_{i_1 2})}_{i = 1}^{b'}$ is a YES-instance $P'$ for the TIMBER problem.    	
    	By our construction and $|\bigcup_{S' \in \mathcal{S}'} S'| \geq X$, merging all pairs in $P'$ makes all the edges among the at least $X$ corresponding groups of $t$-nodes enter the $k$-truss,
    	and the total number is at least $Xd^2$. 
    	
    	\noindent $\Leftarrow)$ Given a YES-instance $P'$ with $|P'| = b' \leq b$ for the TIMBER problem,     	
    	we claim that 
    	(1) those edges entering the $k$-truss are distributed in at least $X$ groups of $t$-nodes corresponding to the elements in $T$, and 
    	(2) there exists $P'' \subseteq \set{(s_{i1}, s_{i2})}_{i = 1}^n$ with $|P''| = b'$ that is also a YES-instance of the TIMBER problem.
    	For (1), assume the opposite, i.e., less than $X$ groups are involved, then the size of the new $k$-truss is at most $(X-1)d^2 + 2(k-3)md + 2mnd < Xd^2$, which contradicts the fact that $P'$ is a YES-instance.
    	For (2), it is easy to see that each non-$(s_{i1}, s_{i2})$-type pair can be replaced an $(s_{i1}, s_{i2})$-type pair without decreasing the size of the $k$-truss.
    	 For $(s_{i1}, s_{j1})$ or $(s_{i2}, s_{j2})$  
    	 or $(t_{ip1}, s_{j1})$ with $i \neq j$,
    	 or a pair containing any $r$-node, there are no edges between two $t$-nodes entering the $k$-truss when we merge such a pair.
    	 For $(s_{i1}, s_{j2})$ with $i \neq j$,    	 
    	 merging such a pair is no better than merging $(s_{i1}, s_{i2})$ or $(s_{j1}, s_{j2})$.
    	 For a pair consisting of an $s$-node and a $t$-node,     	 
    	 it is no better than merging any $(s_{i1}, s_{i2})$ benefiting the same part.
    	Hence we can replace each element in $P'$ by an $(s_{i1}, s_{i2})$-type pair without decreasing the number of groups of edges among $t$-nodes entering the $k$-truss.    	
    	So we can find $P'' \subseteq \set{(s_{i1}, s_{i2})}_{i = 1}^n$ with $|P''| = b'$ and $f(P'') \geq Xd^2$, completing the proof.
    \end{pf}
\end{pf}

\begin{thm}\label{thm:non_mod}
    The function $f(P)$ is not submodular.
\end{thm}
\begin{pf}
	Consider the example in Figure~\ref{fig:NP_example}, but with $k = 5$ (there are $r_1$ and $r_2$ connected to all $t$-nodes).
	Let $X = \set{(s_{11}, s_{12})}$, $Y = \set{(s_{11}, s_{12}), (s_{21}, s_{22})} \supset X$, and $x = (s_{n1}, s_{n2})$,
	$f(X \cup \set{x}) - f(X) = 0 < f(Y \cup \set{x}) - f(Y)$, completing the proof.
\end{pf}

Considering the NP-hardness and non-submodularity of the TIMBER problem, 
we aim to find a practicable and efficient heuristic.

%% file: FIG/NP.tex
\begin{figure}[t!]
    \centering
    \includegraphics[width=0.6\linewidth]{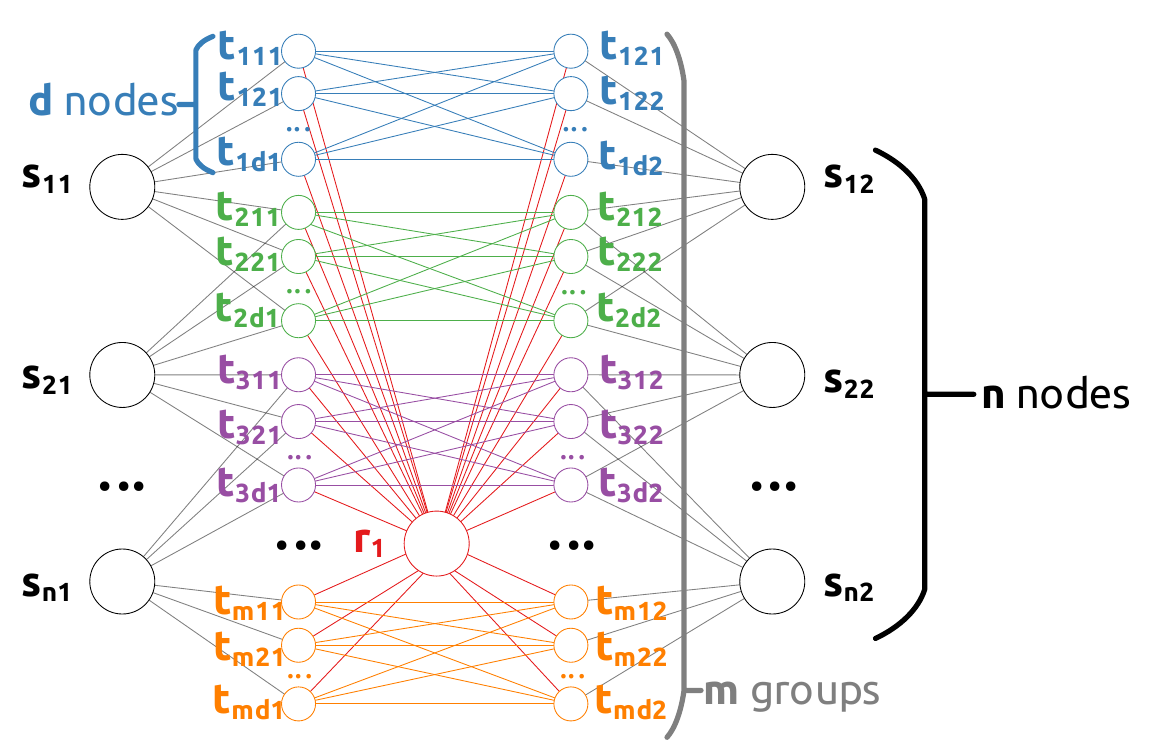}
    \caption{The constructed instance of the TIMBER problem corresponding to the maximum cover problem with $k = 4$, where $S_1 = \set{t_1, t_2}$, $S_2 = \set{t_2, t_3}$, and $S_n = \set{t_3, t_m}$.}
    \label{fig:NP_example}
\end{figure}

%% file: 040mthd.tex
\input{ALG/naive.tex}

\section{Methodology}\label{sec:method}
In this section, starting from the naive greedy algorithm,
we first analyze the changes occurring when we merge a pair of nodes,
and then based on our findings, we improve the computational efficiency while maintaining effectiveness as much as possible.

\subsection{Naive greedy algorithm}
First, we present the naive greedy algorithm in Algorithm~\ref{alg:naive_greedy}.
At each iteration,
we merge each possible pair,
compute the size of the $k$-truss after each merger,
and find and operate the merger with the best performance.
We repeat the above process until $b$ mergers are selected.
Although Algorithm~\ref{alg:naive_greedy} is algorithmically simple 
it suffers from prohibitive complexity, as shown in the following theorem.
\begin{thm}\label{thm:naive_complexity}
    Given an input graph $G = (V, E)$ and budget $b$,
    Algorithm~\ref{alg:naive_greedy} takes $O(b|V|^2|E|^{1.5})$ time and $O(|E|)$ space for any $k$.    
\end{thm}
\begin{pf}
	Truss decomposition algorithm takes $O(|E|^{1.5})$ time and $O(|V| + |E|)$ {space} \citep{wang2012truss}.
	Since we only consider connected graphs, $|E| = O(|V|)$ and thus $O(|V| + |E|) = O(|E|)$.
	Computing the size of the $k$-truss after each merger takes $O(|E|^1.5)$ time.
	Because there are $O(|V|^2)$ pairs and $b$ iterations, the total time complexity is $O(b|V|^2|E|^{1.5})$.    
	The space complexity is determined by that of storing the graphs and truss decomposition, which is $O(|E|)$.
\end{pf}

\begin{rem}\label{rem:naive_time_consuming}
{In the time complexity, $|V|^2$ is from the space of all possible pairs and $|E|^{1.5}$ is from the truss decomposition algorithm.}
\end{rem}

\subsection{Theoretical analyses: changes after mergers}
We shall show several theoretical findings regarding the changes occurring when we merge a pair of nodes.
First of all, in general, {merging two nodes $v_1$ and $v_2$ in $G$ can be viewed as} a two-step process: we (1) remove $v_2$ and all its incident edges (including the edge between $v_1$ and $v_2$ if it exists) and then (2) add edges between $v_1$ and each node that is originally {adjacent} to $v_2$ but not to $v_1$.
The following lemma shows that when we merge two nodes, the trussness of each edge containing neither of them changes (both increase and decrease are possible) by at most $1$.
\begin{lem}\label{lem:limited_truss_change}
    Given any $G$, $v_1$, and $v_2$,
    for any $e \in E(G)$, if $v_1, v_2 \notin e$,    
    then $|t(e; PM(v_1, v_2)) - t(e; G)| \leq 1$.
\end{lem}
\begin{pf}
	Let $G'$ denote $PM(v_1, v_2;G)$.
	First, we show the decrease is limited.    
	For each $k$, for each edge in the current $k$-truss,
	merging a pair of nodes can decrease the support by at most $1$.
	Therefore, each current $k$-truss at least satisfies the condition of $(k-1)$-truss after the merger,
	completing the proof of the limited decrease.
	Regarding the increase, consider the inverse operation of merging two nodes,
	and we shall show the decrease is limited.
	Formally, we split $v_1$ in $G'$ back into two nodes $v_1$ and $v_2$ in $G$,
	with $N(v_1; G') = N(v_1; G) \cup N(v_2; G)$.
	Regarding the trussness of each edge, this operation is no worse than deleting the node.
	Similarly, when we delete a node, for each $k$, for each edge in the current $k$-truss,
	the support decreases by at most $1$, completing the proof.
\end{pf}
Note that (1) the trussness can both increase and decrease and (2)
the above lemma does not apply to the edges incident to the merged nodes.footnote{An example can be found in Figure~\ref{fig:NP_example}, the edges incident to any of the $s$-nodes have trussness $2$ originally, but may have trussness much higher after a merger between two $s$-nodes.}
After a merger, only (1) the edges in the original $(k-1)$-truss and (2) those between a node in the original $(k-1)$-truss and a merged node are \textit{possibly} in the new $k$-truss.
\begin{cor}\label{cor:Tkm1_and_incident_enough}
    Given any $G$, $k$, and {$v_1, v_2\in V(G)$},
    $T_k(PM(v_1, v_2; G))$ $= T_k(G')$, where $V(G')=V(G)$ and
    $E(G') = E(T_{k-1}(G) \setminus \set{v_1, v_2}) \cup \set{(v_1, x): x \in (N(v_1) \cup N(v_2) \setminus \set{v_1, v_2}) \cap V(T_{k-1})})$.    
\end{cor}
\begin{pf}
	Recall that $T_{k-1}(G) \setminus \set{v_1, v_2}$ is defined as the subgraph obtained by removing $v_1$, $v_2$, and all their incident edges from $T_{k-1}(G)$.
	Since $G' \subseteq PM(v_1, v_2)$, $T_k(G') \subseteq T_k(PM(v_1, v_2))$.
	Hence, it suffices to show that $T_k(PM(v_1, v_2)) \subseteq T_k(G')$.
	First, by Lemma~\ref{lem:limited_truss_change}, for $e \in E(G \setminus \set{v_1, v_2})$, if $t(e; G) < k - 1$, then $t(e; PM(v_1, v_2)) < k$ and thus $e \notin E(T_k(PM(v_1, v_2)))$, completing the proof for the first part ($T_{k-1}(G) \setminus \set{v_1, v_2}$).
	Second, for an edge $(v_1, x)$,    
	such an edge exists iff $x \in N(v_1) \cup N(v_2) \setminus \set{v_1, v_2}$;
	if $x \notin V(T_{k-1})$, then $v_1$ will be the only neighbor of $x$ and thus $(v_1, x)$ cannot be in the $k$-truss after the merger, completing the proof.
\end{pf}

The following lemma shows that each edge with trussness larger than that of any merged node cannot lose its trussness.
\begin{lem}\label{lem:safe_edges}
    Given any $G$ {and $v_1, v_2\in V(G)$}, 
    without loss of generality, we assume $t(v_1) \geq t(v_2)$.
    For {any} $e \in E(G)$, if $t(e) > t(v_2)$, then $t(e; PM(v_1, v_2)) \geq t(e; G)$.
\end{lem}
\begin{pf}
	If $t(v_2) < t(e)$, then $v_2 \notin V(T_{t(e)}(G))$.
	So merging $v_1$ and $v_2$ can only bring new edges into the $t(e)$-truss, and thus the trussness of $e$ cannot decrease, completing the proof.
\end{pf}
Notably, mergers between nodes with low trussness can result in an increase in trussness for edges with higher trussness.\footnote{In Figure~\ref{fig:NP_example}, the mergers among the $s$-nodes with trussness $2$ cause trussness increase for the $t$-edges with higher trussness.}

Lemmas~\ref{lem:limited_truss_change} and \ref{lem:safe_edges} reduce the range of edges that we need to check for the $k$-truss after a merger, especially for those edges incident to neither of the merged nodes.
Regarding the edges incident to the merged nodes, Lemma~\ref{lem:core_connection} shows a connection to $k$-cores.
\begin{lem}\label{lem:core_connection}
    Given any $G$, $k$, and $v_1, v_2 \in V(G)$,
    let $N^*$ denote $N(v_1) \cup N(v_2) \setminus \set{v_1, v_2}$.
    For any $x \in N^*$, $(v_1, x)$ is in $T'_k \coloneqq T_k(PM(v_1, v_2))$ {if and only if} $x$ is in the $(k-2)$-core of $T'_k[N^*]$. 
\end{lem}
\begin{pf}
	$\Leftarrow)$ 
	Put $\set{(v_1, x): \text{$x$ is in the $(k-2)$-core of $T'_k[N^*]$}}$ and $E(T'_k[N^*])$ together, each such $(v_1, x)$ is in at least $k-2$ triangles $\triangle_{v_1xx'}$ with $x' \in N^*$, completing the proof.    
	
	\noindent $\Rightarrow)$
	Let $X$ denote $\set{x: (v_1, x) \in T'_k}$.
	For each $x \in X$, we have at least $k - 2$ triangles $\triangle_{v_1xx'}$
	with all three constituent edges in $T'_k$.    
	Hence $d(x; T'_k[N]) \geq k -2, \forall x \in X$, completing the proof.
\end{pf}

Based on the above analyses, we find it useful to consider the nodes \textit{inside and outside} $T_{k-1}$ separately and the neighbors \textit{inside} $T_{k-1}$ of a node need our special attention.
Below, we formally define these concepts that will be frequently used throughout the paper.
\begin{definition}[inside/outside nodes and inside neighbors]\label{def:inside_outside_nodes_inside_nbrs}
    Given a graph $G = (V, E)$ and $k \in \N$, we call a node $v \in V$ an \textbf{inside node} (w.r.t $G$ and $k$) if $v \in V(T_{k-1})$ (i.e., $t(v) \geq k - 1$) and we call $v$ an \textbf{outside node} (w.r.t $G$ and $k$) if $v \notin V(T_{k-1})$ (i.e., $t(v) < k - 1$).
    Given any node $u$, the set of $u$'s \textbf{inside neighbors} (w.r.t $G$ and $k$) is defined as $\tilde{N}_k(u; G) = N(u; G) \cap V(T_{k-1})$.
\end{definition}

Lemma~\ref{lem:comp_outside_nodes} provides a simple way to compare the performance of two outside nodes w.r.t. the considered objective.
\begin{lem}\label{lem:comp_outside_nodes}
	Given $G$ and $k$, for any $u_1, u_2 \notin V(T_{k-1})$, if $\tilde{N}_k(u_1) \subseteq \tilde{N}_k(u_2)$,
	then $T_k(PM(v, u_1)) \subseteq T_k(PM(v, u_2)), \forall v \in V$;
        if further $\tilde{N}_k(u_1) = \tilde{N}_k(u_2)$, then $T_k(PM(v, u_1)) = T_k(PM(v, u_2)), \forall v \in V$.
\end{lem}
\subsection{Proof of Lemma~\ref{lem:comp_outside_nodes}}
\begin{pf}
	Given any $G$, by Lemmas~\ref{lem:limited_truss_change} and \ref{lem:safe_edges}, if $u \notin V(T_{k-1})$, then $T_k \subseteq T_k(PM(v, u)) \subseteq \tilde{T}_k \subseteq PM(v, u)$, where $\tilde{T}_k = \tilde{T}_k(v, u) = T_{k-1} \cup \set{(v, x): x \in N(v) \cup N(u) \setminus \set{u, v}}, \forall v$.
	If $x \notin V(T_{k-1}) \cup \set{v, u}$, then $d(x; T_k(PM(v, u))) \leq d(x; \tilde{T}_k(v, u)) = 0$.
	By Lemma~\ref{lem:core_connection}, $x \notin V(T_k(PM(v, u)))$, and thus
	$T_k(PM(v, u)) = T_k(\tilde{T}_k(v, u)) = \hat{T}_k(v, u)$, where $\hat{T}_k(v, u) = T_{k-1} \cup \set{(v, x): x \in (N(v) \cup N(u) \setminus \set{v, u}) \cap V(T_{k-1})}$.
	For $u_1, u_2 \notin V(T_{k-1})$, if $N(u_1) \cap V(T_{k-1}) \subseteq N(u_2) \cap V(T_{k-1})$, then $\hat{T}_k(v, u_1) \subseteq \hat{T}_k(v, u_2)$.
	If $\Tilde{N}_k(u1) = \Tilde{N}_k(u2)$, i.e., $\Tilde{N}_k(u1) \subseteq \Tilde{N}_k(u2) \land \Tilde{N}_k(u2) \subseteq \Tilde{N}_k(u1)$,
	then $T_k(PM(v, u_1)) \subseteq T_k(PM(v, u_2)) \land T_k(PM(v, u_2)) \subseteq T_k(PM(v, u_1))$, i.e., $T_k(PM(v, u_1)) = T_k(PM(v, u_2))$,
	completing the proof.
\end{pf}

\begin{rem}\label{rem:eq_nbr_eq_res}    
     We can also see that if $\tilde{N}_k(u_1) = \tilde{N}_k(u_2)$, then for each $v \in V(G)$, $T_k(PM(v, u_1)) = T_k(PM(v, u_2))$.
\end{rem}
Below, we shall devise several practical improvements for the naive greedy algorithm (Algorithm~\ref{alg:naive_greedy}) based on the above theoretical findings, in order to increase the time efficiency.

\input{ALG/max_sets.tex}

\subsection{Reduce the number of pairs to consider}\label{subsec:node_heu}
 As mentioned in Remark~\ref{rem:naive_time_consuming}, one reason why the time complexity of the naive algorithm (Algorithm~\ref{alg:naive_greedy}) is high is that the space of all possible mergers is large ($O(|V|^2)$).
We shall first introduce several approaches to reduce the number of pairs to consider for a merger.

\smallsection{Maximal-set-based pruning for outside nodes.}
Lemma~\ref{lem:comp_outside_nodes} shows that for any given outside node $u \notin V(T_{k-1})$, we do not need to consider $u$ if there exists another outside node $u' \notin V(T_{k-1})$ with $N(u') \cap V(T_{k-1}) \supseteq N(u) \cap V(T_{k-1})$.
 It is because, in such a case, for any node $v$, merging $v$ and $u$ cannot be better than merging $v$ and $u'$ w.r.t the considered objective.\footnote{For completeness, we should also consider merging $u$ and $u'$, and it is easy to see that merging $u$ and $u'$ cannot increase the objective that we consider.}
Therefore, we only need to consider those nodes $u$ with maximal set $\tilde{N}(u)$ of inside neighbors.
\begin{lem}\label{lem:maximal_set_enough}
    Given $G$ and $k$, let $V_o = V(G) \setminus V(T_{k-1})$ denote the set of outside nodes, and
    let $\tilde{V}_o = \set{u \in V_o: \nexists u' \in V \setminus V(T_{k-1})~s.t.~\tilde{N}(u') \supsetneq \tilde{N}(u)}$ denote the set of outside nodes with {a} maximal set of inside neighbors.
    Then, $\max \set{|E(T_k(PM(v_1, v_2)))|: v_1, v_2 \in V(G)} = \max \set{|E(T_k(PM(v_1, v_2)))|: v_1, v_2 \in V(T_{k-1}) \cup \tilde{V}_o}$.
\end{lem}
\begin{pf}
	It suffices to show that for any $u \in V_o \setminus \tilde{V}_o$, if a merger includes $u$, then there exists another merger consisting of two nodes in $V(T_{k-1}) \cup \tilde{V}_o$ with no worse performance.
	And this is an immediate corollary of Lemma~\ref{lem:comp_outside_nodes}.
\end{pf}

Moreover, by Lemma~\ref{lem:comp_outside_nodes}, if several outside nodes have the same set of inside neighbors, only one of them needs to be considered.
Finding maximal sets among a given collection of sets is a well-studied theoretical problem \citep{yellin1992algorithms} {with a number of
fast algorithms.}
Based on \citep{so2012maxset}, we present in Algorithm~\ref{alg:max_sets} a simple yet practical way to find the outside nodes with a maximal set of inside neighbors.
Lemma~\ref{lem:maximal_set_complexity} shows the correctness, time complexity, and space complexity of Algorithm~\ref{alg:max_sets}.

\begin{lem}\label{lem:maximal_set_complexity}
   Given the set of outside nodes $V_o$ and the sets of their inside neighbors $\Tilde{N}_k(v), \forall v \in V_o$,
   Algorithm~\ref{alg:max_sets} correctly finds the set of nodes $v' \in V_o$ with maximal $\tilde{N}_k(v')$ in $O(\sum_{v \in V_o} |\tilde{N}_k(v)| |V_o|) = O(|V||E|)$ time and $O(|E|)$ space.
\end{lem}
\begin{pf}
	Lines~\ref{line:unique_nbrs_s} to \ref{line:unique_nbrs_e} remove the outside nodes with duplicate inside neighborhood and take $O(|V_o|)$ times.
	Lines~\ref{line:membership_s} to \ref{line:membership_e} build the membership function $m$ where for a inside node $u$, the $i$-th bit of $m(u)$ indicates the membership relation between $u$ and the $i$-th element of $V'_o$, which takes $O(\sum_{v \in V_o} |\tilde{N}_k(v)|)$ time.
	Lines~\ref{line:maxset_s} to \ref{line:maxset_e} use $m$ to check the maximality of each unique inside neighborhood and take $O(\sum_{v \in V_o} |\tilde{N}_k(v)| |V_o|)$.    
	For the correctness, $r(v)$ consists of the nodes $v'$ with $\Tilde{N}_k(v') \supseteq \Tilde{N}_k(v)$.
	If the final $r$ is a power of $2$, i.e., exactly a single bit of $r$ is $1$, then this bit represents $v$ itself, which means that no other $v'$ satisfies that $\Tilde{N}_k(v') \supseteq \Tilde{N}_k(v)$.
	Regarding the space complexity, the inputs and all the variables ($S$, $V_o'$, and $V_o^*$) take $O(|E|)$ space, $m(v)$ for all $v \in V_o^*$ takes $O(\sum_{v \in V_o'} |\tilde{N}_k(v)|) = O(|E|)$ space if we represent the binary arrays in a sparse way \citep{barrett1994templates}.    
\end{pf}

 \begin{rem}\label{rem:inside_nodes_larger_nbr_not_best}
     This maximal-set-based pruning does not apply to inside nodes.
     Consider again the example in Figure~\ref{fig:example_1} with $k = 3$, both merging $vf$ and $vd$ and merging $vz$ and $vd$ perform worse than merging $vf$ and $vz$, while $\tilde{N}_k(vd) \supsetneq \tilde{N}_k(vf)$ and $\tilde{N}_k(vd) \supsetneq \tilde{N}_k(vz)$.
 \end{rem}
In our implementation, among all the outside nodes with a maximal set of inside neighbors, we further sort the outside nodes by the number of inside neighbors and choose the ones with the most inside neighbors as the candidates.

\smallsection{A heuristic for finding promising inside nodes.}
Notably, our maximal-set-based pruning scheme does not apply to inside nodes, and thus we need different techniques for inside nodes.
We propose and use a heuristic based on \textit{incident prospects} (IPs) to evaluate the inside nodes and select the promising ones.
\begin{definition}[incident prospects]\label{def:incident_prospects}
    Given a graph $G = (V, E)$ and $k \in \N$, for each $v \in V$, the set of the \textbf{incident prospects} (IPs) of $v$ is defined as $\tilde{N}_k(v) \setminus N(v; T_k)$.
\end{definition}
Intuitively, the IPs of a node $v$ correspond to the edges that are not in the current $k$-truss but possibly enter the new $k$-truss after a merger involving $v$ (see Corollary~\ref{cor:Tkm1_and_incident_enough}).
Therefore, if a node $v$ has more IPs, then a merger involving $v$ is preferable since it is more likely that the size of the $k$-truss will increase more because more edges incident to $v$ may enter the new $k$-truss after the merger.
Moreover, the number of the IPs of a node $v$ is a lower bound of the number of inside neighbors of a node $v$, and thus if a node $v$ has a larger number of IPs, then $v$ also has a larger number of inside neighbors, i.e., more non-incident edges may benefit from the merger.
See Section~\ref{subsec:exp_empirical_support} for the empirical support of the proposed heuristic, including the comparison of multiple heuristics.

In our implementation, we sort the inside nodes by the number of IPs of each inside node and choose the ones with the most IPs as our candidate inside nodes.

\smallsection{Exclude outside-outside mergers.}
After dividing nodes into inside nodes and outside nodes, we now have three types of mergers: 
(1) \textit{inside-inside mergers} (IIMs) where two inside nodes are merged, 
(2) \textit{outside-outside mergers} (OOMs) where two outside nodes are merged, and 
(3) \textit{inside-outside mergers} (IOMs) where one inside node and one outside node are merged.
We shall show that OOMs are less desirable than the other two types in general.
Merging two nodes $v_1$ and $v_2$ can equivalently be seen as 
(1) removing all edges incident to $v_2$ and
(2) adding each ``new'' edge $(v_1, x)$ for $x \in N(v_2) \setminus N(v_1) \setminus \set{v_1}$.
Proposition~\ref{prop:no_inside_no_help} shows that if we do not include an inside node in the merger (i.e., for an OOM), then each single ``new'' edge cannot increase the size of $T_k$.
\begin{lem}\label{prop:no_inside_no_help}
    Given any $G = (V, E)$, $k$, and {$v_1, v_2 \notin T_{k-1}$, for any $x \in N(v_1) \cup N(v_2) \setminus \set{v_1, v_2}$, $T_k(G) = T_k(G')$, where $V(G')=V(G)$ and $E(G')=E(G) \cup \set{(v_1, x)}$.}
\end{lem}
\begin{pf}
	If an edge $e_0$ is inserted into $G$ such that the trussness of $e_0$ after the insertion is $l$, then all edges with original trussness at least $l$ will not gain any trussness, and the remaining edges can gain at most $1$ trussness \citep{huang2014querying}.
	Hence, it suffices to show that for each considered $x$, after inserting $(v_1, x)$ into $G$, the trussness of $(v_1, x)$ is at most $k - 1$.
	Indeed, since $v_1 \notin T_{k-1}$, all edges incident to $v_1$ have original trussness at most $k - 2$ and thus have trussness at most $k - 1$ after the insertion.
	Therefore, all triangles containing $(v_1, x)$ will not be in $T_k$ and thus neither will $(v_1, x)$.
\end{pf}

See also Section~\ref{subsec:exp_empirical_support} for the empirical evidence supporting our choice.
Therefore, from now on we assume that we always include at least one inside node in the merger.
Then there are two cases that we need to consider: IOMs and IIMs,
and no one is necessarily better than the other.

\subsection{Promising pairs among promising nodes}\label{subsec:pair_heu}
With the above analyses, we can utilize the maximal-set-based pruning for outside nodes, and use the heuristics for inside and outside nodes to further reduce the number of candidate nodes.
However, even with the above analyses, it is still computationally expensive to compute the size of the new $k$-truss after each possible merger, even when the number of candidate nodes is relatively small.
For example, in {the} \textit{youtube} dataset (to be introduced in Section~\ref{sec:exp}) with $k = 10$, the total number of possible IOMs and IIMs is $6.6$ billion.
Although after pruning the outside nodes, the number is reduced to $374$ million, and even if we only choose $100$ inside nodes and $50$ outside nodes, it still takes more than two hours to check the actual size of the $k$-truss after all the $9,950$ possible mergers.
Therefore, it is still imperative to further reduce the number of times that we check the actual size of the $k$-truss.
To this end, we shall propose and use some heuristics to efficiently find promising mergers (IOMs and IIMs).
For both cases, our algorithmic framework is in the following form:
\begin{enumerate}[leftmargin=*]
    \item We first find the \textit{promising nodes} {as} described above.
    \item Among all the possible pairs between the promising nodes, we {use novel} heuristics to find a small number of \textit{promising pairs}.
    \item We check the increase in the size of the $k$-truss for each of the promising pairs and merge a pair with the greatest increase.
    \item We repeat the above process until we exhaust the budget.
\end{enumerate}

\input{ALG/iom_heu.tex}
\smallsection{Inside-outside mergers.}
We shall deal with IOMs first.
By Corollary~\ref{cor:Tkm1_and_incident_enough}, we know that an IOM between an inside node $v_1$ and an outside node $v_2$, {w.r.t.} the size of the $k$-truss, brings ``new'' edges $(v_1, z)$ for each ``new'' neighbor $z \in (\tilde{N}_k(v_2) \cup \tilde{N}_k(v_1)) \setminus (N(v_1, T_{k-1}) \cup \set{v_1})$ into the current $(k-1)$-truss.
Note that $\tilde{N}_k(v_1) \supsetneq N(v_1, T_{k-1})$ {may hold} since an edge between two nodes in the $(k-1)$-truss may exist in the original graph but not in the $(k-1)$-truss.

To efficiently evaluate the candidate IOMs, we propose to use the concept of \textit{potentially helped shell edges} (PHSEs).
For given $G$ and $k$, we use $\hat{E}_k(G)$ to denote $\hat{E}_k(G) = E(T_{k-1}) \setminus E(T_k)$ (the edges with trussness $k - 1$) and call such edges \textit{shell edges} (w.r.t $G$ and $k$).
\begin{definition}[potentially helped shell edges]\label{def:PHSEs}
    Given a graph $G = (V, E)$, $k \in \N$, an inside node $v_1$, and an outside node $v_2$, the set of the \textbf{potentially helped shell edges} (PHSEs) w.r.t the IOM between $v_1$ and $v_2$, denoted by $\hat{H}_k(v_1, v_2; G)$, consists of the shell edges $(x, y) \in \hat{E}_k(G)$ such that at least a triangle containing $(x, y)$ is newly formed because of the ``new'' edges brought into $T_{k-1}$ by the IOM.
    Formally, $\hat{H}_k(v_1, v_2; G) = \set{e \in \hat{E}_k(G): s(e; G \cup \set{(v_1, z): z \in Z}) > s(e; G)}$, where
    $Z = (\tilde{N}_k(v_2) \cup \tilde{N}_k(v_1)) \setminus (N(v_1, T_{k-1}) \cup \set{v_1})$.    
\end{definition}
In the above definition, we require that $(v_1, z)$ and $(x, y)$ are in the same triangle, thus we have $x \in \set{v_1, z}$ or $y \in \set{v_1, z}$.
Accordingly, there are two ways in which some shell edges $(x, y)$ can be helped:
\textbf{(1)} the IOM brings a ``new'' neighbor $z$ to $v_1$ and thus forms a new triangle $\triangle_{v_1 z z'}$ for some $z'$ that is adjacent to $v_1$ in the original graph, and
\textbf{(2)} the IOM brings two ``new'' neighbors $z_1$ and $z_2$ and thus
forms a new triangle $\triangle_{v_1 z_1 z_2}$.
The first case {(1)} further includes two sub-cases:
\textbf{(1a)} some shell edge $(v_1, z')$ incident to $v_1$ is helped, and
\textbf{(1b)} some shell edge $(z, z')$ not incident to $v_1$ is helped.
In {Figure~\ref{fig:example_2}(a)}, we provide an illustrative example.

We present the whole heuristic-based procedure for choosing IOM candidates in Algorithm~\ref{alg:iom_heu}.
Among the inputs of Algorithm~\ref{alg:iom_heu}, $V_o^*$, $V_i$, $\tilde{N}_k$, $\hat{E}_k$, and $T_k$ are computed from the inherent inputs $G$ and $k$ of the TIMBER problem, while $n_i$, $n_o$, and $n_c$ are set by the user to control the computational cost.

We first choose the most promising inside nodes and outside nodes using some heuristics as {presented} in Section~\ref{subsec:node_heu} (Lines~\ref{line:choose_vi}-\ref{line:choose_vo}).
After choosing the promising nodes, for each chosen inside node $v_i$,
we first compute the incident PHSEs that each ``new'' neighbor may bring (Lines~\ref{line:incident_phse_s}-\ref{line:incident_phse_e}).
Then, for each outside node $v_o$,
we compute the ``new'' neighbors the IOM between $v_i$ and $v_o$ brings to $v_i$ (Line~\ref{line:compute_new_nbrs}),
collect all the incident PHSEs of the ``new'' neighbors (Line~\ref{line:collect_incident_phse}),
compute the non-incident PHSEs (Line~\ref{line:compute_nonincident_phse}),
and take the union to get all the PHSEs (Line~\ref{line:all_phse}).
Finally, we use the computed PHSEs to select the most promising IOMs (Line~\ref{line:return_c_iom}).
See Section~\ref{subsec:exp_empirical_support} for the empirical support of the proposed heuristics.

\begin{lem}\label{lem:iom_time}
    Given pruned outside nodes $V_o^*$, inside nodes $V_i$, inside neighbors $\Tilde{N}_k$, shell edges $\hat{E}_k$, and $k$-truss $T_k$,
    Algorithm~\ref{alg:iom_heu} takes $O(|V_o^*|\log n_o + n_i n_o (|V_i| + |\hat{E}_k| + \log n_c))$ time {to find $n_c$ IOM candidates from $n_i$ and $n_o$ promising inside and outside nodes, respectively.}
\end{lem}
\begin{pf}
	Finding the {top-$n_i$ inside nodes and top-$n_o$ outside nodes} (Lines~\ref{line:choose_vi} and \ref{line:choose_vo}) takes $O(|V_i|\log n_i + |V_o^*|\log n_o)$.
	For all inside nodes and all ``new'' neighbors, computing the incident PHSEs (Lines~\ref{line:incident_phse_s} to \ref{line:incident_phse_e}) takes
	$O(n_i |V(T_{k-1})|)$ {time};
	and computing PHSEs for all pairs (Lines~\ref{line:iom_each_pair_s} to \ref{line:phse_e}) takes
	$O(n_i n_o (|V(T_{k-1})| + |\hat{E}_k|))$ {time}.
	Maintaining the candidate set takes $O(n_i n_o \log n_c)$ {time}.
	Hence, the total time complexity is $O(|V_o^*|\log n_o + n_i n_o (|V_{k-1}| + |\hat{E}_k| + \log n_c))$.
\end{pf}

\smallsection{Inside-inside mergers.}
Now we are going to deal with inside-insider mergers (IIMs).
As we have mentioned, {w.r.t. the size of the $k$-truss, an inside-outside merger (IOM) is equivalent to adding} into the current $(k-1)$-truss new edges incident to the inside node in the IOM.
However, this is not true for inside-insider mergers (IIMs).
Consider an IIM between two nodes $v_1$ and $v_2$.
In {Figure~\ref{fig:example_2}(b)}, we provide an example of an IIM between $v_1$ and $v_2$.
An IIM may incur three kinds of changes that may affect the size of the $k$-truss.
The first kind is \textbf{support gains (SGs)}, which are also caused by IOMs.
For IIMs, SGs further include two sub-cases:
\begin{itemize}[leftmargin=*]
    \item \textbf{SG-n (Support gains of non-incident edges).} It may happen for an edge between a node adjacent to $v_1$ but not to $v_2$ and another node adjacent to $v_2$ but not to $v_1$.
    In {Figure~\ref{fig:example_2}(b)}, $(z_3, z_4)$ gains one support after the IIM between $v_1$ and $v_2$.
    \item \textbf{SG-i (Support gains of incident edges).} Incident edges are the edges incident to either of the merged nodes.    
    In {Figure~\ref{fig:example_2}(b)}, both $(v_1, z_3)$ and $(v_2, z_4)$ gain one support after the IIM.
\end{itemize}
The latter two kinds can only be caused by IIMs but not by IOMs:
\begin{itemize}[leftmargin=*]
    \item \textbf{CL (Collisions).} IIMs can directly make some edges collide and disappear.    
    Specifically, each pair of edges {$(v_1, x)$ and $(v_2, x)$} incident to the same node $x$ and the two merged nodes collide and only one of them remains.
    In {Figure~\ref{fig:example_2}(b)}, {there are collisions between $(z_5, v_1)$ and $(z_5, v_2)$; and between $(z_6, v_1)$ and $(z_6, v_2)$.}    
    \item \textbf{SL (Support losses).} IIMs can reduce the support of some edges in the current $(k-1)$-truss, potentially decreasing their trussness. Specifically, {each edge} {between} the common neighbors of $v_1$ and $v_2$ loses a common neighbor after the merger between $v_1$ and $v_2$.
    In {Figure~\ref{fig:example_2}(b)}, the edge $(z_5, z_6)$ loses one support after the IIM.
\end{itemize}
Due to the new types of changes that we need to consider, there are several noticeable points that we shall discuss below.

\input{FIG/example_phse.tex}

Lemma~\ref{lem:comp_outside_nodes} tells us that for IOMs, outside nodes with large neighborhoods are generally preferable, 
while including inside nodes with large neighborhoods does not always give better performance.
One of the reasons is that including inside nodes with larger neighborhoods may cause more collisions and support losses described above.

For IOMs, we have used the number of all potentially helped shell edges (PHSEs, Definition~\ref{def:PHSEs}) to find the candidate IOMs (Lines~\ref{line:collect_incident_phse} to \ref{line:all_phse} in Algorithm~\ref{alg:iom_heu}).
Specifically, we consider both incident PHSEs (Line~\ref{line:collect_incident_phse}) and non-incident PHSEs (Line~\ref{line:compute_nonincident_phse}).
However, there are two subtleties:
\textbf{(a)} for IIMs, the computation of incident PHSEs becomes tricky due to the collisions mentioned above;
\textbf{(b)} moreover, we also need to {additionally} take the support losses into consideration.
To address the two subtleties, we slightly modified the heuristic we have used for IOMs.
Regarding subtlety (a), since the incident shell edges have been considered when we choose the inside nodes w.r.t the incident prospects (IPs), for simplicity, we only consider the immediate collisions among the edges in the current $k$-truss without computing the support gains and support losses of the incident shell edges.
Regarding subtlety (b), we count both the shell edges with support gains and those with support losses.
To conclude, for each shell edge with support gains, we give $+1$ score (reward) to the corresponding IIM;
for each shell edge {with support losses} and each collision between two edges in the current $k$-truss, we give $-1$ score (penalty).
See Section~\ref{subsec:exp_empirical_support} for the empirical comparisons of different heuristics in choosing candidate IIMs.

\input{ALG/iim_heu.tex}
In Algorithm~\ref{alg:iim_heu}, we present the whole procedure for choosing candidate IIMs.
Among the inputs, $V_i$, $\tilde{N}_k$, $\hat{E}_k$, and $T_k$ are computed from the inherent inputs $G$ and $k$ of the TIMBER problem, while $n_i$ and $n_c$ are set by the user to control the computational cost.

We first choose the most promising inside nodes and outside nodes using the heuristics mentioned in Section~\ref{subsec:node_heu} (Lines~\ref{line:choose_vi_iim}).
After that, for each pair $(v_1, v_2)$ between two chosen inside nodes, we compute its score using the heuristic described above.
Specifically, for each pair, we first initialize the score by giving $-1$ score to each collision between two edges in the current $k$-truss (Line~\ref{line:score_init}), then for each non-incident shell edge $(x, y)$ whose support changes (Line~\ref{line:shell_edge_s_iim}), add $+1$ score for each one whose support increases (Line~\ref{line:reward_iim}), and give $-1$ score for each one whose support decreases (Line~\ref{line:penalty_iim}).
Finally, we use the computed scores to select the most promising IIMs (Line~\ref{line:return_c_iim}).

\begin{lem}\label{lem:iim_time}
    Given inside nodes $V_i$, inside neighbors $\Tilde{N}_k$, shell edges $\hat{E}_k$, and the $k$-truss $T_k$,
    Algorithm~\ref{alg:iim_heu} takes $O(|V_i|\log n_i + n_i^2 (|\hat{E}_k| + \log n_c))$ {time} {to find $n_c$ IIM candidates from $n_i$ promising inside nodes.}
\end{lem}
\begin{pf}
	Finding the top-$n_i$ inside {nodes} (Line~\ref{line:choose_vi_iim}) takes $O(|V_i|\log n_i)$ {time}.
	For all pairs among the chosen inside nodes, computing the scores (Lines~\ref{line:iim_heu_s} to \ref{line:iim_heu_e}) takes
	$O(n_i^2 |\hat{E}_k|)$ {time}.
	Maintaining the set of candidates IOMs takes $O(n_i^2 \log n_c)$ {time}.
	Therefore, the total time complexity is $O(|V_i|\log n_i + n_i^2 (|\hat{E}_k| + \log n_c))$.
\end{pf}

\subsection{Considering both IIMs and IOMs}
Theoretically, merging IOMs is not always better than IIMs, and vice versa.
Indeed, as empirically shown in Section~\ref{subsec:exp_main}, neither IOMs nor IIMs can be consistently superior to the other.
In general, when $k$ is small, IIMs are more desirable, while IOMs gain strength when $k$ increases.
Intuitively, when $k$ increases, the $k$-truss is denser, and thus IIMs inevitably cause more collisions and support losses due to the high overlaps among the neighborhoods of the inside nodes.
Therefore, it is necessary to consider both IOMs and IIMs.

We propose a strategy to take both IIMs and IOMs into consideration without wasting too much computation on the less-promising case.
{The key idea is to \textit{adaptively distribute} the number of candidates between the two cases.}
Specifically, we fix the total number of pairs to choose in each round (i.e., the sum of {$n_c$'s} for Algorithms~\ref{alg:iom_heu} and \ref{alg:iim_heu})
and divide it into two parts for IIMs and IOMs.
Initially, the number is equally divided. 
Then in each round, we shift $1/b$ fraction of the total number,
\textit{to} the case where the best-performing pair in this round belongs \textit{from} the other case. 
We make sure that the $n_c$ for each case does not decrease to zero.
See Section~\ref{subsec:exp_main} for the empirical support for considering both IIMs and IOMs and the adaptive distribution of the number of candidates.
The pseudo-code of the process mentioned above is given in Algorithm~\ref{alg:overall} (see Lines~\ref{line:budget_adjust_overall_s} to \ref{line:budget_adjust_overall_e}), which will be described in detail in Section~\ref{subsec:overall_alg}.

\input{ALG/overall.tex}

\subsection{Check the result after each merger} \label{subsec:check_result_after_merger}
By proposing techniques to reduce the search space and proposing heuristics to find promising pairs efficiently,
we have been addressing the problem of the $O(|V|^2)$ space of all possible pairs.
Another overhead (see Remark~\ref{rem:naive_time_consuming}) is the truss decomposition which takes $O(|E|^{1.5})$ time.

For checking the size of the $k$-truss after each possible merger between two nodes $v_1$ and $v_2$, we do not need to compute it from the whole post-merger graph. 
We use Corollary~\ref{cor:Tkm1_and_incident_enough} by which the computation takes only $O(|E(T_{k-1})|^{1.5})$ time since $|\set{(v_1, x): x \in (N(v_1) \cup N(v_2) \setminus \set{v_1, v_2}) \cap V(T_{k-1})}| = O(|E(T_{k-1})|)$.

\begin{rem}\label{rem:on_incremental_update}
    It is theoretically possible to use incremental algorithms for updating $k$-trusses after edge additions and removals \citep{huang2014querying, luo2020batch}.
    However, their efficiency is limited in our case since even a single node merger can cause a large number of edge additions and removals.    
\end{rem}

\subsection{Overall algorithm (\bmshort)}\label{subsec:overall_alg}
In Algorithm~\ref{alg:overall}, we present the procedure of the proposed algorithm \bmshort~(\bmlong).
The inputs are the inherent inputs of the TIMBER problem ({an} input graph $G$, trussness $k$, and a budget $b$) and the parameters that control the computational cost ($n_i$, $n_o$, and $n_c$).

\input{FIG/main_res.tex}

In each round, we first compute or update the edge trussness (Line~\ref{line:update_truss_edge}), and prepare the information that we need later (Lines~\ref{line:collect_shell_edges} to \ref{line:comp_inside_nbr}).
Then we use Algorithm~\ref{alg:max_sets} to prune the set of outside nodes using the maximal-set-based technique (Line~\ref{line:Vo_max_set_pruning}).
After that, we use Algorithms~\ref{alg:iom_heu} and \ref{alg:iim_heu} to obtain the candidate IOMs and IIMs, respectively (Lines~\ref{line:get_C_iom} and \ref{line:get_C_iim}).
Then we check the performance of all the candidate mergers and find the best one (Line~\ref{line:check_all_candidates}),
and update the graph together with its edge trussness accordingly if not all budget has been exhausted (Line~\ref{line:if_budget_remains}).
Regarding the distribution of the number of pairs to check in each round, initially the number is equally distributed between IOMs and IIMs (Line~\ref{line:budget_init_overall}), and in each round we increase the number of the case where the best-performing pair belongs and decrease that of the other case (Lines~\ref{line:budget_adjust_overall_s} to \ref{line:budget_adjust_overall_e}).
We make sure that both cases are considered throughout the process.

\begin{thm}\label{thm:overall_time}
Given {an} input graph $G$, trussness $k$, a budget $b$, and the parameters $n_i$, $n_o$, and $n_c$, Algorithm~\ref{alg:overall} takes $O(b(|E|^{1.5} + n_c |E(T_{k-1})|^{1.5} + |V_o^*|\log n_o + n_i n_o (|V_i| + |\hat{E}_k| + \log n_c) + n_i^2 (|\hat{E}_k| + \log n_c)))$ time and $O(|E| + n_c)$ space to find $b$ pairs to be merged.
\end{thm}
\begin{pf}
	In each round, truss decomposition (Line~\ref{line:update_truss_edge}) takes $O(|E|^{1.5})$ time.
	Collecting all the information (Lines~\ref{line:collect_shell_edges} to \ref{line:Vo_max_set_pruning}) takes $O(|E|)$ time.
	By Lemmas~\ref{lem:iom_time} and \ref{lem:iim_time}, obtaining the candidate mergers (Lines~\ref{line:get_C_iom} and \ref{line:get_C_iim}) takes $O(|V_o^*|\log n_o + n_i n_o (|V_i| + |\hat{E}_k| + \log n_c) + n_i^2 (|\hat{E}_k| + \log n_c))$ time.
	Checking the results after all candidates (Line~\ref{line:check_all_candidates}) takes $O(n_c |E(T_{k-1})|^{1.5})$ time.
	Updating the graph (Line~\ref{line:update_graph}) takes $O(|E|)$ time.
	Hence, it takes $O(b(|E|^{1.5} + n_c |E(T_{k-1})|^{1.5} + |V_o^*|\log n_o + n_i n_o (|V_i| + |\hat{E}_k| + \log n_c) + n_i^2 (|\hat{E}_k| + \log n_c)))$ time in total.
	All the inputs and variables take $O(|E| + n_c)$ space, including the intermediate ones in Algorithms~\ref{alg:iom_heu} and \ref{alg:iim_heu} (note that we only maintain the set of best candidate nodes and pairs).
	By Lemma~\ref{lem:maximal_set_complexity}, Algorithm~\ref{alg:max_sets} takes $O(|E|)$ space.
	Hence, the total space complexity is $O(|E| + n_c)$.
\end{pf}

\begin{rem}\label{rem:time_dominant_truss}
	{The dominant terms in the time complexity are the two $1.5$-order terms from the truss decomposition and checking the truss size after each candidate merger.
		When we fix $n_i$ and $n_o$ as constants, the time complexity becomes $O(b(|E|^{1.5} + n_c |E(T_{k-1})|^{1.5})$;
		If we further fix $n_c$ as a constant, then the time complexity becomes $O(b|E|^{1.5})$ totally dominated by that of truss decomposition.		 
		Empirically, we do observe that for different heuristics, the differences in the running time are small as long as the times checking the truss size are the same (see Section~\ref{subsec:exp_main}).}
\end{rem}

%% file: ALG/naive.tex
\begin{algorithm}[t!]
    \small
    \caption{Naive greedy algorithm} \label{alg:naive_greedy}
	\DontPrintSemicolon
	\SetKwInOut{Input}{Input}
    \SetKwInOut{Output}{Output}
    \SetKwComment{Comment}{\blue{$\triangleright$ }}{}
    
    \Input{graph $G = (V, E)$;
        trussness $k$;
        budget $b$}
    \Output{$P$: the pairs of nodes to be merged}
    
    $P \leftarrow \emptyset$ \\
    \While{$|P| < b$}{
        $f({\{p\}}) \leftarrow |E(T_k(PM(p; G)))|, \forall p \in \binom{V'}{2}$\\
        $p^* \leftarrow \arg \max_{p} f({\{p\}})$; $P \leftarrow P \cup \set{p^*}$\\        
        \textbf{if} $|P| < b$ \textbf{then} $G = (V, E) \leftarrow PM(p^*; G)$\\}
    \Return{$P$}
\end{algorithm}

%% file: ALG/max_sets.tex
\begin{algorithm}[t!]
    \small
    \caption{Prune outside nodes (based on \cite{so2012maxset})} \label{alg:max_sets}
	\DontPrintSemicolon
	\SetKwInOut{Input}{Input}
    \SetKwInOut{Output}{Output}
    \SetKwComment{Comment}{\blue{$\triangleright$ }}{}
    
    \Input{outside nodes $V_o$; inside neighbors $\Tilde{N}_k(v), \forall v \in V_o$}
    \Output{$V_o^*$: the outside nodes with maximal set of inside neighbors}
    
    $S, V'_o \leftarrow \emptyset$ \label{line:unique_nbrs_s}  \\ 
    \For{$v \in V_o$}{
        \textbf{if} $\Tilde{N}_k(v) \notin S$ \textbf{then} \{$S \leftarrow S \cup \set{\Tilde{N}_k(v)}$; $V'_o \leftarrow V'_0 \cup \set{v}$\} \\
    } \label{line:unique_nbrs_e}
    $i \leftarrow 0$; $m(v) \leftarrow 0, \forall v \in V_o$; $V_o^* \leftarrow \emptyset$ \\ \label{line:membership_s}
    \For{$v \in V'_o$}{
        \textbf{for} $u \in \Tilde{N}_k(v)$ \textbf{do} $m(u) \leftarrow \texttt{BitwiseOr}(m(u), 2^i)$ \\
        $i \leftarrow i + 1$ \\
    } \label{line:membership_e}    

    $r(v) \leftarrow \texttt{BitwiseAnd}(\set{m(u): u \in \Tilde{N}_k(v)}), \forall v \in V_o'$ \\ \label{line:maxset_s}
    $V_o^* \leftarrow \set{v: \text{$r(v)$ is a power of $2$}}$ \\ \label{line:maxset_e}    
    \Return{$V_o^*$}
\end{algorithm}

%% file: ALG/iom_heu.tex
\begin{algorithm}[t!]
    \small
    \caption{Find IOM candidates} \label{alg:iom_heu}
	\DontPrintSemicolon
	\SetKwInOut{Input}{Input}
    \SetKwInOut{Output}{Output}
    \SetKwComment{Comment}{\blue{$\triangleright$ }}{}
    
    \Input{pruned outside nodes $V_o^*$;
        inside nodes $V_i$;        
        inside neighbors $\Tilde{N}_k(v), \forall v \in V_o^* \cup V_i$;
        shell edges $\hat{E}_k$;        
        $k$-truss $T_{k}$;        
        number of inside nodes to check $n_i$;
        number of outside nodes to check $n_o$;
        number of pairs to choose $n_c$
    }
    \Output{$C_{IOM}$: the chosen IOM candidates}
    $\hat{V}_i \leftarrow$ the $n_i$ inside nodes $v_i$ in $V_i$ with most incident prospects 
    \label{line:choose_vi} \\
    $\hat{V}_o \leftarrow$ the $n_o$ outside nodes $v_o$ in $V_o^*$ with most inside neighbors 
    \label{line:choose_vo} \\    
    \For{$v_i \in \hat{V}_i$} { \label{line:incident_phse_s}        
        $H(t_i) \leftarrow \tilde{N}_k(t_i) \cup \tilde{N}_k(v_i), \forall t_i \in V_i \setminus \tilde{N}_k(v_i) \setminus \set{v_i}$ \\ \label{line:incident_phse_e}

        \For{$v_o \in \hat{V}_o$} { \label{line:iom_each_pair_s}
            $Z = Z(v_i, v_o) \leftarrow 
                (\tilde{N}_k(v_i) \cup \tilde{N}_k(v_o)) \setminus
                (N(v_i; T_k) \cup \set{v_i})$\label{line:compute_new_nbrs}\\
            $H_i \leftarrow \bigcup_{z \in Z} H(z)$ \label{line:collect_incident_phse}\\
            $H_n \leftarrow \set{(x, y) \in \hat{E}_k: (x \in Z \lor y \in Z) 
            \land x \in Z \cup \tilde{N}_k(v_i) 
            \land y \in Z \cup \tilde{N}_k(v_i)}$ \label{line:compute_nonincident_phse}\\
            $\hat{H}_k(v_i, v_o) \leftarrow H_i \cup H_n$ \label{line:all_phse} \\
        }
    } \label{line:phse_e}
    $C_{IOM} \leftarrow$ the $n_c$ IOMs $(v_i, v_o) \in \hat{V}_i \times \hat{V}_o$ with largest $|\hat{H}_k(v_i, v_o)|$ 
    (tie broken by $|Z(v_i, v_o)|$) 
    \label{line:return_c_iom} \\
    \Return{$C_{IOM}$}
\end{algorithm}

%% file: FIG/example_phse.tex
\begin{figure}[t!]
    \centering
    \begin{subfigure}[b]{0.23\textwidth}
        \centering
        \includegraphics[scale=0.1]{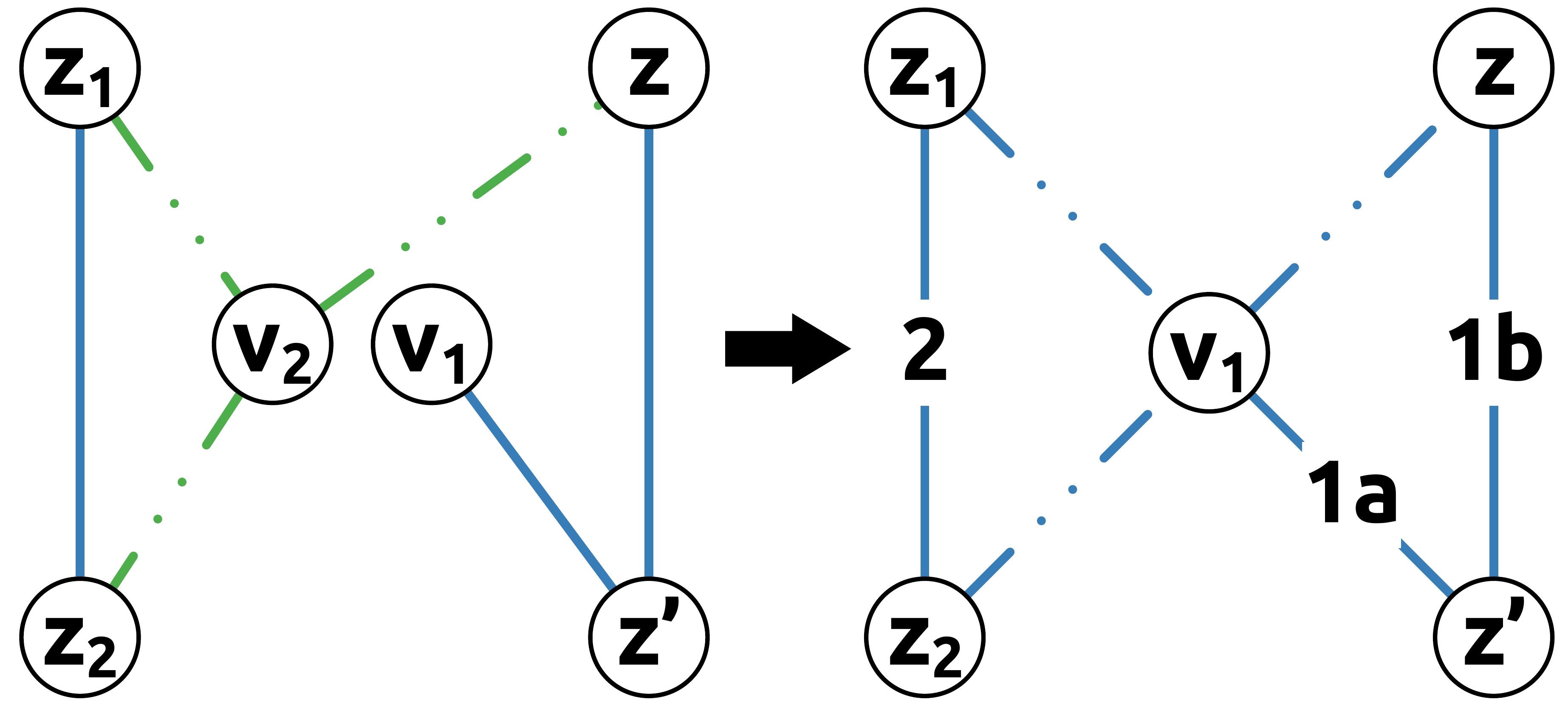}
        \caption{an example of an IOM} 
    \end{subfigure}
    \begin{subfigure}[b]{0.23\textwidth}
        \centering
        \includegraphics[scale=0.1]{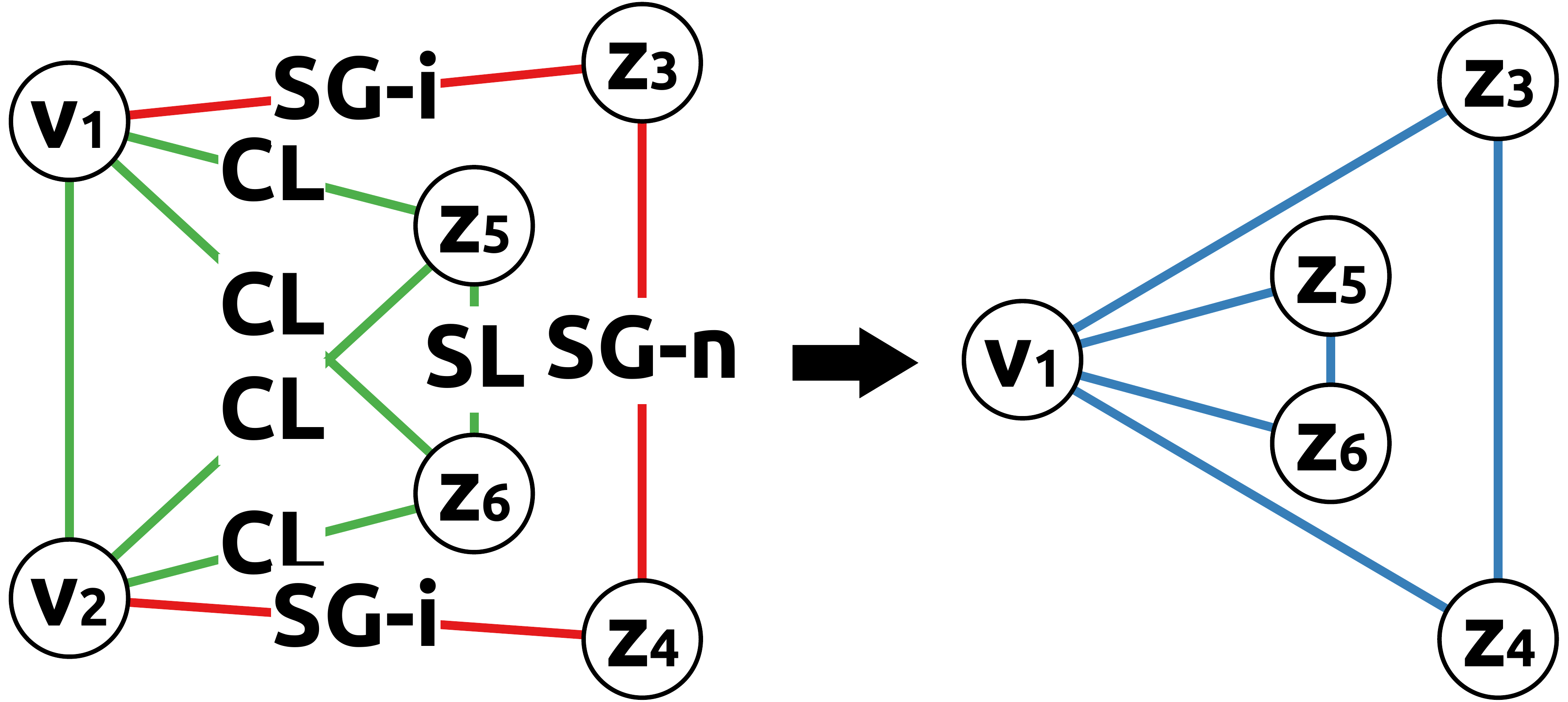}
        \caption{an example of an IIM}        
    \end{subfigure}
    \caption{Illustrative examples of the changes caused by an IOM (left) or an IIM (right) between $v_1$ and $v_2$.}
    \label{fig:example_2}
\end{figure}

%% file: ALG/iim_heu.tex
\begin{algorithm}[t!]
    \small
    \caption{Find IIM candidates} \label{alg:iim_heu}
	\DontPrintSemicolon
	\SetKwInOut{Input}{Input}
    \SetKwInOut{Output}{Output}
    \SetKw{Continue}{continue}
    \SetKwComment{Comment}{\blue{$\triangleright$ }}{}
    
    \Input{inside nodes $V_i$;        
        inside neighbors $\Tilde{N}_k(v), \forall v \in V_i$;
        shell edges $\hat{E}_k$;
        $k$-truss $T_{k}$;
        number of inside nodes to check $n_i$;
        number of pairs to choose $n_c$
    }
    \Output{$C_{IIM}$: the chosen IIM candidates}
    $\hat{V}_i \leftarrow$ the $n_i$ inside nodes $v_i$ in $V_i$ with most incident prospects \label{line:choose_vi_iim} \\
    \For{$(v_1, v_2) \in \binom{\hat{V}_i}{2}$} { \label{line:iim_heu_s}
        $h(v_1, v_2) \leftarrow -|\set{u \in V(T_k): \set{(v_1, u), (v_2, u)} \subseteq E(T_k)}|$ \\ \label{line:score_init}
        \For{$(x, y) \in \hat{E}_k$ with $x, y \notin \set{v_1, v_2}$ and $\set{x, y} \subseteq \Tilde{N}_k(v_1) \cup \Tilde{N}_k(v_2)$}{ \label{line:shell_edge_s_iim}
            \If{$x, y \notin \Tilde{N}_k(v_1) \cap \Tilde{N}_k(v_2)$}{
                $h(v_1, v_2) \leftarrow h(v_1, v_2) + 1$ \\ \label{line:reward_iim}
            } \ElseIf{$\set{x, y} \subseteq \Tilde{N}_k(v_1) \cap \Tilde{N}_k(v_2)$}{
                $h(v_1, v_2) \leftarrow h(v_1, v_2) - 1$ \\ \label{line:penalty_iim}
            }
        }
    } \label{line:iim_heu_e}
    $C_{IIM} \leftarrow$ the $n_c$ IIMs $(v_1, v_2) \in \binom{\hat{V}_i}{2}$ with largest $h(v_1, v_2)$ \label{line:return_c_iim} \\
    \Return{$C_{IIM}$}
\end{algorithm}

%% file: ALG/overall.tex
\begin{algorithm}[t!]    
    \small
    \caption{\bmshort: final proposed algorithm} \label{alg:overall}
	\DontPrintSemicolon
	\SetKwInOut{Input}{Input}
    \SetKwInOut{Output}{Output}
    \SetKw{Continue}{continue}
    \SetKwComment{Comment}{\blue{$\triangleright$ }}{}
    
    \Input{graph $G = (V, E)$;
        trussness $k$;
        budget $b$;        
        number of inside nodes to check $n_i$;
        number of outside nodes to check $n_o$;
        number of pairs to choose $n_c$
    }
    \Output{$P$: the pairs of nodes to be merged}
    $P \leftarrow \emptyset$; $n_{io} \leftarrow \lfloor n_c / 2 \rfloor$ \\ \label{line:budget_init_overall}
    \While{$|P| < b$}{
        compute or update $t(e)$ using truss decomposition \\ \label{line:update_truss_edge}
        $\hat{E}_k \leftarrow \set{e \in E: t(e) = k - 1}$ \\ \label{line:collect_shell_edges}
        $t(v) \leftarrow \max_{e \ni v} t(e), \forall v \in V$ \\ \label{line:comp_truss_nodes}
        $V_i \leftarrow \set{v \in V: t(v) \geq k - 1}$; $V_o \leftarrow \set V \setminus V_i$ \\ \label{line:collect_Vo}
        $\tilde{N}_k(v) \leftarrow N(v) \cap V(T_{k-1(G)}), \forall v \in V$ \\ \label{line:comp_inside_nbr}
        $V_o^* \leftarrow$ Alg.~\ref{alg:max_sets} with inputs $V_o$ and $\tilde{N}_k$ \\ \label{line:Vo_max_set_pruning}
        $C_{IOM} \leftarrow$ Alg.~\ref{alg:iom_heu} with inputs $V_o^*, V_i, \tilde{N}_k, \hat{E}_k, T_{k}(G), n_i, n_o, n_{io}$ \\ \label{line:get_C_iom}    
        $C_{IIM} \leftarrow$ Alg.~\ref{alg:iim_heu} w/ inputs $V_i, \tilde{N}_k, \hat{E}_k, T_k(G), n_i, n_c - n_{io}$ \\ \label{line:get_C_iim}
        $p^* \leftarrow \argmax_{c = (v_1, v_2) \in C_{IOM} \cup C_{IIM}} T_k(PM(v_1, v_2))$ 
        \Comment{Corollary~\ref{cor:Tkm1_and_incident_enough} is used for simplification}
        \label{line:check_all_candidates}
        \setadd{P}{p^*} \\        
        \If{$|P| < b$}{ \label{line:if_budget_remains}
            $G = (V, E) \leftarrow PM(p^*; G)$ \\ \label{line:update_graph}
            \If{$p^* \in C_{IOM}$}{ \label{line:budget_adjust_overall_s}
                $n_{io} \leftarrow \min(n_{io} + \lfloor n_c / b \rfloor, \lceil n_c (b-1) / b \rceil)$
            } \Else{
                $n_{io} \leftarrow \max(n_{io} - \lfloor n_c / b \rfloor, \lfloor n_c / b \rfloor)$
            } \label{line:budget_adjust_overall_e}
        }
    }
    \Return{$P$} \\ \label{line:return_P_overall}
\end{algorithm}

%% file: FIG/main_res.tex
\begin{figure*}[t!]    
    \centering    
    \includegraphics[scale=0.3]{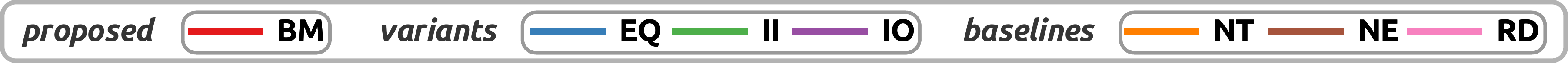} \\
    \includegraphics[height=3.2cm]{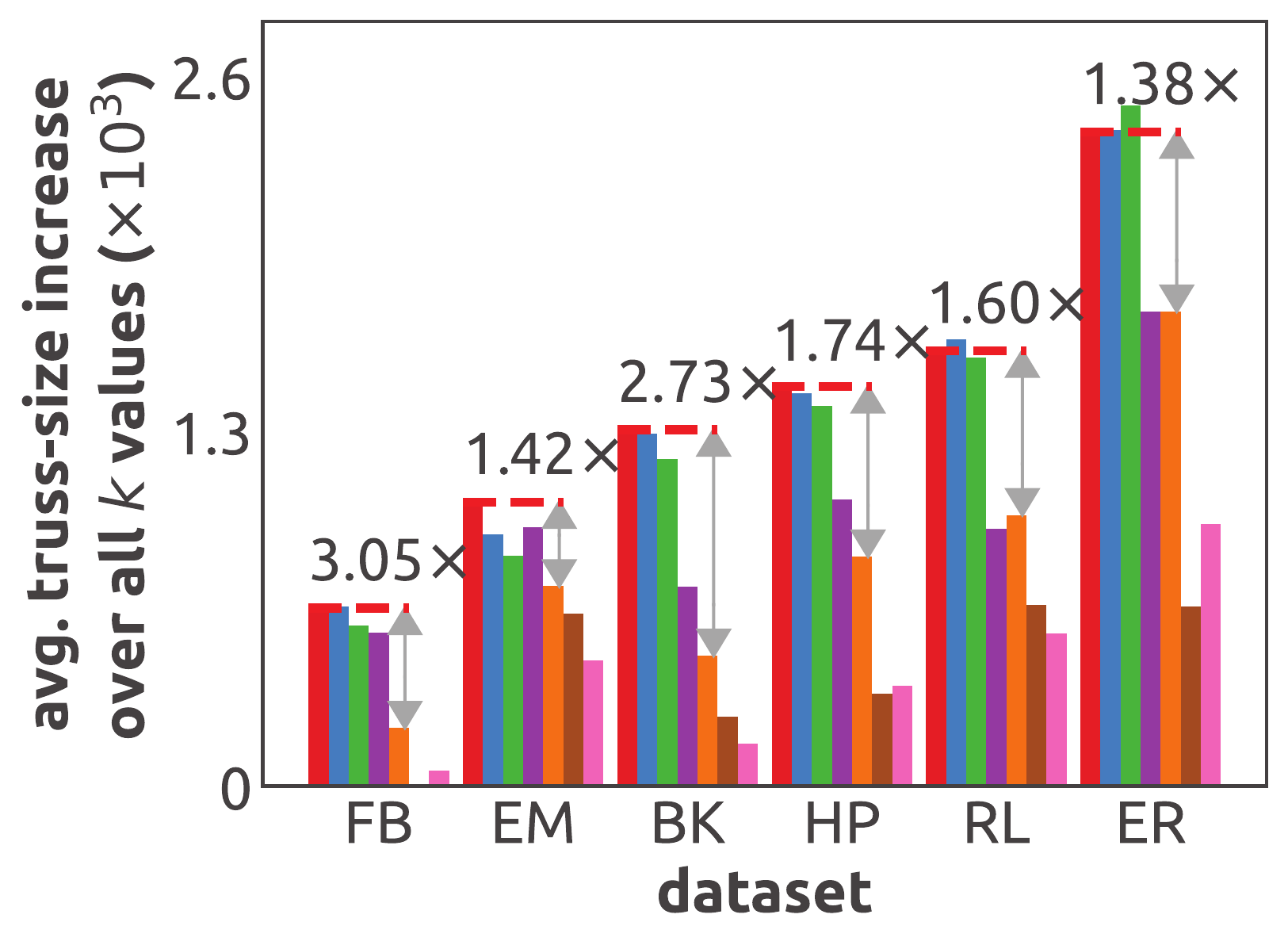}
    \includegraphics[height=3.2cm]{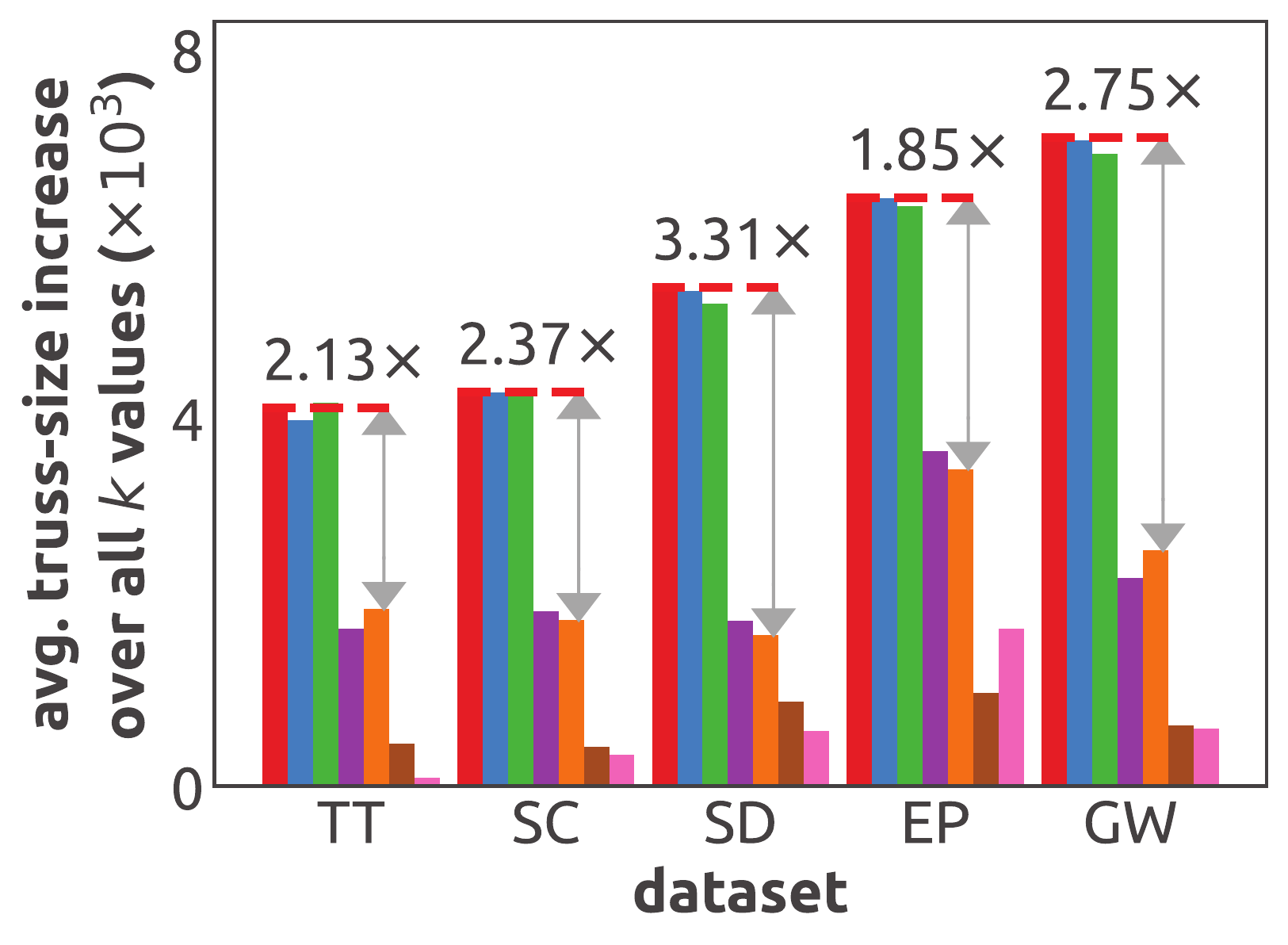}
    \includegraphics[height=3.2cm]{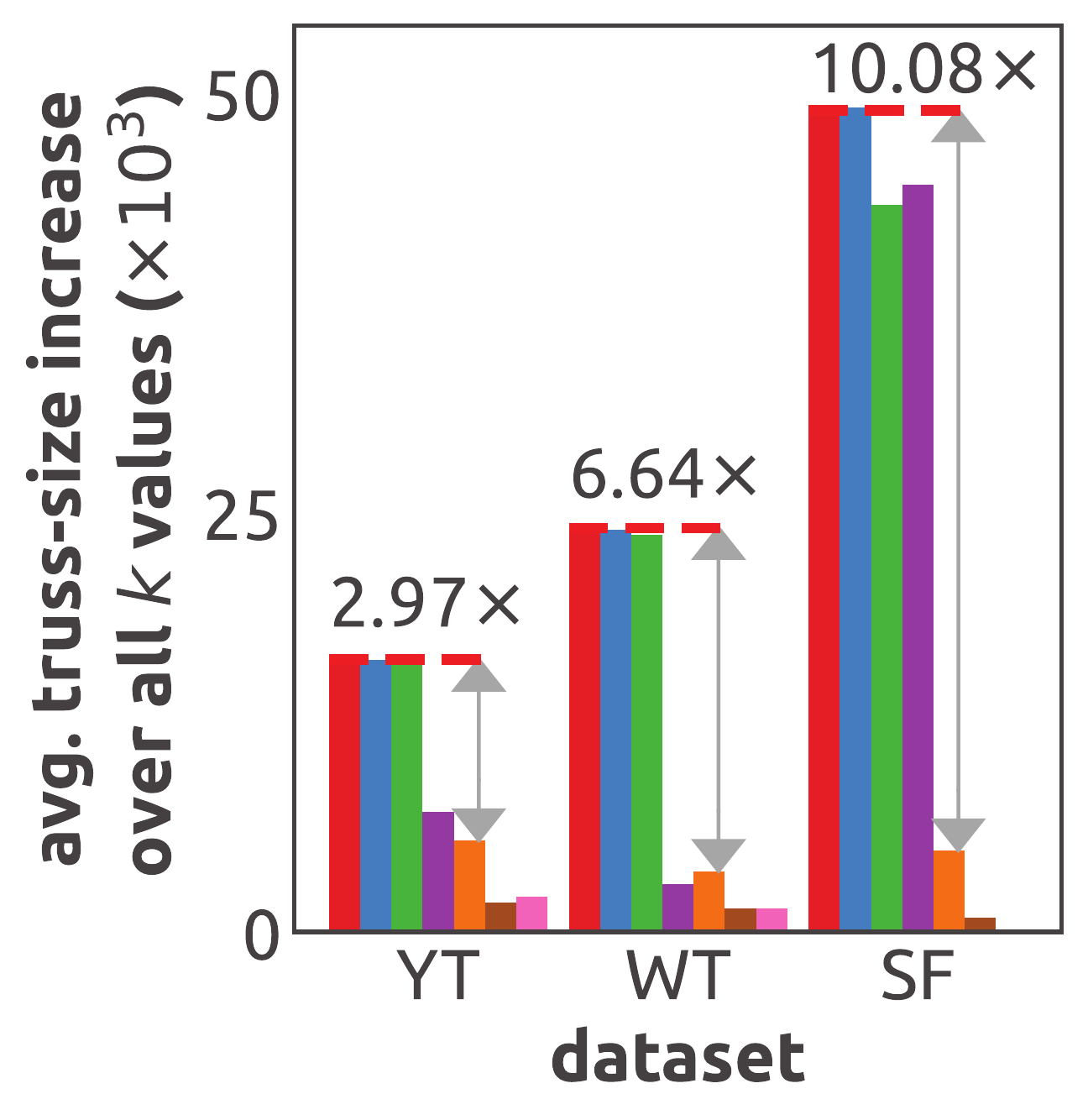}        
    \includegraphics[height=3.3cm]{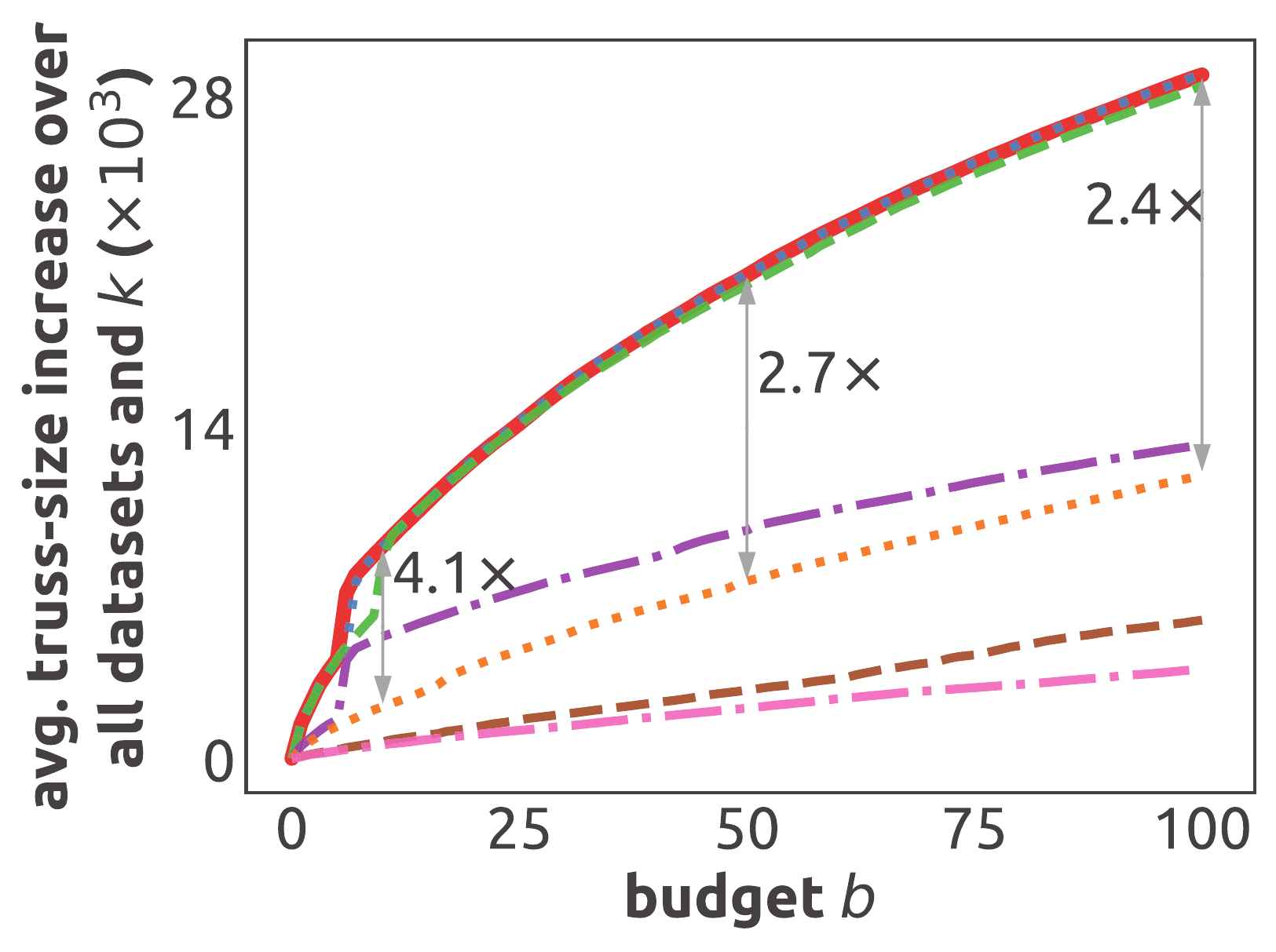}    
    \caption{\underline{\smash{The first three subfigures on the left:}} The average increase in the truss size of each considered algorithm on each dataset over all the considered $k$ values.
    The proposed algorithm (BM), with its variants (EQ, II, and IO) consistently outperforms the baseline methods (NT, NE, and RD) by $1.38\times$ to $10.08\times$.
    Overall, BM performs better than its variants, showing the usefulness of our algorithmic designs.
    \underline{\smash{The rightmost subfigure:}} The results with each algorithm with different budget $b$.
    The proposed method (BM) and its variants consistently outperform the baseline methods by $2.4\times$ to $4.1\times$. \label{fig:main_res_ds}
    \label{fig:b_avg}}
\end{figure*}

%% file: 050exper.tex
\section{Experimental Evaluation}\label{sec:exp}
In this section, through extensive experiments on fourteen real-world graphs, we shall show the effectiveness and the efficiency of \bmshort, the proposed algorithm.
Specifically, we shall answer each of the following questions:
\begin{itemize}[leftmargin=*]
    \item \textbf{Q1}: how effective are merging nodes and maximizing the size of a $k$-truss in enhancing graph robustness?
    \item \textbf{Q2}: how effective and computationally efficient is \bmshort in maximizing the size of a $k$-truss by merging nodes?
    \item \textbf{Q3}: how effective is each algorithmic choice in \bmshort?
\end{itemize}

\smallsection{Experimental settings.}
For each dataset, we conduct experiments for each $k \in \set{5, 10, 15, 20}$.
We use $b = 10$, check $100$ inside nodes and $50$ outside nodes ($n_i = 100$, $n_o = 50$ in Algorithm~\ref{alg:overall}), and the number of pairs to check in each round ($n_c$ in Algorithm~\ref{alg:overall}) is set to $10$ by default.
We conduct all the experiments on a machine with i9-10900K CPU and $64$GB RAM.
All algorithms are implemented in C++, and complied by G++ with O3 optimization.

\input{TAB/datasets.tex}
\smallsection{Datasets.}
In Table~\ref{tab:datasets}, we report some statistics (the number of nodes/edges, max trussness, and sizes of $k$-trusses for different $k$ values) of the real-world graphs~\citep{yin2017local,leskovec2007graph,leskovec2012learning,leskovec2009community,cho2011friendship,relato2013,richardson2003trust,leskovec2005graphs,leskovec2010signed,nr,yang2015defining,leskovec2010predicting} used for the experiments.

\subsection{\red{Q1: Effectiveness of merging nodes and truss-size maximization}}\label{subsec:effective_study}
We shall first show that merging nodes is an effective operation to enhance graph robustness. Then, we show that when we maximize the size of a $k$-truss, we effectively improve graph robustness.

\input{eff_three_random_models}
\smallsection{Effectiveness of merging nodes.}
First, we show that merging nodes is an effective way to enhance cohesiveness and robustness, specifically, compared to adding edges.
We consider different robustness measures~\citep{freitas2021evaluating,ellens2013graph}:
\textbf{(1)} VB (average vertex betweenness),
\textbf{(2)} EB (average edge betweenness),
\textbf{(3)} ER (effective resistance)~\citep{ghosh2008minimizing, ellens2011effective},
\textbf{(4)} SG (spectral gap)~\citep{malliaros2012fast}, and
\textbf{(5)} NC (natural connectivity)~\citep{chan2014make}.
On the Erd\"{o}s-R\'{e}nyi model~\citep{erdHos1960evolution} with $n = 50$ and $p = 0.1$, for each measure, we use greedy search to find the mergers or new edges that improve the measure most.
In Figure~\ref{fig:effective_merge}, we report the change of the measure in each setting when we merge nodes or add edges 10 times, where merging nodes is much more effective.
Mean values over five independent trials are reported.

We also consider other random graph models. The considered random graph models are:
\begin{itemize}[leftmargin=*]
	\item the Erd\"{o}s-R\'{e}nyi model ($n = 50$ and $p = 0.1$),
	\item the Watts–Strogatz small-world model ($n = 50$, $k = 7$, and $p = 0.1$), and         
	\item the Holme-Kim powerlaw-cluster model ($n = 50$, $m = 3$, and $p = 0.1$).
\end{itemize}
In Tables~\ref{tab:effective_merging_models}, we provide the results using the three different random graph models.       

\smallsection{Effectiveness of enlarging a $k$-truss.}
Second, we conduct a case study on the \textit{email} dataset.
In Figure~\ref{fig:effective_truss}, we show how the five robustness measures mentioned above change along with the truss size, when we apply our proposed algorithm \bmshort on the \textit{email} dataset to maximize the size of its $10$-truss by $100$ mergers.
The measures are linearly normalized so that all the original values correspond to $1$.
The chosen mergers increase the robustness even though \bmshort only directly aims to increase the size of a $k$-truss, showing that maximizing the size of a $k$-truss is indeed a reasonable way to reinforce graph cohesiveness and robustness.

\input{TAB/robust_correlation_truss_core}
We also empirically compare $k$-trusses with $k$-cores in measuring cohesiveness and robustness.
We have conducted experiments where we apply our proposed algorithm on the email dataset to enlarge the size of its $10$-truss by 100 mergers. 
We observe that the size of the $10$-truss increases and so does the robustness of the whole graph w.r.t multiple robustness measures, where high correlations between the truss size and the robustness measures exist.
However, the size (specifically, both the number of nodes and the number of edges) of $11$-core ($10$-truss is a subgraph of $11$-core) actually decreases, i.e., the size of a $k$-core decreases while the robustness of the whole graph increases, surprisingly showing a negative correlation between the size of a $k$-core and the robustness of the whole graph.
We have also tried merging nodes to maximize the size of a $k$-core (i.e., the counterpart problem using k-cores).
We adapt an algorithm for the anchored k-core problem for this purpose because the counterpart problem using $k$-cores is technically similar to the anchored $k$-core problem (see Section~\ref{subsec:on_counterpart_k_core}).
We observe that the correlations between the core size and the robustness measures (although also positive) are clearly weaker than the correlations between the truss size and the robustness measures (see Table~\ref{tab:robust_correlation_truss_core}).

\subsection{Q2: Effectiveness \& efficiency of \bmshort}\label{subsec:exp_main}
We shall compare \bmshort with several baseline methods, showing \bmshort's high effectiveness and high efficiency.

\smallsection{Considered algorithms.}
Since the TIMBER problem is formulated for the first time by us, no existing algorithms are directly available.
Therefore, we use several intuitive baseline methods as the competitors and also compare several variants of the proposed algorithm.
For all algorithms, the maximal-set-based pruning for outside nodes described in Section~\ref{subsec:node_heu} is always used.
In each round, all the algorithms find candidate mergers and operate the best one after checking all the candidates.
The considered algorithms are:
\begin{itemize}[leftmargin=*]
	     \item \textbf{RD (Random)}: uniform random sampling among all the IIMs and IOMs.
	     {Average performances over five trials are reported.}    
	     \item \textbf{NE (Most new edges)}: among all the IOMs,\footnote{Note that IIMs cannot increase the number of such edges.}
        choosing the ones that increase the number of \textit{edges} among the nodes in the current $(k-1)$-truss most.
	     \item  \textbf{NT (Most new triangles)}: among all the IIMs and IOMs, choosing ones that increase the number of \textit{triangles} consisting of the nodes in the current $(k-1)$-truss most.
	     \item \textbf{BM (\bmshort)}: the proposed method (Algorithm~\ref{alg:overall}).
	     \item \textbf{EQ (\bmshort-EQ)}: a \bmshort variant always \textbf{equally distributing} the number of pairs to check between IIMs and IOMs.
	     \item \textbf{II (\bmshort-II)}: a \bmshort variant considering \textbf{IIMs} only.
	     \item \textbf{IO (\bmshort-IO)}: a \bmshort variant considering \textbf{IOMs} only.
\end{itemize}

\begin{rem}\label{rem:on_adding_nodes}
	We acknowledge that there exist algorithms proposed for maximizing/minimizing the size of a $k$-truss via adding/deleting nodes/edges, and have discussed them in Section~\ref{sec:relwk}.
	However, since those algorithms greedily find a single node/edge as the objective and do not consider the combination of multiple nodes/edges, those algorithms cannot be directly applied to our problem.
	Moreover, we compared the best pairs to merge and connect, and the best choices for merging and connecting are very different.
	Specifically, merging the best pair for connecting usually does not increase the size of the $k$-truss for the same $k$.
\end{rem}

\input{FIG/trade_off.tex}

\smallsection{Evaluation metric.}
We evaluate the \textbf{performance} of each algorithm by the  increase in the size of the $k$-truss, {i.e., we measure $|E(T_k(PM(P;G)))|-|E(T_k(G))|$, for each output $P$ for a graph $G$.}

\smallsection{Results on each dataset.}
In Figure~\ref{fig:main_res_ds} (the first 3 subfigures), for each dataset, we report the average performance over all $k \in \set{5, 10, 15, 20}$ of each algorithm. The proposed algorithm \bmshort with its variants consistently outperforms the baseline methods, and the overall performance of \bmshort is better than that of its variants. 
Specifically, compared to the best baseline method on each dataset, \bmshort gives $1.38 \times$ to $10.08 \times$ better performance,
and the ratio is above $2\times$ on 9 out of 14 datasets.
Overall, \bmshort performs better than its variants that consider only IIMs or IOMs, or always equally distribute the number of candidate mergers to check.
This shows the usefulness of considering both
IIMs and IOMs and adaptive distribution of the number of candidate mergers.

\smallsection{Results on different budgets $b$.}
In Figure~\ref{fig:b_avg} (the 4th subfigure), for each $1 \leq b \leq 100$, we report the average performance of each algorithm over all datasets and all $k$ values.
As shown in the results, \bmshort clearly outperforms the baseline methods regardless of $b$ values. 
When $b = 10$, $50$, and $100$, \bmshort performs $4.1\times$, $2.7\times$, and $2.4\times$ better than the best baseline method, respectively.

\smallsection{Results on different \# candidates.}
In Figure~\ref{fig:tradeoff_avg} (the first 3 subfigures), for each algorithm checking different numbers ($1,2,5,10,15,20$) of candidates in each round, we report the average running time and the average performance over all datasets and all $k$ values.
The proposed algorithm \bmshort clearly outperforms the baseline methods, even when the baseline methods check more candidate mergers than \bmshort in each round;
\bmshort is also more effective and more stable than its variants, especially when we check a larger number of candidate pairs.
Also, for different algorithms (except NT),  the running time is similar when we check the same number of candidate pairs in each round, validating the theoretical analyses on the time complexity of \bmshort.

\smallsection{Results on different $k$.}
We divide the considered $k$ values into two groups: \textit{low} (5/10) and \textit{high} (15/20), and compare the performance of the algorithms in each group.
In Figure~\ref{fig:k_avg} (the 4th subfigure), for each group, we report the average increase in the truss size of each algorithm over all the datasets and over the two $k$ values in the group.
Again, \bmshort consistently outperforms the baseline methods, regardless of the $k$ value.
Notably, when $k$ is low, IIMs perform much better than IOMs w.r.t the increase in the truss size;
but when $k$ is high, this superiority is decreased, even reversed, and thus considering both IIMs and IOMs shows higher necessity.

The above results show from different perspectives that \bmshort overall outperforms the baselines as well as its variants.
The \textbf{full results} in each considered setting (datasets and parameters) are in the online appendix~\cite{onlineSuppl}.

\smallsection{Case Study.} Here, we provide a case study on the \textit{relato} dataset showing which nodes (companies) are merged together by \bmshort.  
In the \textit{relato} dataset, each node represents a company (mainly in the IT field) and each edge represents some business-partner relationship between two companies.
We use BATMAN on the relato dataset with $(k, b, n_i, n_o, n_c) = (10, 100, 100, 50, 10)$.
After $b = 100$ rounds, the size of $(k = 10)$-truss increases from $89041$ to $94944$.
Only 110 nodes participate in the 100 mergers.
There are 7 groups of companies consisting of more than two companies are merged together:
\begin{itemize}[leftmargin=*]
	\item (size = 28) Apple Inc., Experian, Ameriprise, Peavey, REC Solar, Salesforce, Cubic, Kirin, RSA, Hanold Associates, Novatek, SKS, Interbrand, Hewlett Packard, Cemex, Beijing Enterprises, space150, NuGen, InSite, Wesco, Thomas \& Betts, Bloom Energy, Ashland, Oshkosh Corporation, Azul Systems, ADC, BTG, Palantir; 
	This group consists of a giant company (Apple Inc.) and 27 relatively small companies in various fields; This kind of mergers usually happen in real-world situations.
	\item (size = 22) SAP, CloudBees, DragonWave, Klocwork, SunTrust, Basho, Merry Maids, Signal, Xcerra, SGI, Veeva, SWIFT, Mitsui \& Co, Hologic, Comdata, Martin Agency, Spectra Energy, Zensar Technologies, United Rentals, ThreatMetrix, IMG College, NAVTEQ; Similarly, this group consists of a giant company SAP and 21 relatively small companies in various fields.
	\item (size = 19) Oracle, Airwatch, Amylin Pharmaceuticals, E2open, BMO Financial Group, Cyber-Ark Software, UC Berkeley, MedImmune, Petco, Piper Aircraft, Wheel Pros, Aker Solutions, Swiss Life, Torch, Brooks Brothers, RWE Group, Bell Mobility, Calabrio, Compal Electronics; Similar to the previous two groups, where the giant company is Oracle.
	\item (size = 12) Google, Databricks, SimpliVity, Azul, Henry Schein, Apple, Newport News, HCL Technologies America, HP, AES, Hewlett Packard Enterprise, Marshall Aerospace; Similarly, the giant company is Google.
	\item (size = 11) IBM, Nine Entertainment, Gores, OneSpot, TALX, Kaiser Aluminum, Coty, JC Penney, Scivantage, Ch2m Hill, State bank of India; Similarly, the giant company is IBM; It is interesting to see State bank of India here.
	\item (size = 8) Facebook, VCE, SDL, Reval, MAXIM INTEGRATED, NEC, ThyssenKrupp, Commonwealth Bank of Australia; Similarly, the giant company is Facebook; Again, we see a foreign bank here (Commonwealth Bank of Australia).
	\item (size = 4) Amazon Web Services, Hewlett-Packard, Xignite, Gainsight; This group is a bit different, both Amazon Web Services and Hewlett-Packard giant companies, while Xignite and Gainsight are two relatively small data and market companies.
	\item (size = 2) Cisco, YourEncore; This is a combination of Cisco, which corresponds to the node with the highest node degree in the dataset, and a fairly small company YourEncore.
	\item (size = 2) Intel, Mimecast; This is also a combination between a large company and a small company.
	\item (size = 2) Amazon, Purdue University; In this dataset, Amazon Web Services and Amazon are two separate nodes; It is interesting to see that Purdue University can help Amazon through a merger between these two entities.            
\end{itemize}
Here, mergers represent business mergers and the size of each merger represents the scale of the corresponding business merger.
We conclude that the main case is that a giant company gets merged with a large number of companies in various fields, which is also the most common case in real-world situations~\citep{lamoreaux1988great}; it is fairly uncommon for two relatively large companies to be merged together, but not totally impossible.

\subsection{Q3: Effectiveness of the algorithmic choices} \label{subsec:exp_empirical_support}
We shall show several results that empirically support our three algorithmic choices: 
(1) excluding outside-outside mergers, and 
(2 \& 3) the heuristics for choosing promising inside nodes and mergers.

\input{TAB/Q3_summary.tex}
\input{FIG/random_one_round.tex}

\smallsection{Exclude outside-outside mergers.}
For each dataset and each $k \in \set{5, 10, 15, 20}$, we randomly sample $10,000$ \textbf{inside-inside (II)/inside-outside (IO)/outside-outside (OO)} mergers, and report the highest increase in the truss size among all the $10,000$ sampled mergers in each of the three cases.
For each experimental setting (dataset and $k$) and each case (II/IO/OO), we do five independent trials and take the average.
We do this for one round (i.e., the budget $b = 1$).
For each dataset, we compute the truss-size increase and the number of possible mergers of each case {averaged over} all the considered $k$ values.
Compared to II and IO mergers, the number of OO mergers is much higher ($38\times$ on average), while their performance is much lower ($1.8\times$ on average).
{Thus, excluding OO mergers in our proposed method, \bmshort, for speed is justified.}

\smallsection{Heuristics for choosing nodes and mergers.}
Specifically, there are three kinds of heuristics we shall compare: the heuristics for choosing inside nodes, those for choosing IOMs, and those for choosing IIMs.
The considered \textbf{heuristics for choosing inside nodes} are:
\begin{itemize}[leftmargin=*]
\item \textbf{IP (Most inside prospects)}: choosing the inside nodes with most inside prospects (Definition~\ref{def:incident_prospects});
\item \textbf{IN (Most inside neighbors)}: choosing the inside nodes with most inside neighbors (Definition~\ref{def:inside_outside_nodes_inside_nbrs});
\item \textbf{RD (Random)}: sampling inside nodes uniformly at random. We report the average performance over three independent trials.
\end{itemize}
The considered \textbf{heuristics for choosing IOMs} are:    
\begin{itemize}[leftmargin=*]
    \item \textbf{SE (Most potentially helped shell edges)}: choosing the mergers with most potentially helped shell edges (Definition~\ref{def:PHSEs});
    \item \textbf{NN (Most new neighbors)}: choosing the mergers that bring most ``new'' neighbors to the inside node in the mergers;
    \item \textbf{RD (Random)}: sampling mergers uniformly at random.\footnote{\label{footnote:random_avg}We report the average performance over three independent trials.}
\end{itemize}
The considered \textbf{heuristics for choosing IIMs} are:    
\begin{itemize}[leftmargin=*]
    \item \textbf{SE\footnote{The SE for IOMs can be seen as a special case of the SE for IIMs since the -1 scores are only possible for IIMs, and thus we use the same abbreviation for both cases.} (Scoring using shell edges)}: choosing the mergers with highest scores that are described in Section~\ref{subsec:pair_heu},\footnote{+1 / -1 for each non-incident shell edge with support gains / losses; also -1 for each collision between two edges in the current $k$-truss;}
    \item \textbf{AE (Scoring using all edges in the $(k-1)$-truss)}: choosing the mergers that with highest scores {that are measured as in {SE} but based on {all edges in the $(k-1)$-truss} instead of shell edges;}
    \item \textbf{RD (Random)}: sampling mergers uniformly at random.\footref{footnote:random_avg}
\end{itemize}
For each kind of heuristics, for each dataset and each $k \in \set{5, 10, 15, 20}$, we use each considered heuristic to choose the inside nodes or mergers, and report the best performance among all the mergers using the chosen inside nodes (or all the chosen mergers).
We choose $100$ inside nodes using each heuristics for choosing inside nodes, and choose $10$ mergers using each heuristic for choosing mergers.
We always use $50$ outside nodes with most inside neighbors.
We do this for one round (i.e., the budget $b = 1$).

We compare the best performance among all the possible 
IIMs (the $4950$ pairs among the $100$ chosen inside nodes) or 
IOMs (the $5000$ pairs between the $100$ chosen inside nodes and the $50$ outside nodes with most inside neighbors)
using the $100$ inside nodes chosen by each heuristic.
On all datasets, IP outperforms the random baseline, with an average superiority of $1.4\times$ and $4.3\times$, for IOMs and IIMs, respectively.
Overall, the performance of IP is better than that of IN.
Hence, {our proposed algorithm, \bmshort, employs} the heuristic IP to choose the inside nodes.
Below, for the heuristics choosing pairs, the inside nodes are chosen by IP.

We compare the best performance among the $10$ IIMs/IOMs chosen by each heuristic.
{We also include the \textbf{best (BE)} performance among all possible mergers from the considered inside and outside nodes (i.e., the results of IP in Figure~\ref{fig:random_one_round}(b)), as a reference.}
In both cases, SE outperforms the other competitors consistently.
Notably, in both bases, SE achieves results close (93.2\% for IOMs and 96.6\% for IIMs) to the best possible performance (BE).
{Thus, our proposed algorithm, \bmshort, uses the SE heuristic.}
See Figure~\ref{fig:random_one_round} for the detailed comparison among the algorithmic choices.
We provide summarized results in Table~\ref{tab:q3_summary}.
We summarize the results in the table as follows:
\begin{itemize}[leftmargin=*]
   \item
    In Table~\ref{tab:q3_summary}(a), we show the best performance (Perf.) among $10000$ random inside-inside (II) / inside-outside (IO) / outside-outside (OO) mergers, and the total number of mergers (\#) of each case.
    Compared to IIMs and IOMs, the number of OOMs is much higher, while their performance is much lower, which justifies excluding them in BATMAN.  
    \item
    In Table~\ref{tab:q3_summary}(b), we show the best performance among all the IOMs / IIMs using the $100$ inside nodes chosen by each heuristic. Overall, the heuristic used in \bmshort for choosing inside nodes, IP, outperforms the competitors.     
    \item
    In Table~\ref{tab:q3_summary}(c), we show the best performance among all the IOMs / IIMs using the $10$ outside nodes chosen by each heuristic and the $100$ inside nodes chosen by IP. Overall, the heuristic used in \bmshort for choosing mergers, SE, outperforms the competitors, achieving a performance close to the best possible (BE) results achievable using the $100$ inside nodes chosen by IP. 
\end{itemize}

\input{TAB/stat_test}
\subsection{Tests for statistical significance}\label{subsec:stat_test}
We conduct tests for the statistical significance of the empirical superiority of our proposed method \bmshort, especially over the variants.
We generate 400 random graphs using the Watts-Strogatz small-world random graph model~\citep{watts1998collective} with $n = 1000$ nodes, $d \in \set{10, 20, 30, 40}$, and $p = 0.5$, where $d$ is the number of nearest neighbors each node is joined with, and $p$ is the probability of rewiring each edge. 
For each $d$ value, we generate 100 random graphs.
We apply the proposed method (BM) over the variants (EQ, II, and IO) to the 400 random graphs and obtain the 400 final truss sizes $(s^{M}_{1}, s^{M}_2, \ldots, s^{M}_{400})$ for each method $M \in \{\text{BM}, \text{EQ}, \text{II}, \text{IO}\}$.
We apply the proposed method and the variants to maximize $(k = 5)$-truss with $b = 10 $ mergers, $n_i = 100$ candidate inside nodes, $n_o = 50$ candidate outside nodes, and $n_c = 100$ pairs to check in each round.
For each variant $T \in \{EQ, II, IO\},$ we compute the differences in the final truss size ${D}^T = (s^{BM}_{1} - s^{T}_{1}, s^{BM}_{2} - s^{T}_{2}, \ldots, s^{BM}_{400} - s^{T}_{400})$, and apply a one-tailed one-sample $t$-test to this population $D^T$ with the null hypothesis $H_0$: the mean value of $D_T \leq 0$.
See Table~\ref{tab:stat_test} for the details ($t$-statistics and $p$-values) of the tests, where we observe that \bmshort outperforms the variants with statistical significance.

Note that the only difference between the proposed method \bmshort and the variants is the adaptive candidate-distribution component. Intuitively, such a candidate-distribution component enhances the stability of our method since a wider range of candidate mergers is considered (as shown in Fig~6), which is validated by the cases when the variants (EQ, II, and IO) significantly perform worse than the proposed method. The candidate-distribution component is used mainly to improve the worst-case performance, and such a target is achieved as shown by the cases where the variants without such a component perform much worse than \bmshort with such a component.

It is definitely possible that in some scenarios, e.g., inside-inside mergers (IIMs) are consistently better than inside-outside mergers (IOMs). In such cases, \bmshort would increase the proportion of IIMs in the candidate mergers and can perform comparably to II (as we observe in the experimental results). We will further clarify this in the revised manuscript.

\section{Additional experiments on real-world bus station datasets}\label{app:add_exp_bus}
We conduct additional experiments on real-world bus station datasets, where real-world distance constraints are considered.

\subsection{Necessary algorithmic modifications}
As mentioned in Section~\ref{sec:on_rw_examples}, we can make modifications in Algorithms 2 and 3 to incorporate distance constraints (or other real-world constraints).
Here, provide more details.

Suppose that we are given a distance threshold $d_{th}$ and we are allowed to merge stations within the distance threshold only.
For each pair of nodes, $v_1$ and $v_2$, let $\operatorname{dist}(v_1, v_2)$ denote the distance between $v_1$ and $v_2$.
In Algorithm~2, after Line~5 and before Line~6, we can check the distance between $v_i$ and $v_o$, and skip the pair $(v_i, v_o)$ if the distance between the two nodes exceeds the given threshold.
Similarly, in Algorithm~3, after Line~2 and before Line~3,
we can check the distance between $v_1$ and $v_2$, and skip the pair $(v_1, v_2)$ if the distance between the two nodes exceeds the given threshold.
By doing so, we make sure that the candidate mergers all satisfy the distance-threshold constraints.

\input{TAB/bus_datasets.tex}
\subsection{Experimental results}
We conduct experiments on real-world bus station datasets~\citep{kujala2018collection}.
Among the 25 available datasets, four datasets are not included because they are too sparse.
See Table~\ref{tab:datasets_bus} for the basic statistics of the datasets.
Note that the number of possible mergers can be considerably large in real-world bus station networks, and thus it is nontrivial to find good mergers.
We set the distance threshold as one kilometer for all the datasets.
We take the largest connected component of each dataset.

We compare the proposed method \bmshort with 3 baseline methods:
\begin{itemize}[leftmargin=*]
	\item \textbf{BM (\bmshort)}: the proposed method with necessary modifications to consider distance constraints;
	\item \textbf{CR (core)}: a method for the counterpart problem considering $k$-cores;
	\item \textbf{CS (constraints)}: a method considering distance constraints only, and pricking pairs satisfying the constraints uniformly at random (the average performance over 10 random trials are reported);
	\item \textbf{CL (closest)}: a method greedily merging the pairs with the smallest distance.
\end{itemize}

\input{FIG/bus_visual_single.tex}

\input{TAB/bus_results.tex}
The performance is evaluated by 8 robustness measures that have been considered for transportation networks:
\textbf{(1)} VB (average vertex betweenness)~\citep{lordan2014robustness,wandelt2021estimation,peng2018fine,cats2020metropolitan},
\textbf{(2)} EB (average edge betweenness)~\citep{lordan2014robustness,wandelt2021estimation,peng2018fine,cats2020metropolitan},
\textbf{(3)} ER (effective resistance)~\citep{yang2018designing,wang2017multi,yamashita2021effective},
\textbf{(4)} SG (spectral gap)~\citep{candelieri2019vulnerability,wang2017multi,estrada2006network},
\textbf{(5)} NC (natural connectivity)~\citep{wang2017multi,frutos2019study},
\textbf{(6)} AD (average distance)~\citep{frutos2019study,mouronte2012structural},
\textbf{(7)} TS (transitivity)~\citep{julliard2015robustness,frutos2019study,wang2017multi}, and
\textbf{(8)} LC (average local clustering coefficient)~\citep{julliard2015robustness,frutos2019study,wang2017multi}.
The proposed method performs best overall.
See Tables~\ref{tab:bus_results_z_score} and \ref{tab:bus_results_avg_rank} for the detailed results.
For each dataset, we apply the mergers output by each considered method, and compute each considered robustness metric of the post-merger graph. 
We compute for each metric the Z-score (standard score) and the average ranking (the arithmetic mean of the ranking w.r.t the corresponding metric, over all datasets).
We also report the average Z-scores and the average (again, the arithmetic mean) of the average rankings over all the considered metrics.

In Figures~\ref{fig:bus_visual_kuopio}-\ref{fig:bus_visual_rome}, we visualize the results of different methods on several datasets.
For each dataset, we draw each node (bus station) with the location information and the topology information (i.e., routes).
In each plot, each point represents a node (station), and each black line segment between two nodes represents an edge (route).
We also mark the merged nodes (bus stations) in different colors, where the nodes merged together are in the same color.
The visualization of graphs helps us obtain a direct and intuitive understanding of the structure of the bus system and helps us to more easily analyze the pattern of merged stations.

%% file: TAB/datasets.tex
\begingroup
\setlength{\tabcolsep}{10.5pt}
\renewcommand{\arraystretch}{0.85}
\begin{table*}[t!]     
	\begin{center}
		\caption{The basic statistics of the $14$ real-world datasets.  
  Notations: $n$ denotes the number of nodes, $n_k$ the number of nodes in $T_k$, $m$ the number of edges, $m_k$ the number of edges in $T_k$, and $k_{max}$ the maximum $k$ such that $T_k$ is non-empty.}
		\label{tab:datasets}
			\resizebox{\linewidth}{!}{%
			\begin{tabular}{ l*{11}{r}}
			\toprule
			\textbf{Dataset}& $n$ & $m$ & $k_{max}$ &
                    $n_{5}$ & $m_{5}$ & $n_{10}$ & $m_{10}$ &
                    $n_{15}$ & $m_{15}$ & $n_{20}$ & $m_{20}$ \\
			\midrule
			email (EM) & 986 & 16,064 & 23 & 743 & 14,771 & 492 & 10,494 & 257 & 5,308 & 73 & 1,622 \\
			facebook (FB) & 4,038 & 87,887 & 97 & 3,599 & 85,336 & 2,509 & 74,436 & 1,707 & 62,567 & 1,196 & 52,884 \\
			enron (ER) & 33,696 & 180,811 & 22 & 13,983 & 139,351 & 2,159 & 53,913 & 769 & 21,837 & 192 & 4,441 \\
			brightkite (BK)  & 56,739 & 212,945 & 43 & 8,009 & 74,498 & 1,454 & 27,742 & 544 & 15,950 & 353 & 12,274 \\
			relato (RL)  & 54,007 & 251,370 & 44 & 6,897 & 144,787 & 2,386 & 89,041 & 1,282 & 60,093 & 781 & 41,808 \\
			epinions (EP)  & 75,877 & 405,739 & 33 & 9,706 & 218,990 & 3,138 & 111,694 & 1,357 & 55,560 & 593 & 25,679 \\
			hepph (HP)  & 34,401 & 420,784 & 25 & 22,760 & 298,416 & 5,011 & 75,343 & 864 & 14,065 & 124 & 2,109 \\
			slashdot (SD) & 77,360 & 469,180 & 35 & 4,048 & 72,554 & 638 & 19,174 & 372 & 13,036 & 237 & 9,554 \\
			syracuse (SC)  & 13,640 & 543,975 & 59 & 12,274 & 484,914 & 8,696 & 301,374 & 5,446 & 185,365 & 3,672 & 128,992 \\
			gowalla (GW)  & 196,591 & 950,327 & 29 & 42,860 & 434,483 & 7,163 & 140,993 & 2,060 & 52,009 & 531 & 16,381 \\
			twitter (TT)  & 81,306 & 1,342,296 & 82 & 61,162 & 1,255,418 & 35,354 & 961,958 & 21,911 & 697,239 & 13,592 & 479,795 \\
			stanford (SF) & 255,265 & 1,941,926 & 62 & 151,955 & 1,569,406 & 49,199 & 934,901 & 33,980 & 694,205 & 16,157 & 383,159 \\
			youtube (YT)  & 1,134,890 & 2,987,624 & 19 & 42,508 & 543,739 & 4,061 & 120,055 & 998 & 33,637 & 0 & 0 \\
			wikitalk (WT) & 2,388,953 & 4,656,682 & 53 & 34,509 & 811,728 & 6,577 & 405,501 & 3,349 & 281,684 & 2,259 & 214,676 \\
			\bottomrule
		\end{tabular}
  		}
	\end{center}
\end{table*}
\endgroup

%% file: eff_three_random_models.tex
    \begin{table}[t!]        
    \begin{subtable}[t]{\columnwidth}        
    	\begin{center}
    		\caption*{(a) The Erd\"{o}s-R\'{e}nyi model.}      
    			\resizebox{0.8\linewidth}{!}{%
    			\begin{tabular}{ c | c | *{10}{c}}
    				\toprule
    				measure & \# operations & 1 & 2 & 3 & 4 & 5 & 6 & 7 & 8 & 9 & 10 \\
    				\midrule                
                    VB & merging & 83.1 & {\textbf{79.2}} & 75.5 & 71.9 & 68.4 & 65.4 & 62.5 & 59.9 & 57.5 & 55.4 \\
                    87.3 & adding & 86.2 & 85.4 & 84.7 & 84.1 & 83.5 & 83.0 & 82.5 & 82.1 & 81.6 & 81.2 \\
                    \midrule
                    EB & merging & 23.3 & 21.6 & {\textbf{20.0}} & 18.6 & 17.4 & 16.3 & 15.2 & 14.2 & 13.3 & 12.4 \\
                    25.3 & adding & 24.6 & 24.1 & 23.6 & 23.2 & 22.8 & 22.5 & 22.1 & 21.7 & 21.4 & 21.1 \\
                    \midrule
                    ER & merging & 714.8 & 633.0 & {\textbf{570.2}} & 515.7 & 468.5 & 427.0 & 389.3 & 355.2 & 325.1 & 297.5 \\
                    834.8 & adding & 776.1 & 739.5 & 712.3 & 689.9 & 671.1 & 654.0 & 639.7 & 626.0 & 613.1 & 601.3 \\
                    \midrule
                    NC & merging & 3.1 & 3.6 & {\textbf{4.0}} & 4.3 & 4.7 & 5.0 & 5.4 & 5.7 & 6.0 & 6.2 \\
                    2.7 & adding & 2.8 & 2.9 & 3.0 & 3.1 & 3.2 & 3.4 & 3.5 & 3.6 & 3.7 & 3.9 \\
                    \midrule
                    SG & merging & 3.1 & 3.7 & {\textbf{4.3}} & 4.7 & 5.2 & 5.5 & 5.9 & 6.2 & 6.5 & 6.8 \\
                    2.3 & adding & 2.5 & 2.7 & 2.9 & 3.1 & 3.2 & 3.4 & 3.6 & 3.7 & 3.9 & 4.0 \\
    				\bottomrule %
    			\end{tabular}
    		}
    	\end{center}
    \end{subtable} 
    \begin{subtable}[t]{\columnwidth}
        \begin{center}
    		\caption*{(b) the Watts-Strogatz small-world model.}
    			\resizebox{0.8\linewidth}{!}{%
    			\begin{tabular}{ c | c | *{10}{c}}
    				\toprule
    				measure & \# operations & 1 & 2 & 3 & 4 & 5 & 6 & 7 & 8 & 9 & 10 \\
    				\midrule                
                    VB & merging & {\textbf{70.1}} & 67.7 & 65.5 & 63.5 & 61.7 & 60.0 & 58.5 & 56.9 & 55.3 & 53.7 \\
                    72.58 & adding & 72.3 & 72.1 & 71.9 & 71.7 & 71.4 & 71.3 & 71.1 & 70.9 & 70.7 & 70.5 \\
                    \midrule
                    EB & merging & 10.2 & {\textbf{9.6}} & 9.1 & 8.6 & 8.1 & 7.6 & 7.2 & 6.8 & 6.4 & 6.0 \\
                    10.8 & adding & 10.6 & 10.6 & 10.5 & 10.4 & 10.3 & 10.2 & 10.1 & 10.0 & 9.9 & 9.9 \\
                    \midrule
                    ER & merging & 343.1 & {\textbf{319.2}} & 296.3 & 274.3 & 253.4 & 233.9 & 215.3 & 198.2 & 182.1 & 167.2 \\
                    369.0 & adding & 364.6 & 360.7 & 356.8 & 353.1 & 349.6 & 346.1 & 342.8 & 339.7 & 336.8 & 333.9 \\
                    \midrule
                    NC & merging & 8.0 & 8.4 & {\textbf{8.8}} & 9.2 & 9.5 & 9.8 & 10.1 & 10.4 & 10.7 & 11.0 \\
                    7.5 & adding & 7.6 & 7.8 & 7.9 & 8.0 & 8.2 & 8.3 & 8.4 & 8.5 & 8.6 & 8.7 \\
                    \midrule
                    SG & merging & 3.1 & 3.7 & {\textbf{4.3}} & 4.7 & 5.2 & 5.5 & 5.9 & 6.2 & 6.5 & 6.8 \\
                    2.3 & adding & 2.5 & 2.7 & 2.9 & 3.1 & 3.2 & 3.4 & 3.6 & 3.7 & 3.9 & 4.0 \\
    				\bottomrule %
    			\end{tabular}
    		}
    	\end{center}
    \end{subtable}
    \begin{subtable}[t]{\columnwidth}
\begin{center}
    		\caption*{(c) the Holme-Kim powerlaw-cluster model.}
    			\resizebox{0.8\linewidth}{!}{%
    			\begin{tabular}{ c | c | *{10}{c}}
    				\toprule
    				measure & \# operations & 1 & 2 & 3 & 4 & 5 & 6 & 7 & 8 & 9 & 10 \\
    				\midrule                
                    VB & merging & {\textbf{73.4}} & 69.5 & 65.8 & 63.0 & 60.8 & 59.1 & 57.4 & 55.8 & 54.2 & 52.6 \\
                    77.5 & adding & 76.9 & 76.4 & 76.0 & 75.6 & 75.2 & 74.9 & 74.5 & 74.2 & 73.8 & 73.5 \\
                    \midrule
                    EB & merging & 9.7 & {\textbf{8.8}} & 8.1 & 7.5 & 7.0 & 6.6 & 6.3 & 5.9 & 5.6 & 5.3 \\
                    10.6 & adding & 10.4 & 10.3 & 10.2 & 10.1 & 10.0 & 9.8 & 9.7 & 9.6 & 9.5 & 9.4 \\
                    \midrule
                    ER & merging & 276.1 & {\textbf{254.7}} & 236.3 & 219.8 & 204.4 & 190.1 & 176.7 & 164.0 & 152.0 & 140.7 \\
                    300.2 & adding & 296.4 & 293.1 & 289.9 & 287.1 & 284.3 & 281.8 & 279.3 & 277.0 & 274.7 & 272.6 \\
                    \midrule
                    NC & merging & 7.0 & {\textbf{7.7}} & 8.3 & 8.7 & 9.3 & 9.7 & 10.1 & 10.5 & 10.9 & 11.2 \\
                    6.5 & adding & 6.6 & 6.6 & 6.7 & 6.8 & 6.9 & 6.9 & 7.0 & 7.1 & 7.2 & 7.3 \\
                    \midrule
                    SG & merging & 2.8 & {\textbf{3.6}} & 4.3 & 4.8 & 5.6 & 6.1 & 6.6 & 7.1 & 7.6 & 8.1 \\
                    1.7 & adding & 1.9 & 2.0 & 2.1 & 2.3 & 2.4 & 2.5 & 2.6 & 2.7 & 2.9 & 3.0 \\
    				\bottomrule %
    			\end{tabular}
    		}
    	\end{center}
    \end{subtable}
\caption{Results of the effectiveness study using three different random graph models. For each measure, we report the change of it when we do 10 times of merging nodes or adding edges, over five independent trials, with the original value below the measure name. For each measure, the first result of merging nodes that is better than adding 10 edges is marked in bold.\label{tab:effective_merging_models}}
    \end{table}

%% file: TAB/robust_correlation_truss_core.tex
\begingroup
\setlength{\tabcolsep}{10pt}
\renewcommand{\arraystretch}{0.85}
\begin{table}[t!]
  \centering
  \caption{The correlations between the core size and the robustness measures are clearly weaker than the correlations between the truss size and the robustness measures. For each robustness metric, we list Pearson’s $r$ between the truss size and the metric when we try to maximize the size of a k-truss, and Pearson’s $r$ between the core size and the metric when we try to maximize the size of a $k$-core.}
  \scalebox{1.0}{%
    \begin{tabular}{lrr}
    \toprule
    metric & truss & core \\
    \midrule
    VB    & 0.99 & 0.55 \\
    EB    & 0.98 & 0.57 \\
    ER    & 0.97 & 0.51 \\
    SG    & 0.99 & 0.50 \\
    NC    & 0.99 & 0.55 \\
    \bottomrule
    \end{tabular}%
    }
  \label{tab:robust_correlation_truss_core}%
\end{table}%
\endgroup

%% file: FIG/trade_off.tex
\begin{figure*}[t!]
    \centering
    \includegraphics[scale=0.3]{FIG/main_legend.pdf} \\
    \vspace{1mm}    
    \includegraphics[height=3.2cm]{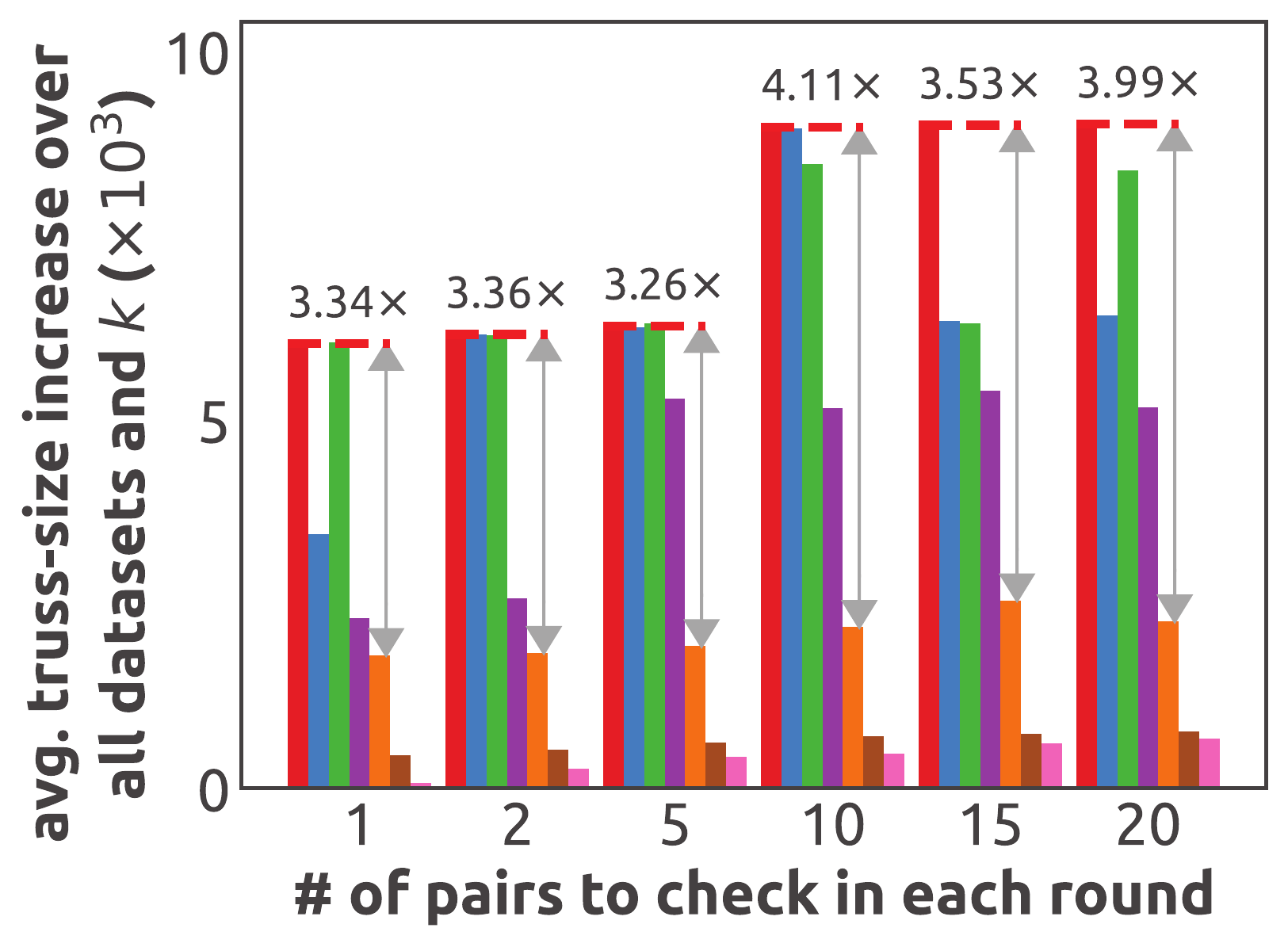}
    \includegraphics[height=3.2cm]{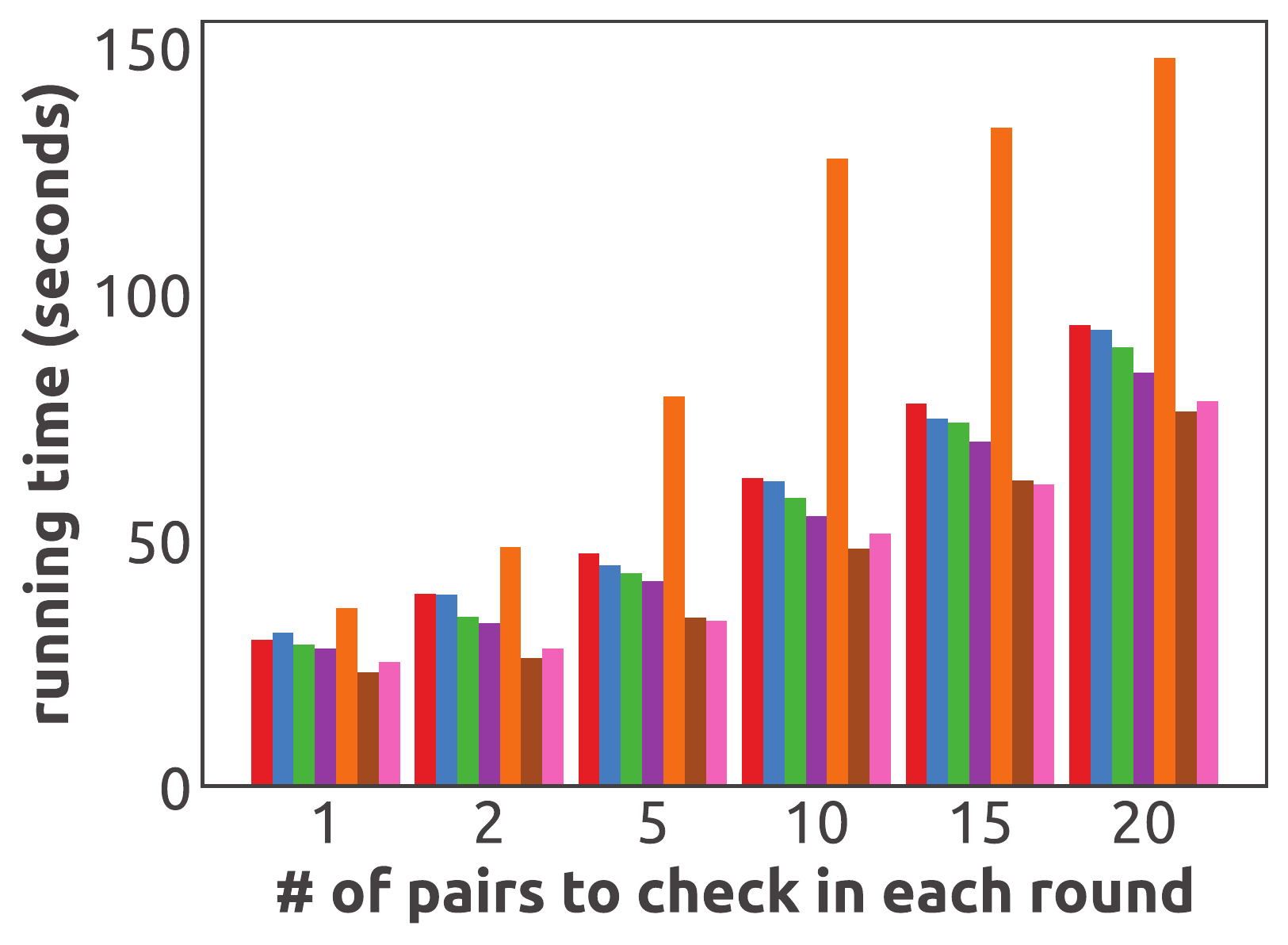}
    \includegraphics[height=3.2cm]{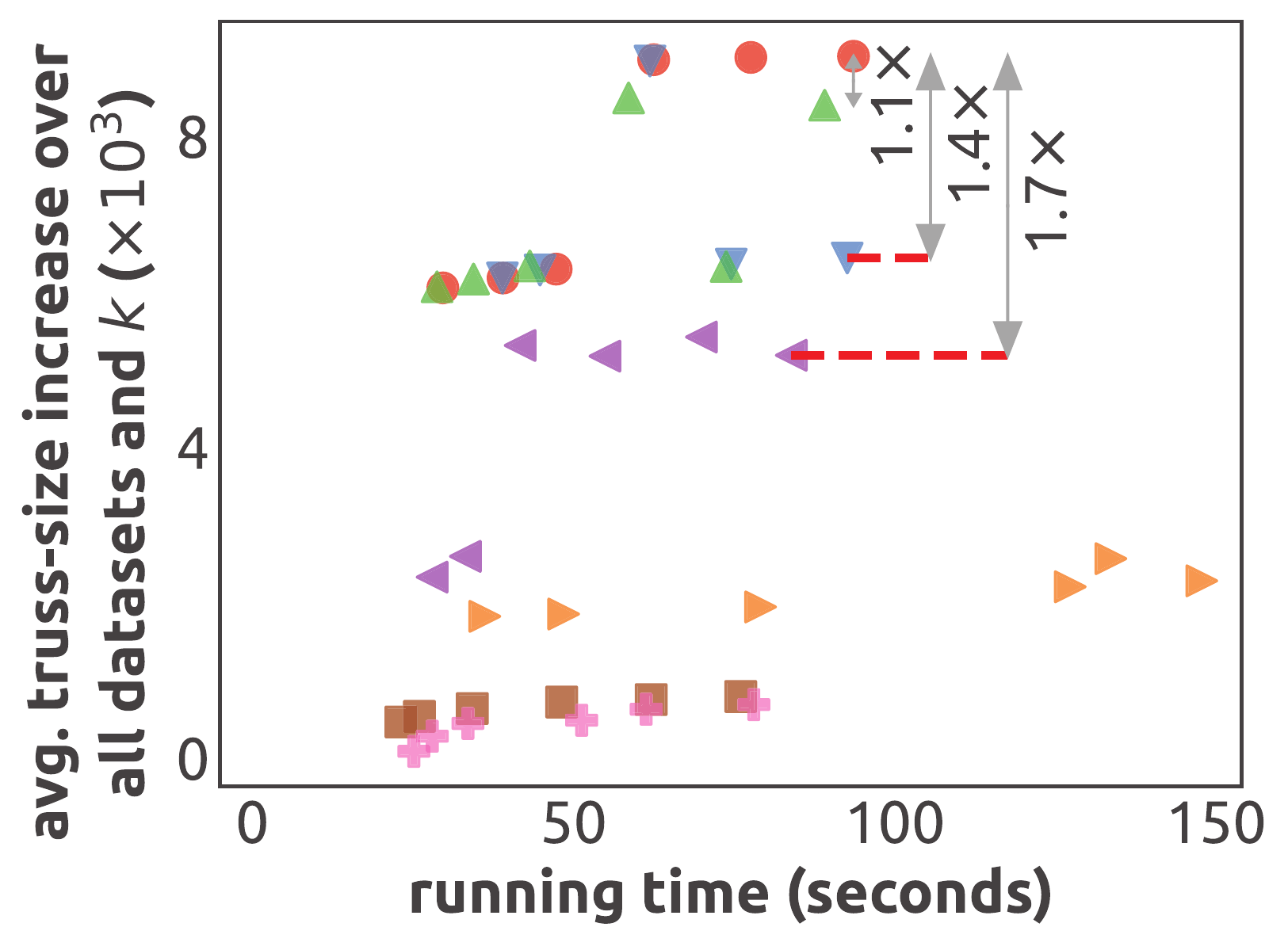}
    \includegraphics[height=3.2cm]{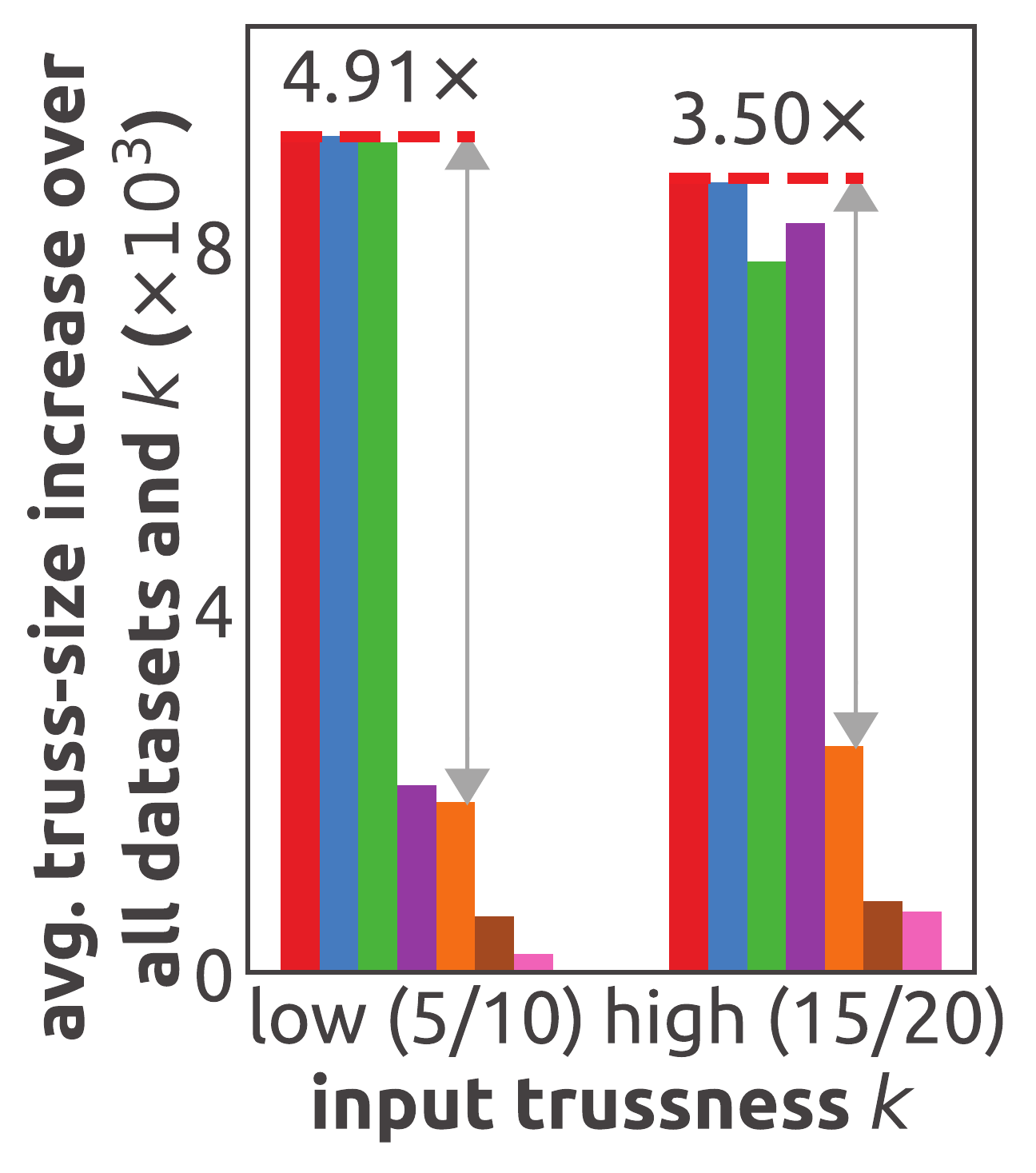}
    \vspace{1mm}
    \caption{\underline{\smash{The first three subfigures on the left:}} The average performance and running time of each considered algorithm over all datasets and $k$ values when the number of candidate pairs checked in each round varies.
    The proposed algorithm (BM) and its variants (EQ, II, and IO) clearly outperform the baseline methods (NT, NE, and RD).
    BM outperforms the baseline methods even when the baseline methods check more candidates;
    and BM is more effective and more stable than its variants, especially when we check more candidates (see Appendix~J for the tests for statistical significance of the superiority of the proposed method over the variants).
    \underline{\smash{The rightmost subfigure:}} The average performance of each algorithm overall datasets when the trussness $k$ varies.
    The proposed method (BM) and its variants (EQ, II, and IO) outperform the baseline methods (NT, NE, and RD) for both low and high $k$ values.}
    \vspace{1mm}    
    \label{fig:tradeoff_avg}
    \label{fig:k_avg}
\end{figure*}

%% file: TAB/Q3_summary.tex
\begingroup
\setlength{\tabcolsep}{3pt}
\begin{table}[t!]
	\caption{\label{tab:q3_summary} Empirical support of our algorithmic choices: excluding outside-outside (OO) mergers and the proposed heuristics (IP and SE). The results are averaged over all the datasets. See Appendix~H for the results on each individual dataset.}
	\begin{subtable}[t]{0.30\textwidth}        
    \caption*{(a) Justification of excluding outside-outside mergers.}
		\centering           
        \scalebox{1.0}{  
		\begin{tabular}[t]{c c c}        
        \toprule        
        Type & Perf. & \# \\
        \midrule                
		II* & 408.9 & $10^{8.76}$ \\
        IO* & 269.1 & $10^{9.81}$ \\
        OO & 152.1 & $10^{11.39}$ \\
        \bottomrule
        \multicolumn{3}{l}{* used in \bmshort} \\
		\end{tabular}        
        } 
	\end{subtable}
	\hfill
	\begin{subtable}[t]{0.30\textwidth}
        \caption*{(b) Justification of using the heuristic IP to choose inside nodes.}
		\centering
        \scalebox{1.0}{ 
		\begin{tabular}[t]{c c c}
        \toprule
        & \multicolumn{2}{c}{Performance} \\
        Heur. & IOM & IIM \\
        \midrule        
		IP* & 562.3 & 1496.3 \\
        IN & 547.9 & 1407.9 \\
        RD & 397.9 & 326.3 \\
        \bottomrule
        \multicolumn{3}{l}{* used in \bmshort}
		\end{tabular}}
	\end{subtable}
	\hfill
	\begin{subtable}[t]{0.30\textwidth}
        \caption*{(c) Justification of using the heuristic SE to choose mergers.}
		\centering
  \scalebox{1.0}{ 
		\begin{tabular}[t]{c c c}
        \toprule
         & \multicolumn{2}{c}{Performance} \\
        Heur. & IOM & IIM \\
        \midrule
		SE*      & 524.1     & 1445.5 \\
        NN/AE   & 203.9     & 1348.9 \\
        RD      & 182.8     & 339.7 \\
        BE      & 562.3     & 1496.3 \\
        \bottomrule
        \multicolumn{3}{l}{* used in \bmshort}
		\end{tabular}}
	\end{subtable} \\
\end{table}
\endgroup

%% file: FIG/random_one_round.tex
\begin{figure*}[htb]
    \centering
    \begin{subfigure}[t]{0.32\textwidth}
        \centering
        \includegraphics[scale=0.3]{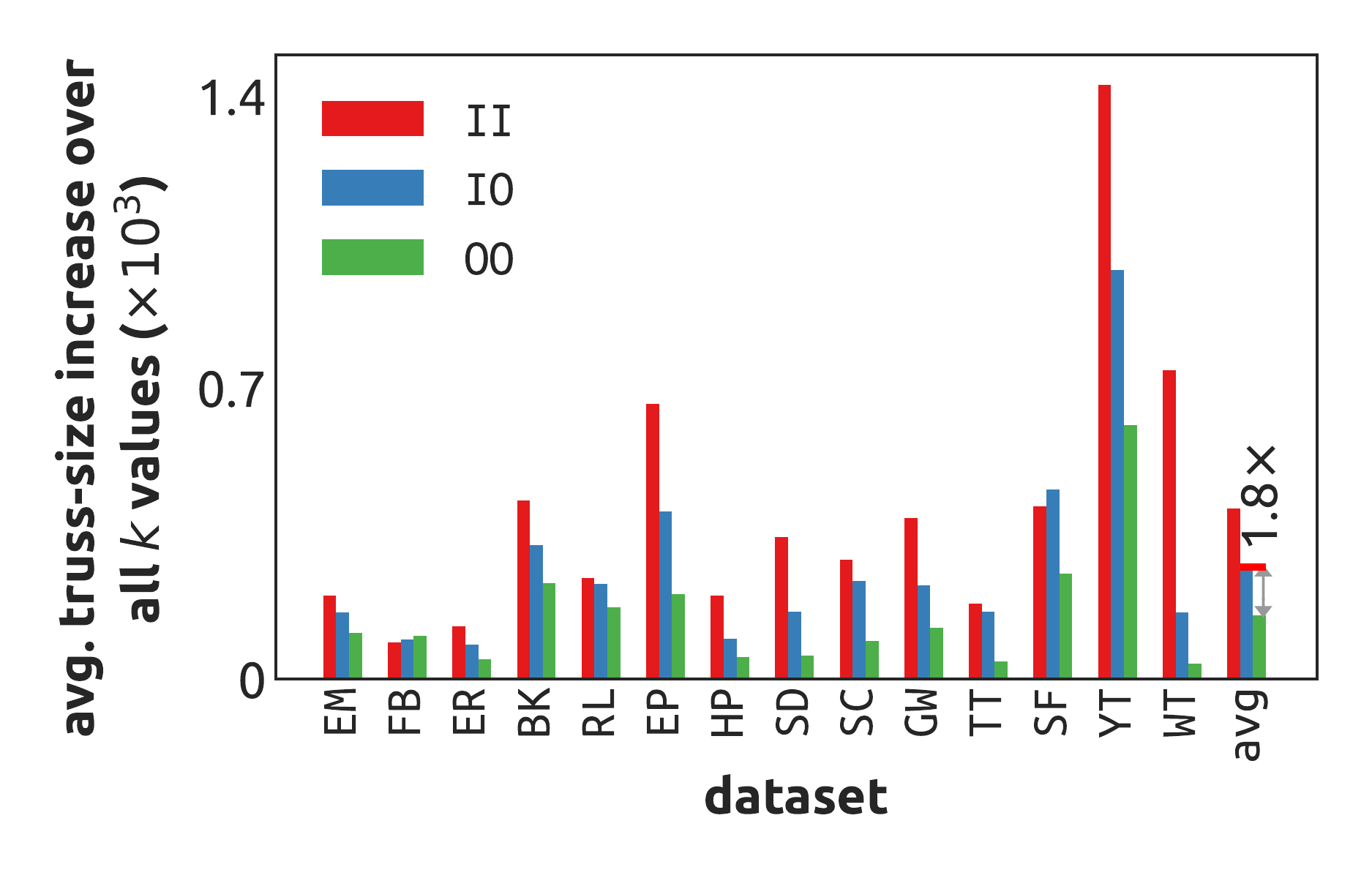}            
        \includegraphics[scale=0.3]{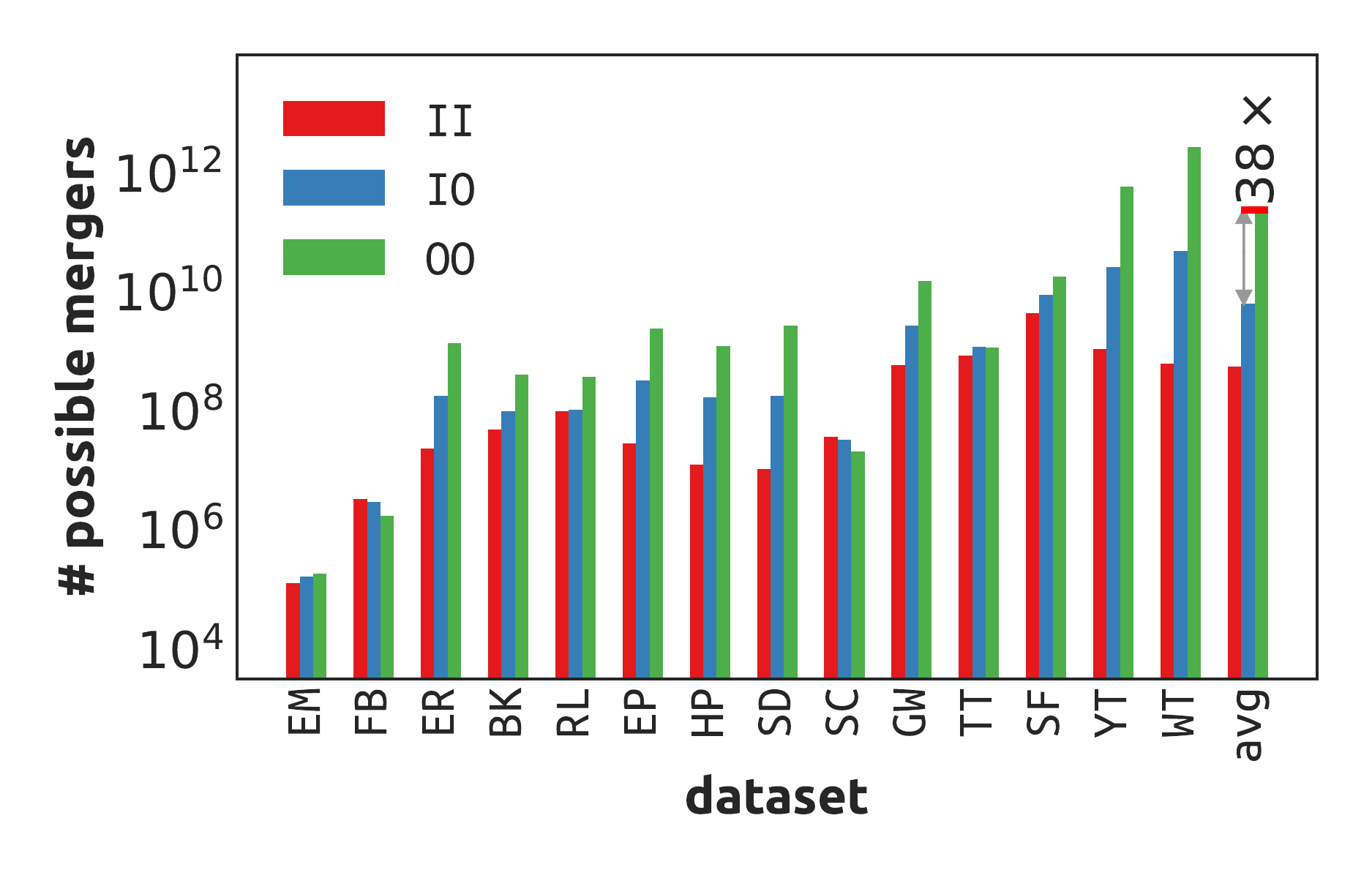}                    
        \vspace{-6mm}
        \caption{Top: the best performance among $10000$ random inside-inside (II) / inside-outside (IO) / outside-outside (OO) mergers.
        Bottom: the total number of possible mergers of each case.
        Overall, OO mergers have the highest number and worse performance.}
    \end{subfigure}
    \hfill
    \begin{subfigure}[t]{0.32\textwidth}
        \centering        
        \includegraphics[scale=0.3]{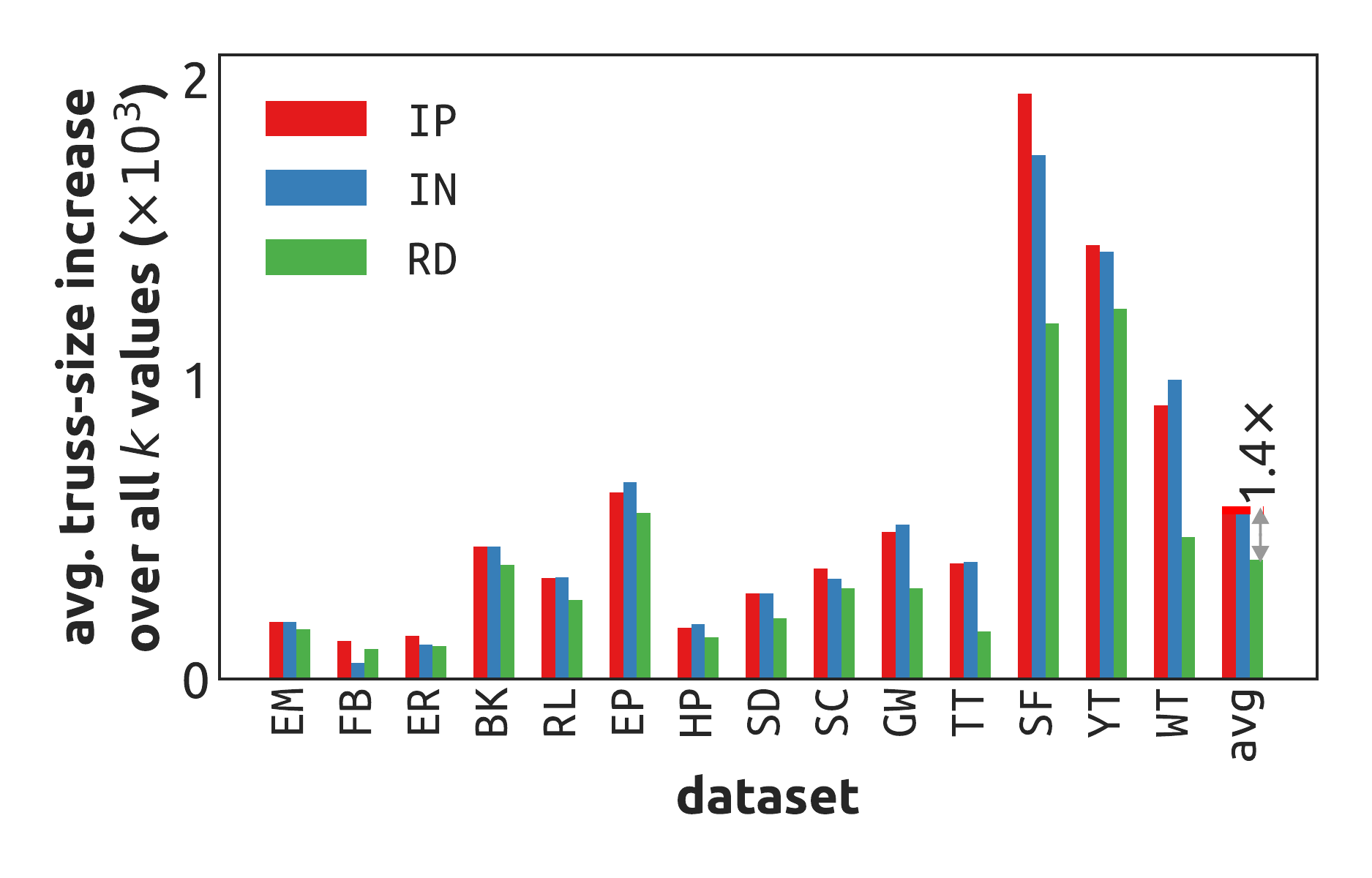}            
        \includegraphics[scale=0.3]{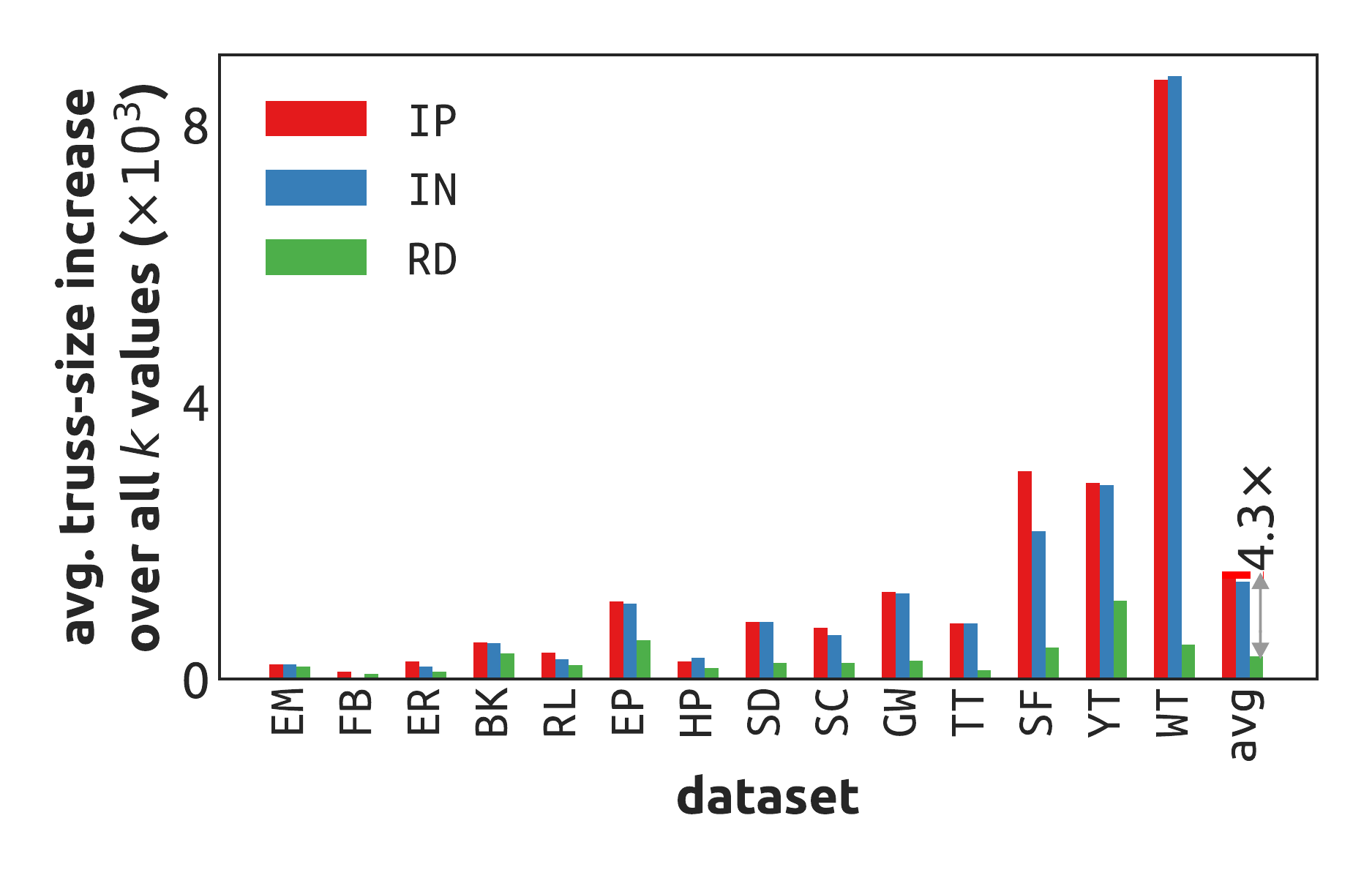}            
        \vspace{-6mm}
        \caption{Top/Bottom: the best performance among all the IOMs / IIMs using the $100$ inside nodes chosen by each heuristic.
        The heuristic used in \bmshort, IP, outperforms the random baseline consistently and performs more stably than IN with better overall performance.}
    \end{subfigure}
    \hfill
    \begin{subfigure}[t]{0.32\textwidth}
        \centering
        \includegraphics[scale=0.3]{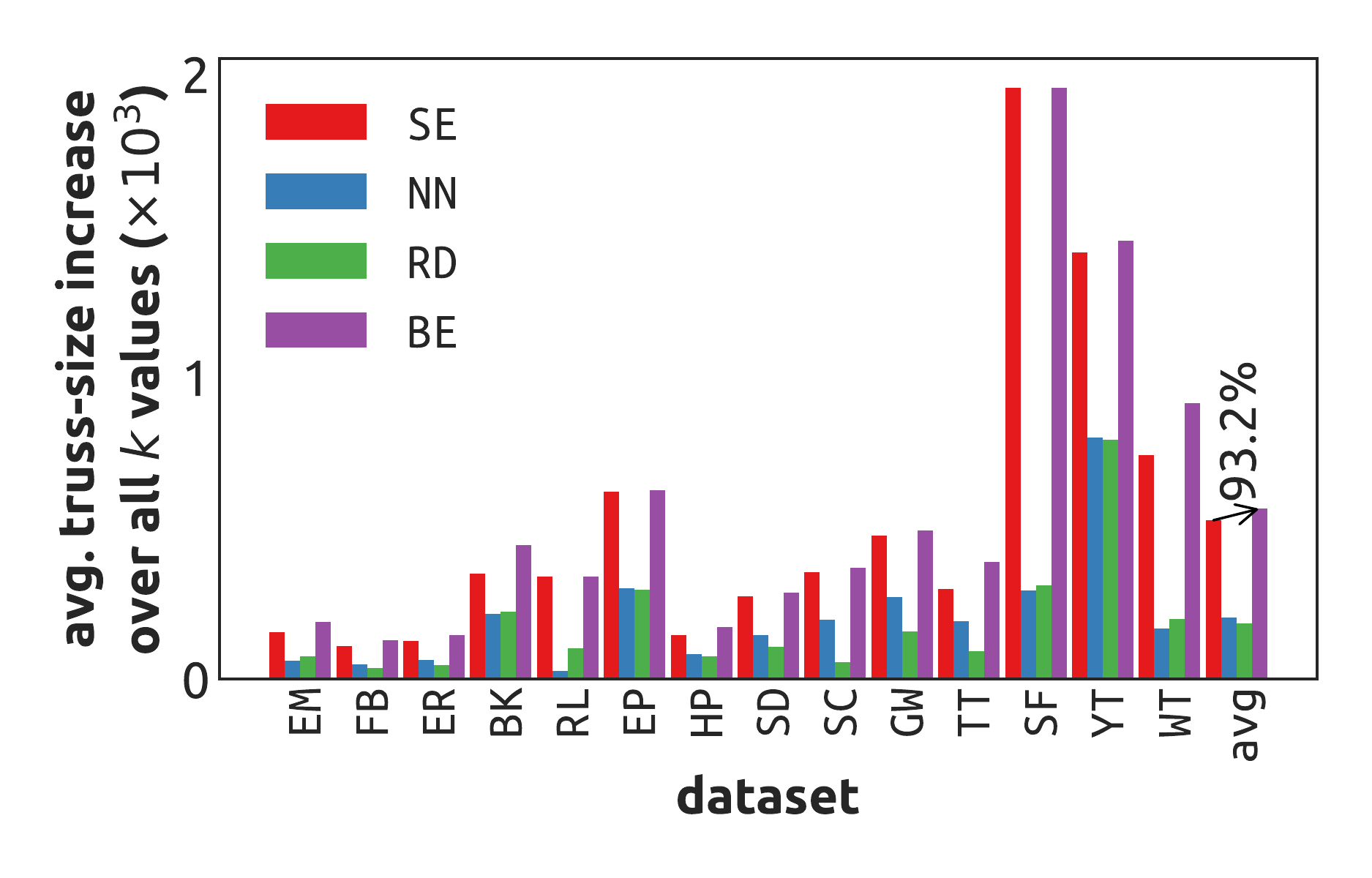}            
        \includegraphics[scale=0.3]{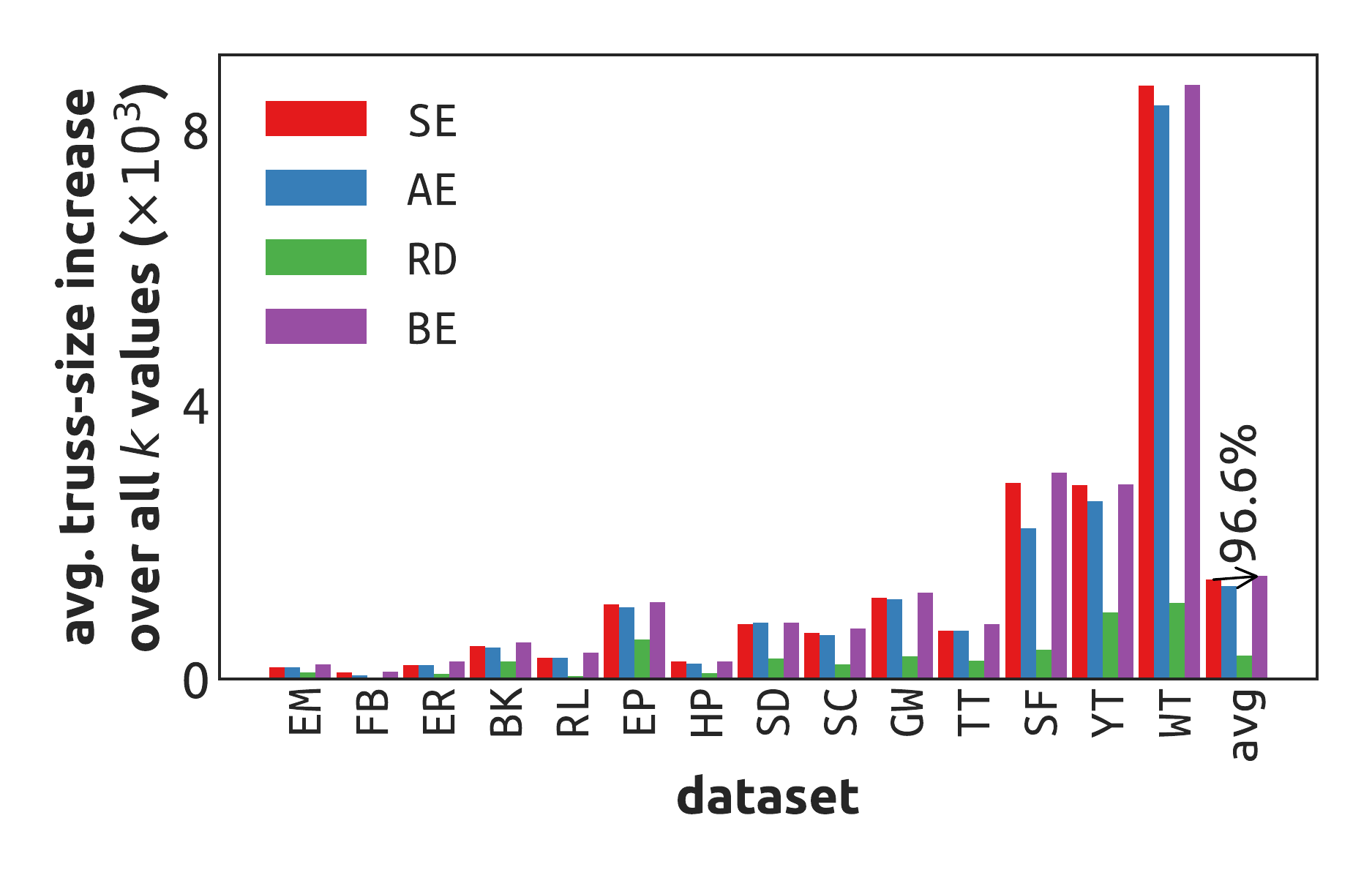}   
       \vspace{-6mm}
        \caption{Top/Bottom: the best performance among all the IOMs / IIMs using the $10$ outside nodes chosen by each heuristic and the $100$ inside nodes chosen by IP.
    {The performance of the heuristic used in \bmshort, SE, is consistently higher than that of the competitors, being very close (IOMs: 93.2\%; IIMs: 96.6\%) to the best possible (BE)}.}
    \end{subfigure}    
    \caption{Empirical support of our algorithmic choices: excluding outside-outside mergers and the proposed heuristics.
    }
    \label{fig:random_one_round}
\end{figure*}


%% file: TAB/stat_test.tex
\begingroup
\setlength{\tabcolsep}{10pt}
\renewcommand{\arraystretch}{0.85}
\begin{table}[t!]
  \centering
  \caption{The results of the tests for statistical significance. Smaller p-values indicate the superiority of \bmshort over the corresponding variant is more obvious. For each variant, the null hypothesis is clearly rejected with any reasonable significance level.}
  \scalebox{1.0}{%
    \begin{tabular}{lrr}
    \toprule
    variant & $t$-statistic & $p$-value \\
    \midrule    
    EQ    & 3.28  & 5.65e-04 \\
    II    & 4.75  & 1.41e-06 \\
    IO    & 25.85 & 1.40e-87 \\
    \bottomrule
    \end{tabular}%
    }
  \label{tab:stat_test}%
\end{table}%
\endgroup

%% file: TAB/bus_datasets.tex
\begingroup
\setlength{\tabcolsep}{15pt}
\renewcommand{\arraystretch}{0.85}
\begin{table}[t!]
	\begin{center}
		\caption{The basic statistics of the $21$ bus station datasets.
  Notations: $n$ denotes the number of nodes, $m$ the number of edges, 
  and $p_{th}$ the number of pairs within the $1$-km distance threshold.}
\label{tab:datasets_bus}
\resizebox{0.6\linewidth}{!}{%
\begin{tabular}{ l*{3}{r}}
\toprule
\textbf{Dataset}& $n$ & $m$ & $p_{th}$ \\
\midrule
kuopio & 528   & 676   & 6,695 \\
luxembourg & 1,350  & 1,850  & 9,382 \\
venice & 1,622  & 2,250  & 20,177 \\
turku & 1,817  & 2,304  & 37,094 \\
palermo & 2,176  & 2,559  & 72,907 \\
nantes & 2,206  & 2,579  & 40,198 \\
canberra & 2,520  & 2,908  & 42,143 \\
lisbon & 2,730  & 3,362  & 181,852 \\
bordeaux & 3,212  & 3,798  & 64,786 \\
berlin & 4,316  & 5,869  & 26,764 \\
dublin & 4,361  & 5,271  & 97,764 \\
prague & 4,441  & 5,862  & 58,002 \\
winnipeg & 5,079  & 5,846  & 159,139 \\
detroit & 5,683  & 5,946  & 140,450 \\
helsinki & 6,633  & 8,592  & 145,802 \\
adelaide & 7,210  & 8,827  & 131,111 \\
rome & 7,457  & 9,616  & 219,972 \\
brisbane & 9,279  & 11,242 & 181,933 \\
paris & 10,644 & 12,309 & 372,037 \\
melbourne & 17,250 & 19,071 & 363,122 \\
sydney & 22,659 & 26,720 & 555,262 \\
\bottomrule
\end{tabular}
}
\end{center}
\end{table}
\endgroup

%% file: FIG/bus_visual_single.tex
\begin{figure*}[t!]    
    \centering
    \begin{subfigure}[b]{0.45\textwidth}
         \centering
         \includegraphics[width=\textwidth]{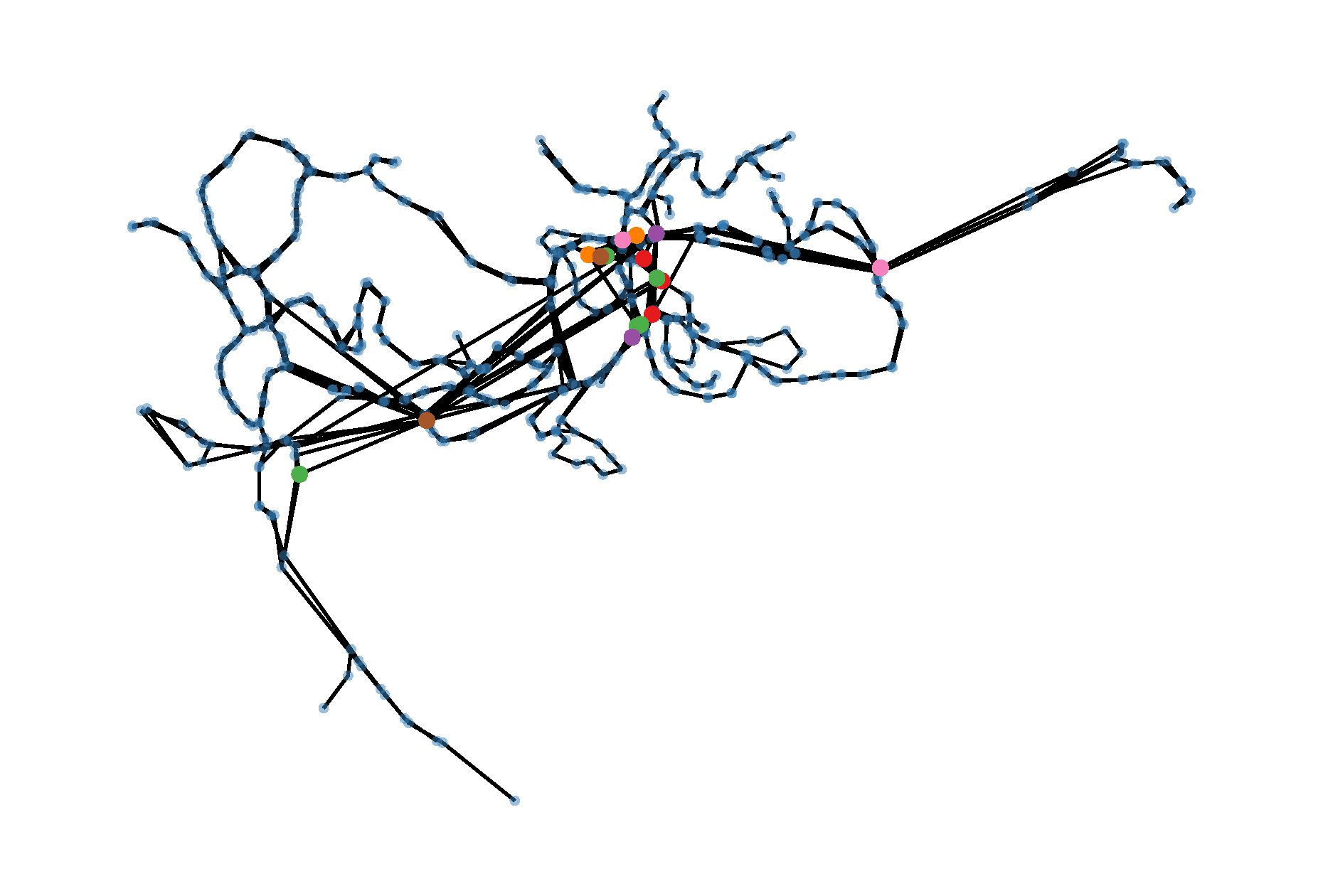}
         \caption{BM (\bmshort)}
     \end{subfigure}
     \hfill
     \begin{subfigure}[b]{0.45\textwidth}
         \centering
         \includegraphics[width=\textwidth]{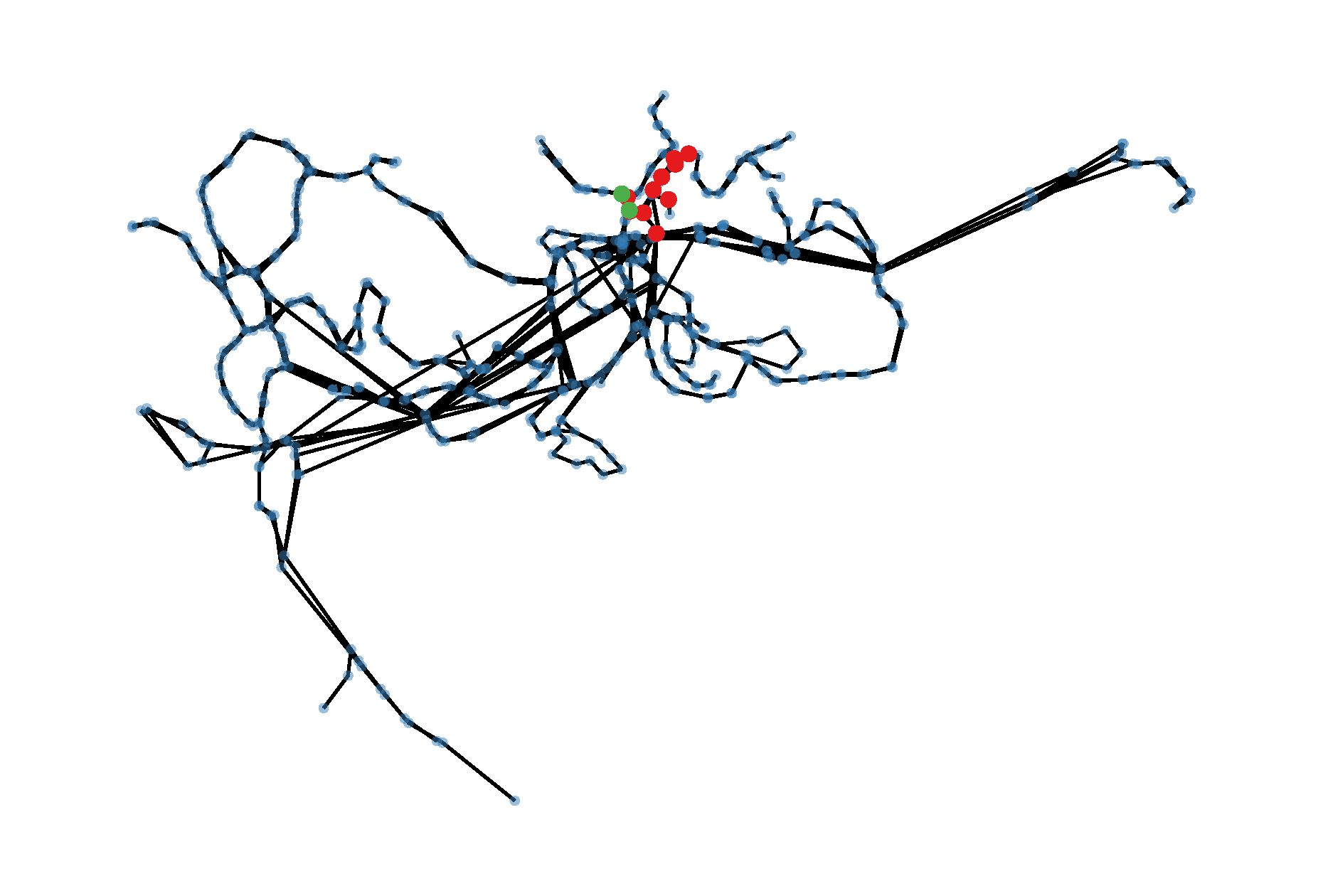}
         \caption{CR (core)}
     \end{subfigure}
     \hfill        
     \begin{subfigure}[b]{0.45\textwidth}
         \centering
         \includegraphics[width=\textwidth]{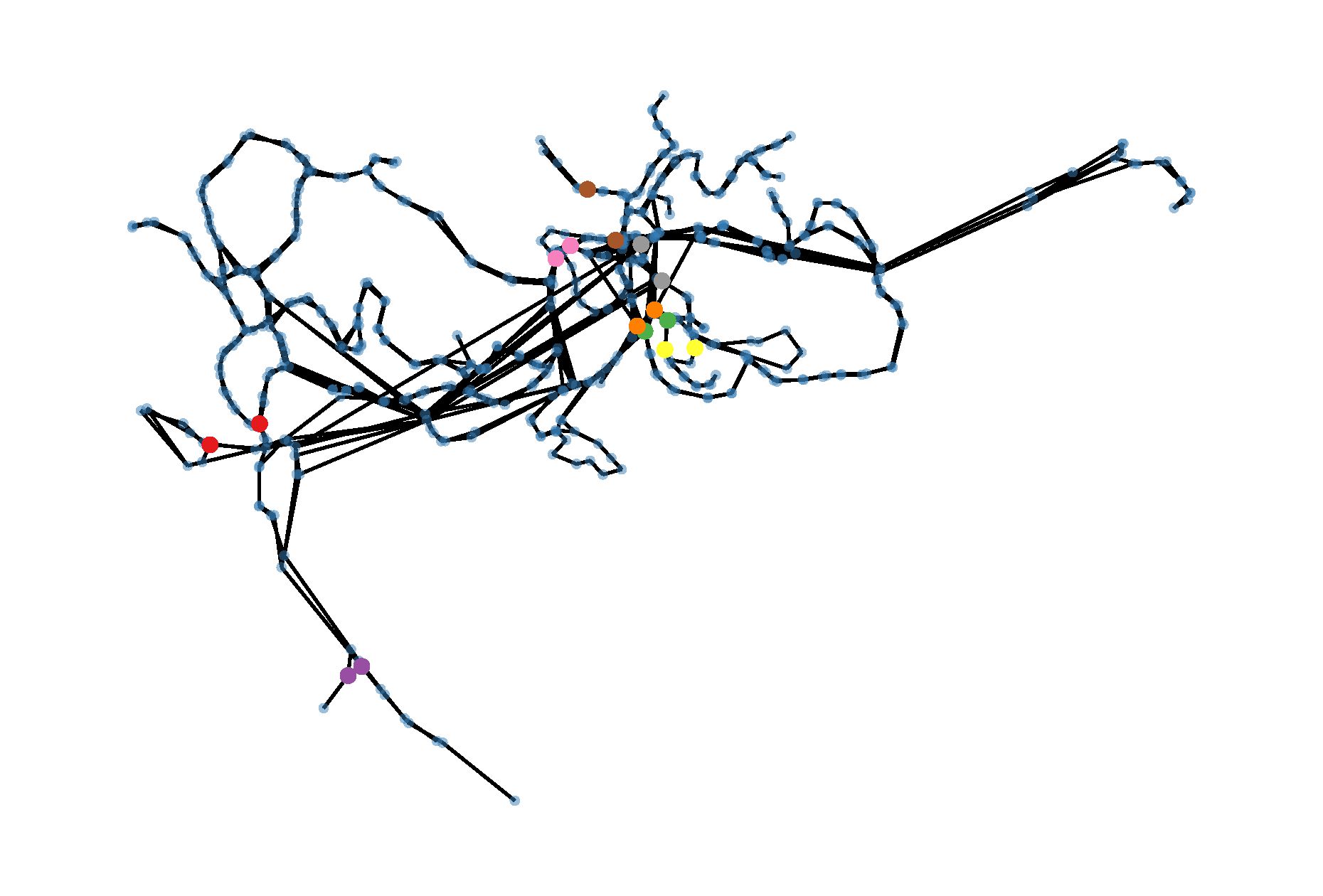}
         \caption{CS (constraints)}         
     \end{subfigure}
     \hfill       
     \begin{subfigure}[b]{0.45\textwidth}
         \centering
         \includegraphics[width=\textwidth]{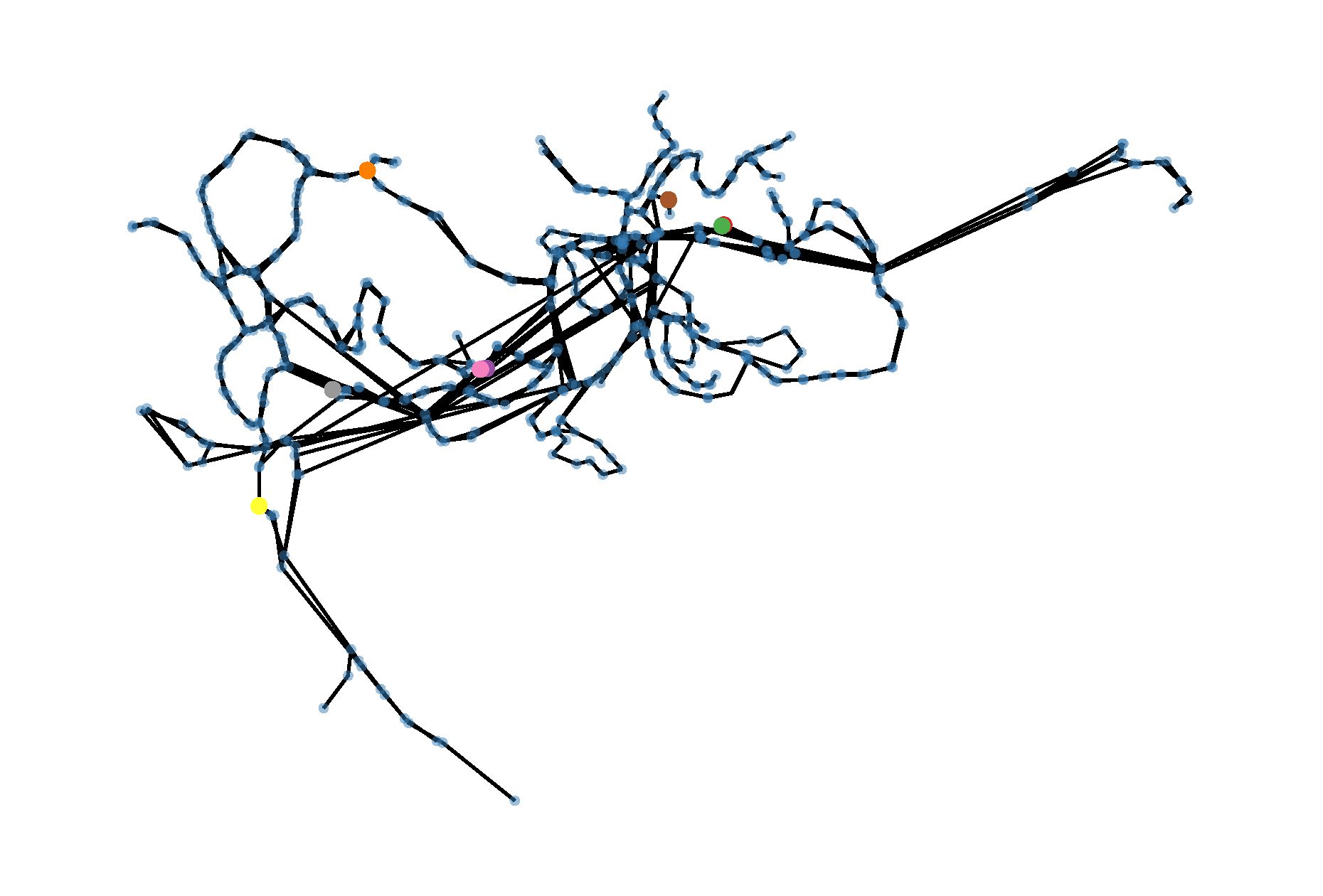}
         \caption{CL (closed)}
     \end{subfigure}
    \caption{The results on the \textit{kuopio} dataset in the additional experiments on real-world bus station datasets.}
    \label{fig:bus_visual_kuopio}
\end{figure*}

\begin{figure*}[t!]    
    \centering
    \begin{subfigure}[b]{0.45\textwidth}
         \centering
         \includegraphics[width=\textwidth]{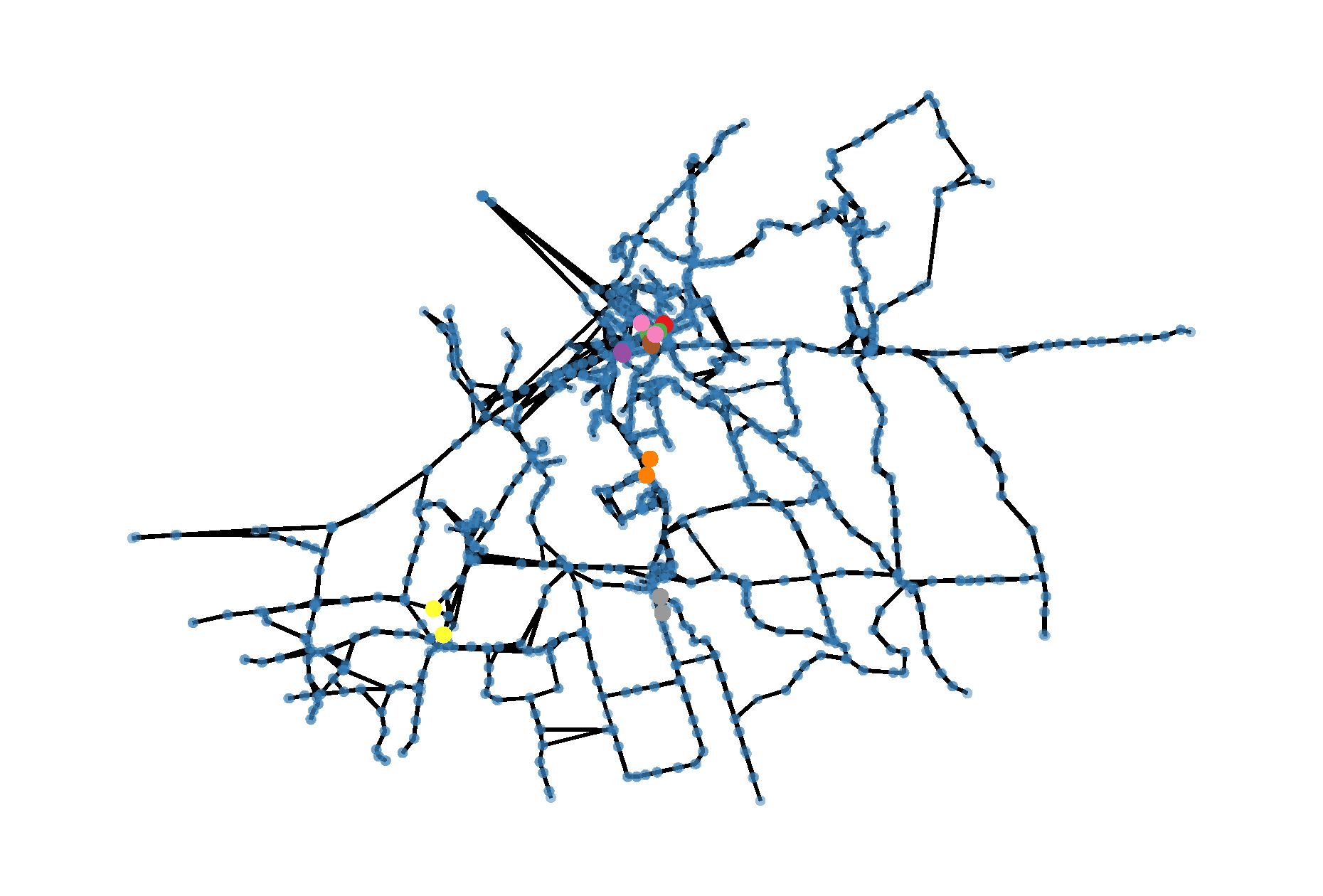}
         \caption{BM (\bmshort)}
     \end{subfigure}
     \hfill
     \begin{subfigure}[b]{0.45\textwidth}
         \centering
         \includegraphics[width=\textwidth]{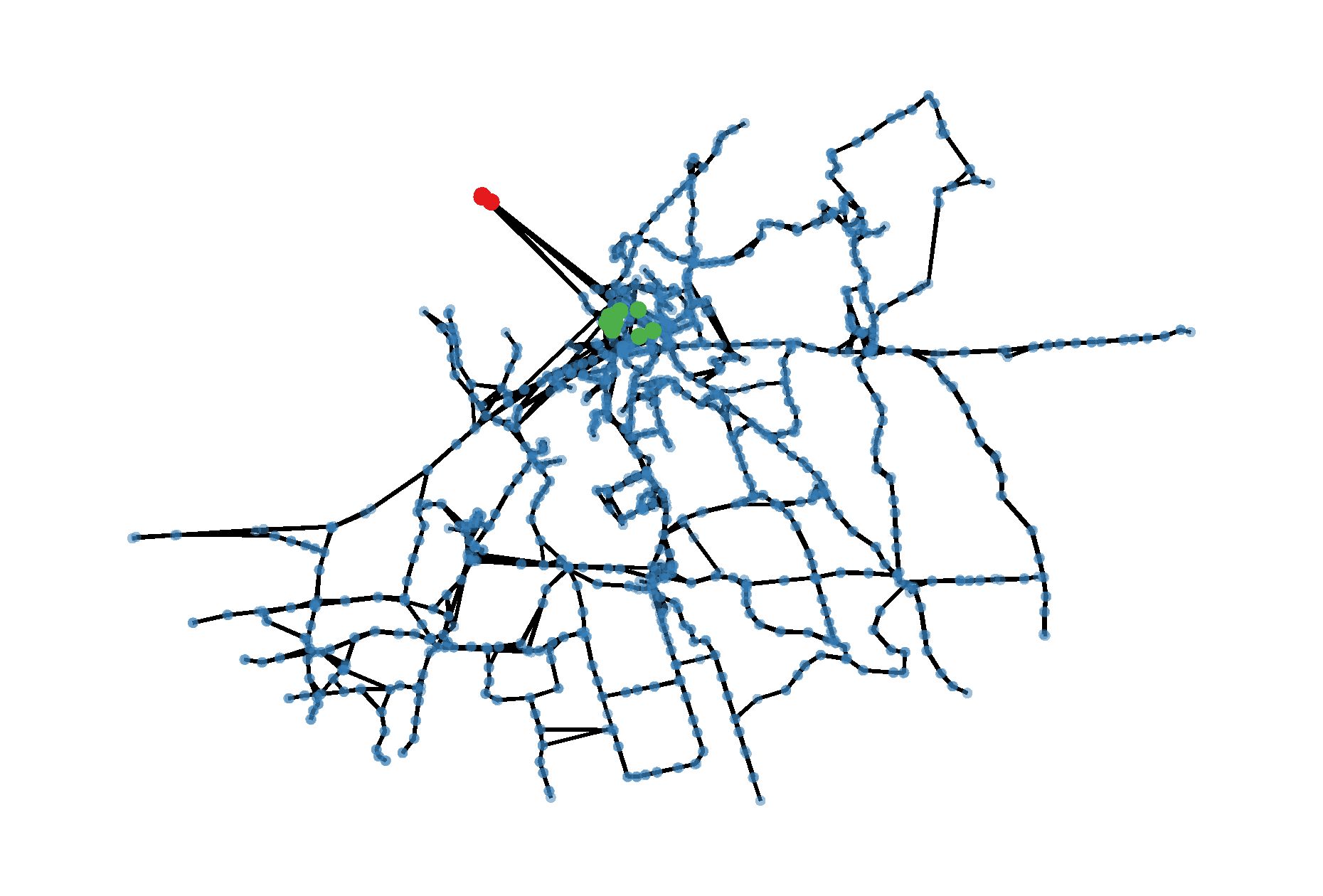}
         \caption{CR (core)}
     \end{subfigure}
     \hfill        
     \begin{subfigure}[b]{0.45\textwidth}
         \centering
         \includegraphics[width=\textwidth]{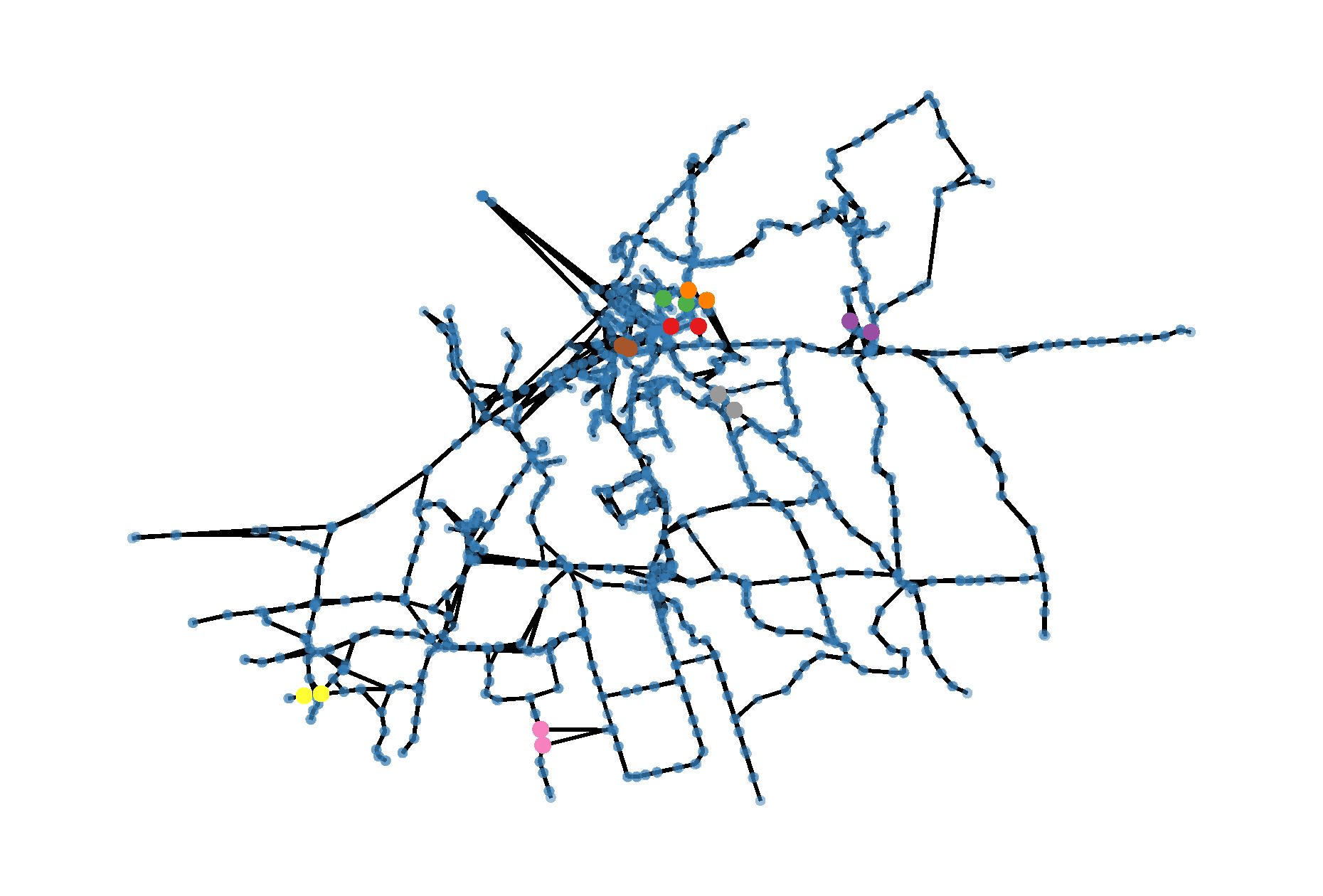}
         \caption{CS (constraints)}         
     \end{subfigure}
     \hfill       
     \begin{subfigure}[b]{0.45\textwidth}
         \centering
         \includegraphics[width=\textwidth]{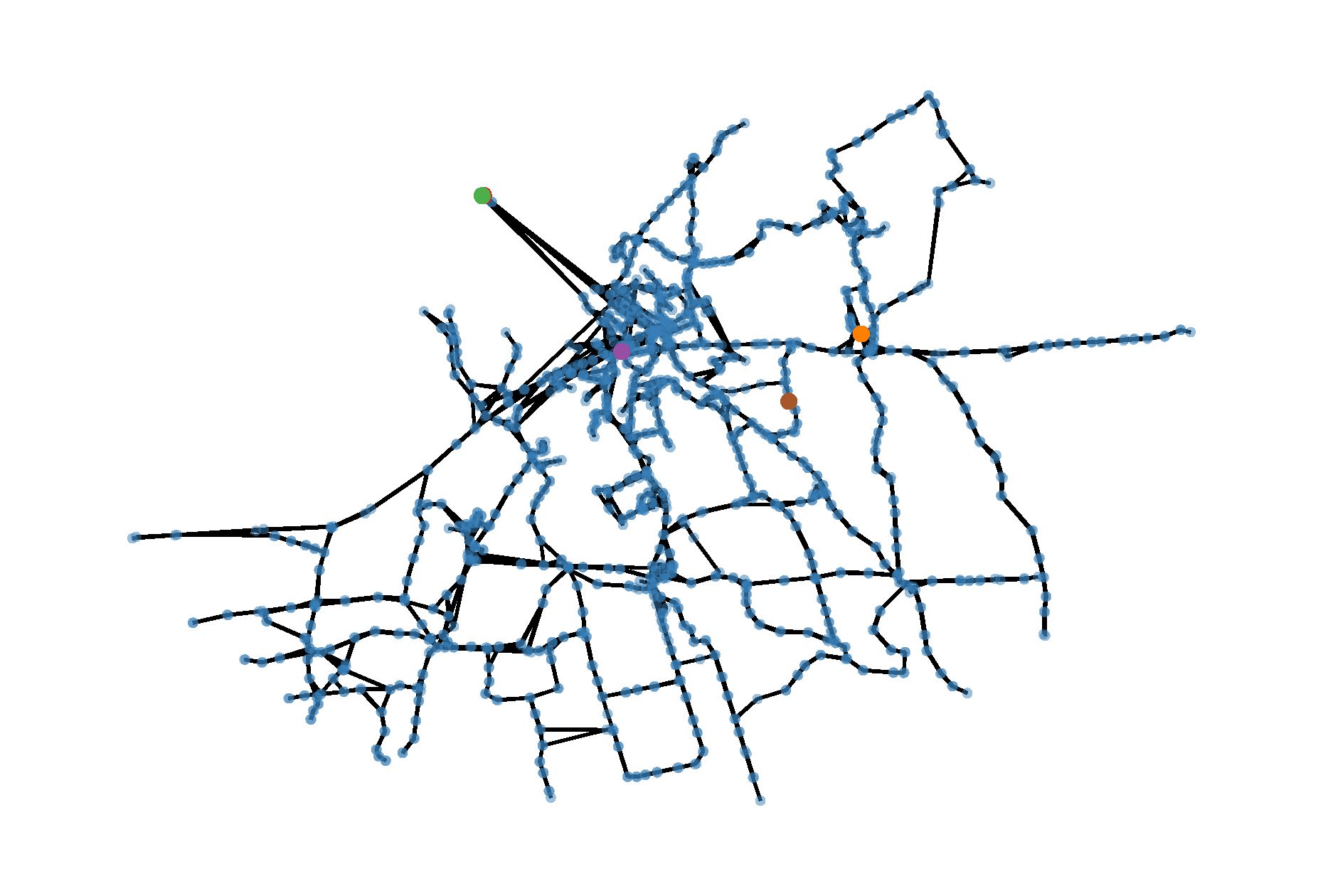}
         \caption{CL (closed)}
     \end{subfigure}
    \caption{The results on the \textit{venice} dataset in the additional experiments on real-world bus station datasets.}
    \label{fig:bus_visual_venice}
\end{figure*}

\begin{figure*}[t!]    
    \centering
    \begin{subfigure}[b]{0.45\textwidth}
         \centering
         \includegraphics[width=\textwidth]{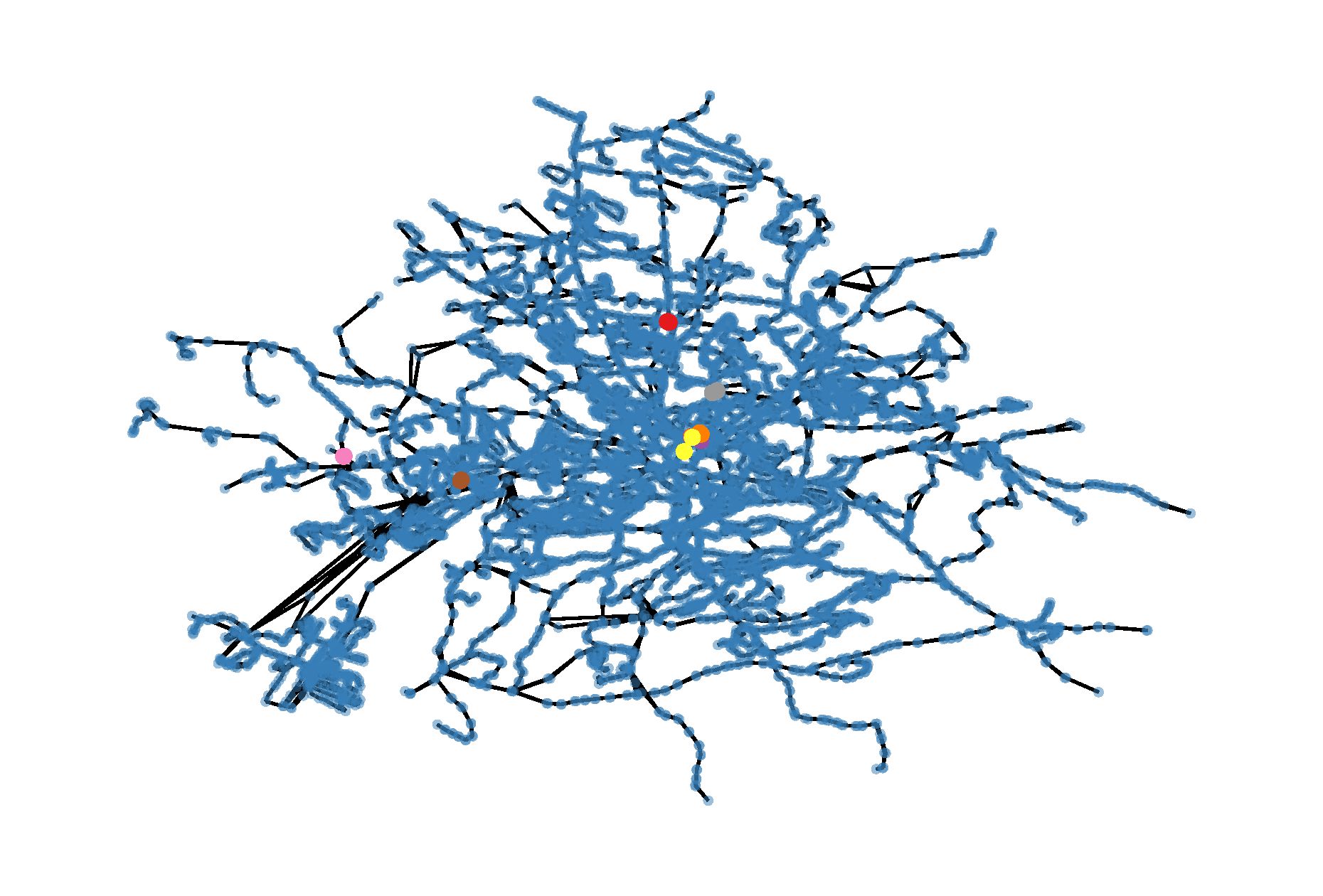}
         \caption{BM (\bmshort)}
     \end{subfigure}
     \hfill
     \begin{subfigure}[b]{0.45\textwidth}
         \centering
         \includegraphics[width=\textwidth]{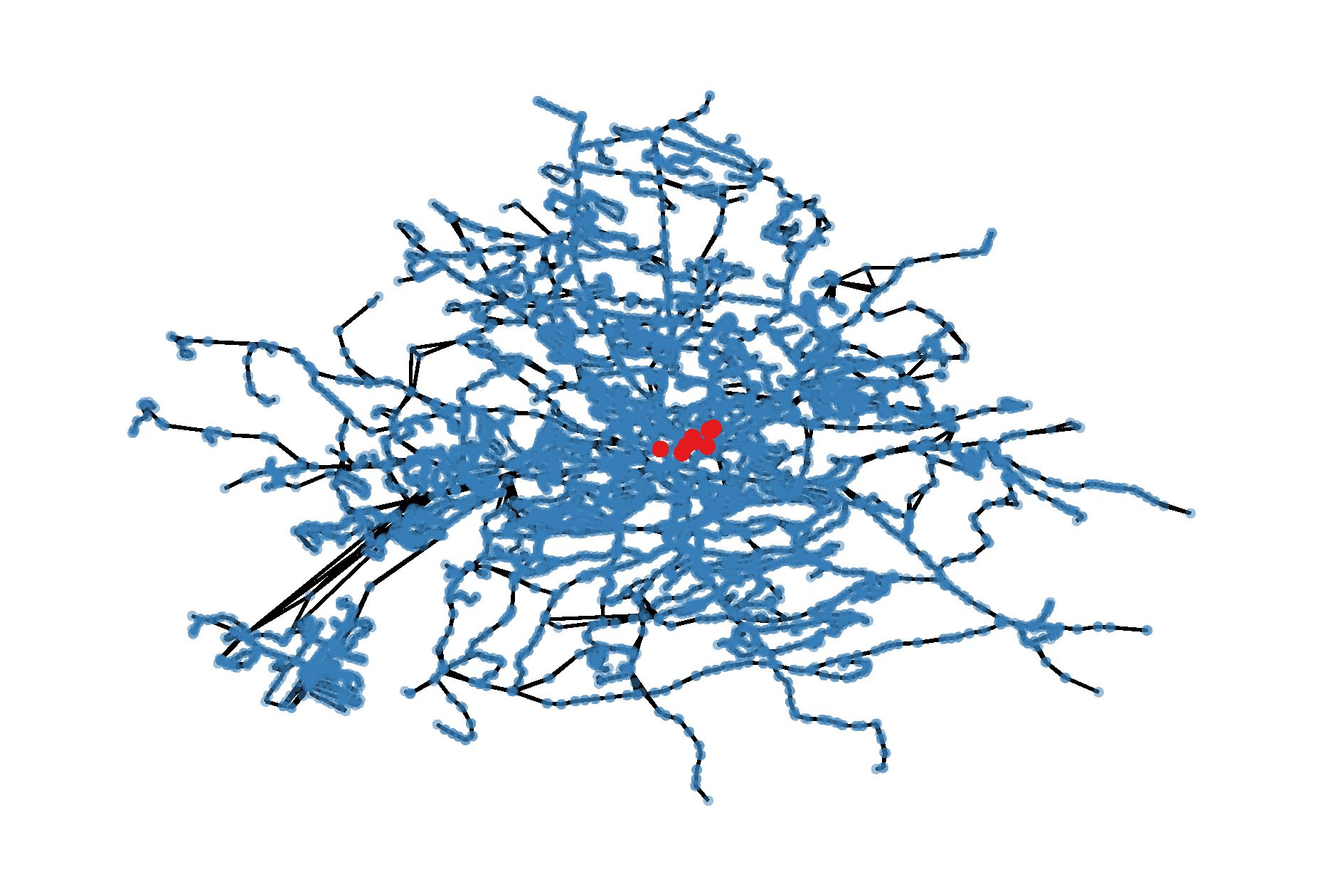}
         \caption{CR (core)}
     \end{subfigure}
     \hfill        
     \begin{subfigure}[b]{0.45\textwidth}
         \centering
         \includegraphics[width=\textwidth]{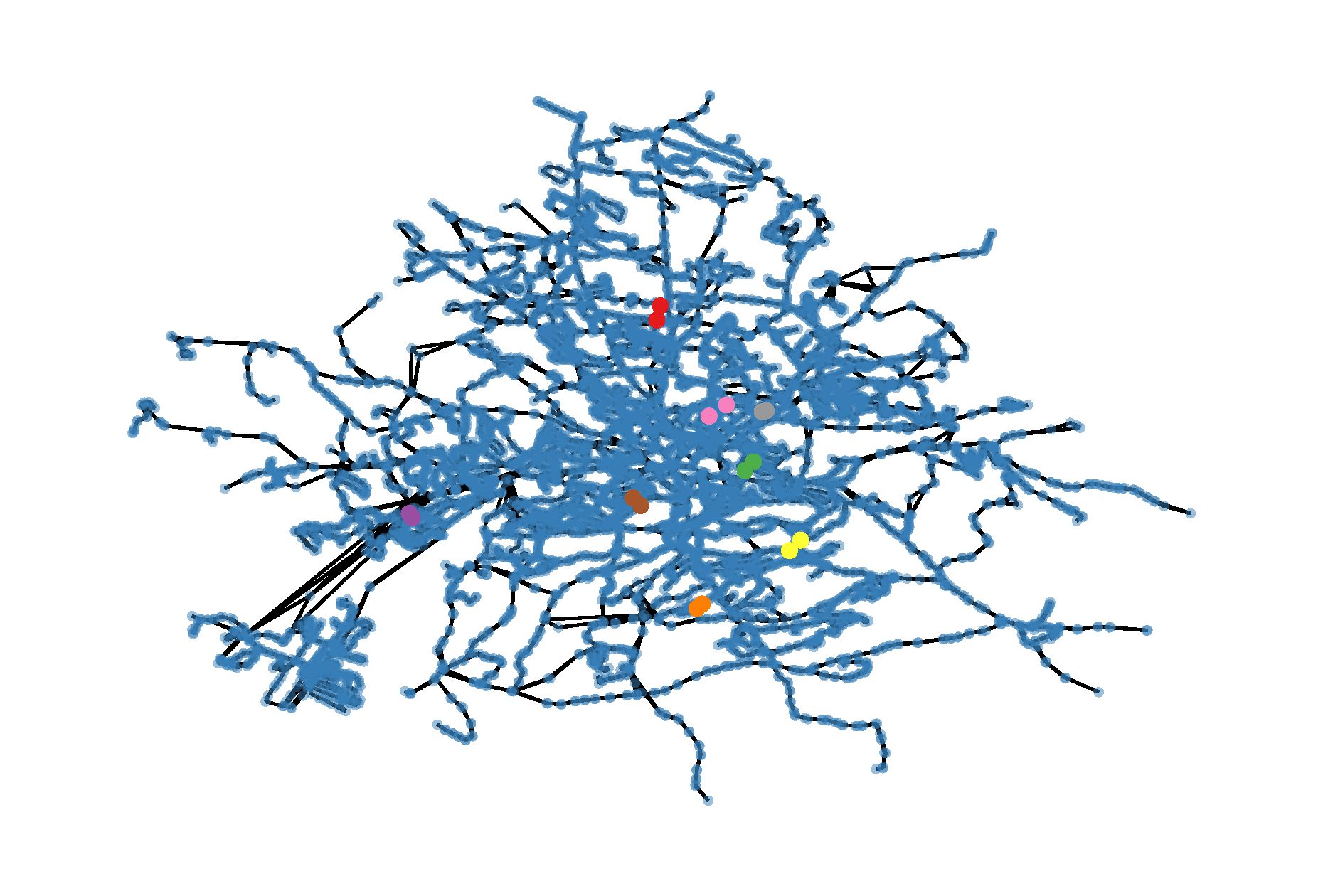}
         \caption{CS (constraints)}         
     \end{subfigure}
     \hfill       
     \begin{subfigure}[b]{0.45\textwidth}
         \centering
         \includegraphics[width=\textwidth]{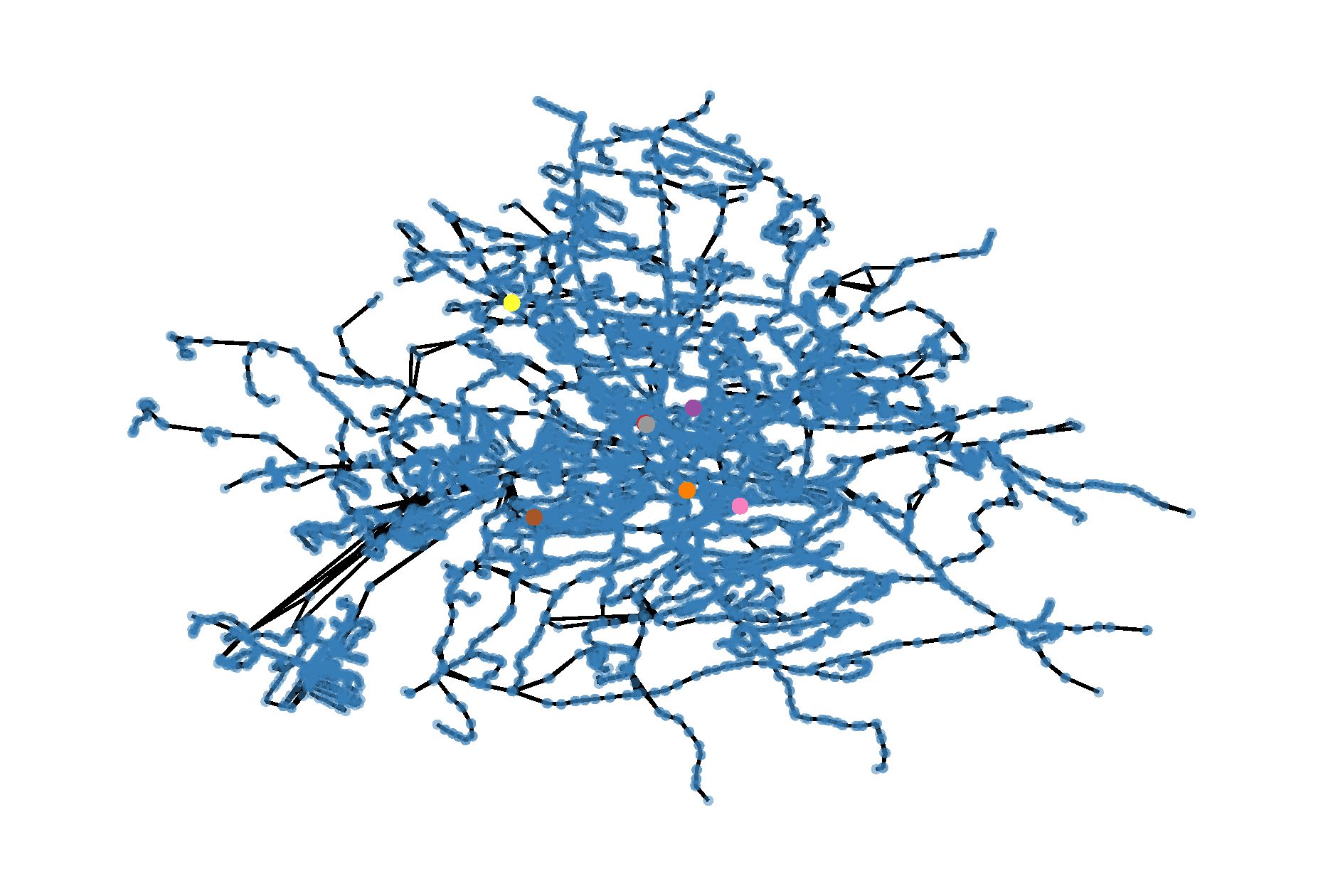}
         \caption{CL (closed)}
     \end{subfigure}
    \caption{The results on the \textit{rome} dataset in the additional experiments on real-world bus station datasets.}
    \label{fig:bus_visual_rome}
\end{figure*}


%% file: TAB/bus_results.tex
\begingroup
\setlength{\tabcolsep}{10pt}
\renewcommand{\arraystretch}{0.85}
\begin{table}[t!]
  \centering
  \caption{The results of additional experiments on real-world bus station datasets measured by the Z-score (the larger the better). The best performance is marked in bold.}
  \scalebox{1.0}{%
    \begin{tabular}{lrrrr}
    \toprule
    metric & {BM} & {CR} & {CS} & {CL} \\
    \midrule
    VB    & 0.2961 & \textbf{0.4953} & 0.2996 & -1.0910 \\
    EB    & 0.2686 & \textbf{0.5820} & 0.3109 & -1.1615 \\
    ER    & -0.8604 & \textbf{0.9090} & 0.6313 & -0.6798 \\
    SG    & \textbf{0.7637} & -0.2035 & -0.2932 & -0.2670 \\
    NC    & \textbf{1.6303} & -0.3605 & -0.5380 & -0.6366 \\
    AD    & -0.2857 & -0.1408 & \textbf{1.2968} & -0.8702 \\
    TS    & \textbf{1.6837} & -0.6469 & -0.3103 & -0.7266 \\
    LC    & \textbf{0.8895} & -0.7762 & 0.8599 & -0.9732 \\
    \midrule
    average   & \textbf{0.5482} & -0.0177 & 0.2821 & -0.8008 \\
    \bottomrule
    \end{tabular}%
    }
  \label{tab:bus_results_z_score}%
\end{table}%

\begin{table}[t!]
  \centering
  \caption{The results of additional experiments on real-world bus station datasets measured by the average rank (the smaller the better). The best performance is marked in bold.}
  \scalebox{1.0}{%
    \begin{tabular}{lrrrr}
    \toprule
    metric & {BM} & {CR} & {CS} & {CL} \\
    \midrule
    VB    & 2.0952 & \textbf{2.0476} & 2.0952 & 3.7619 \\
    EB    & 2.0476 & \textbf{2.0000} & 2.1429 & 3.8095 \\
    ER    & 3.2857 & \textbf{1.6190} & 1.8095 & 3.2857 \\
    SG    & \textbf{1.7619} & 2.5238 & 2.3810 & 2.6667 \\
    NC    & \textbf{1.0000} & 2.3333 & 2.7619 & 2.9048 \\
    AD    & 2.6190 & 2.2381 & \textbf{1.2381} & 3.9048 \\
    TS    & \textbf{1.0000} & 3.3810 & 2.2381 & 3.3810 \\
    LC    & \textbf{1.5714} & 3.2381 & 1.6190 & 3.5238 \\
    \midrule
    average   & \textbf{1.9226} & 2.4226 & 2.0357 & 3.4048 \\
    \bottomrule
    \end{tabular}%
    }
  \label{tab:bus_results_avg_rank}%
\end{table}%

\endgroup

%% file: 0B0concl.tex
\section{Conclusion}\label{sec:concl}
In this work, motivated by real-world scenarios and applications,
we formulated and studied the problem of improving the connectivity and robustness of graphs by merging nodes (Problem~\ref{prob:best_mergers}), 
for which we used the number of edges in the $k$-truss for some given $k$ as the objective.
Then, we proved the NP-hardness and non-submodularity of the problem (Theorems~\ref{thm:NP_hard} and \ref{thm:non_mod}).
For the problem,
based on our theoretical findings regarding mergers between nodes and $k$-trusses (Lemmas~\ref{lem:limited_truss_change}-\ref{lem:maximal_set_enough}),
we proposed \bmshort (Algorithm~\ref{alg:overall}), a fast and effective algorithm equipped with strong search-space-pruning schemes (Algorithms~\ref{alg:iom_heu}-\ref{alg:iim_heu} and \ref{alg:max_sets}) and analyzed its time and space complexity (Theorem~\ref{thm:overall_time}).
Through experiments on real-world graphs, we demonstrated the superiority of \bmshort over several baselines and the effectiveness of every component of \bmshort (Figures~\ref{fig:main_res_ds}-\ref{fig:k_avg}). 
For reproducibility, we made the code and datasets publicly available at \citep{onlineSuppl}.
We plan to consider this problem on weighted/uncertain graphs and explore other cohesive models.

\vspace{1mm}
\smallsection{Acknowledgements.}
This work was supported by National Research Foundation of Korea (NRF) grant funded by the Korea government (MSIT) (No. NRF-2020R1C1C1008296) and Institute of Information \& Communications Technology Planning \& Evaluation (IITP) grant funded by the Korea government (MSIT) (No. 2022-0-00871, Development of AI Autonomy and Knowledge Enhancement for AI Agent Collaboration) (No. 2019-0-00075, Artificial Intelligence Graduate School Program (KAIST)). \\